\documentclass[11pt, oneside]{article}   	
\usepackage{jheppub}
\usepackage{amsmath}
\usepackage{amssymb}
\usepackage{amsfonts}
\usepackage{amsthm}
\usepackage{amsbsy}
\usepackage{array}
\usepackage{mathtools}
\usepackage[usenames,dvipsnames]{xcolor}
\usepackage{subcaption}
\usepackage{dsdshorthand}
\usepackage{graphicx}
\usepackage[vcentermath]{youngtab}

\usepackage{tikz}
\usetikzlibrary{decorations.pathmorphing}
\usetikzlibrary{positioning,decorations.pathreplacing}
\tikzset{snake it/.style={decorate, decoration=snake}}

\newcommand\wL{\mathbf{L}}
\newcommand\wS{\mathbf{S}}
\renewcommand\vol{\mathop{\mathrm{vol}}}

\newcommand{\dDisc}{\text{dDisc}}
\newcommand{\Li}{\text{Li}}

\newcommand{\Jo}{J_{\triangleleft}}  

\makeatletter
\def\@fpheader{\ }
\makeatother

\title{Detectors in weakly-coupled field theories}
\author{Simon Caron-Huot$^1$, Murat Kolo\u{g}lu$^{2,3,4}$, Petr Kravchuk$^{2,5,6,7}$, David Meltzer$^{2,8}$, David Simmons-Duffin$^2$}
\affiliation{
${}^1$Department of Physics, McGill University, 3600 Rue University, Montréal, QC Canada \\
${}^2$Walter Burke Institute for Theoretical Physics, Caltech, Pasadena, California 91125, USA \\
${}^3$Mathematical Institute, University of Oxford, Oxford, OX2 6GG, UK \\
${}^4$Department of Physics, Yale University, New Haven, CT 06520, USA\\
${}^5$School of Natural Sciences, Institute for Advanced Study, Princeton, New Jersey 08540, USA \\
${}^6$Simons Center for Geometry and Physics, Stony Brook University, Stony Brook, NY
11794, USA\\
${}^7$Department of Mathematics, King's College London, Strand, London, WC2R 2LS, UK
\\
${}^8$Department of Physics, Cornell University, Ithaca, NY 14850, USA
}

\emailAdd{schuot@physics.mcgill.ca}
\emailAdd{murat.kologlu@yale.edu}
\emailAdd{petr.kravchuk@kcl.ac.uk}
\emailAdd{dm694@cornell.edu}
\emailAdd{dsd@caltech.edu}

\date{}
\abstract{We initiate a study of asymptotic detector operators in weakly-coupled field theories. These operators describe measurements that can be performed at future null infinity in a collider experiment. In a conformal theory they can be identified with light-ray operators, and thus have a direct relation to the spectrum of the theory. After a general discussion of the underlying physical picture, we show how infrared divergences of general detector operators can be renormalized in perturbation theory, and how they give rise to detector anomalous dimensions. We discuss in detail how this renormalization can be performed at the intersections of the Regge trajectories where non-trivial mixing occurs, which is related to the poles in anomalous dimensions at special values of spin. Finally, we discuss novel horizontal trajectories in scalar theories and show how they contribute to correlation functions. Our calculations are done in the example of $\f^4$ theory in $d=4-\e$ dimensions, but the methods are applicable more broadly. At the Wilson-Fisher fixed point our results include an explicit expression for the Pomeron light-ray operator at two loops, as well as a prediction for the value of the Regge intercept at five loops.}

\preprint{CALT-TH 2022-031}

\renewcommand\beforetochook{\pagestyle{myplain}\pagenumbering{roman}\renewcommand{\baselinestretch}{0.947}}

\begin{document}

\maketitle
\renewcommand{\baselinestretch}{1}
\pagenumbering{roman}
\setcounter{page}{2}
\newpage
\pagenumbering{arabic}
\setcounter{page}{1}

\section{Introduction}
In quantum field theory, the theory itself dictates which observables are well-defined. For example, consider a local operator in a perturbative field theory,
like the product $\mathcal{O}_{\rm bare}=\phi(x)^2$ in $\lambda \phi^4$ theory.
When calculating with this operator, one quickly finds that the bare operator is ultraviolet divergent, and is not itself a good observable, even after renormalizing the couplings of the theory.
To obtain finite quantities, one must choose an ultraviolet cutoff and define a renormalized operator $\mathcal{O}_\mathrm{ren}\equiv Z \mathcal{O}_{\rm bare}$,
where $Z$ is a wave-function renormalization factor that cancels the divergences.
The renormalized operator $\mathcal{O}_\mathrm{ren}$ is then a ``good'' observable, interpreted as a measurement of the square of $\phi$ smeared over a region of size the renormalization scale.
Furthermore, its scaling dimension, and thus its units, depend on the dynamics of the theory.

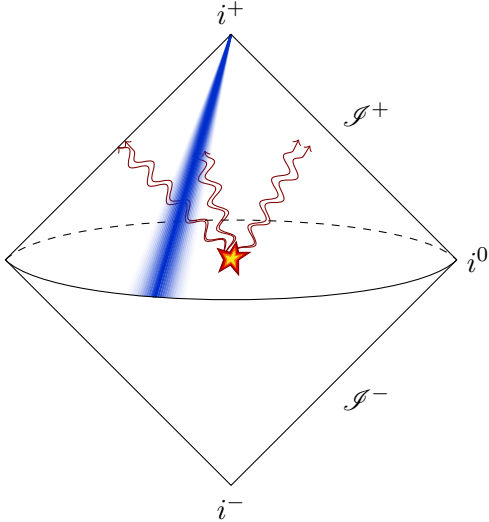
\begin{figure}[h!]
\centering
\begin{tikzpicture}
	\draw [dashed] (-3,0) to[out=45,in=135,distance=1.0cm] (3,0);
	\draw [black!50!red,snake it,->] (0.05,0) -- (-0.35,1.45);
	\draw [black!50!red,snake it,->] (0.05,0) -- (-0.48,1.45);
	\draw [black!50!red,snake it,->] (0.1,0) -- (1.05,1.52);
	\draw [black!50!red,snake it,->] (0.1,0) -- (0.9,1.6);
	\draw [black!50!red,snake it,->] (0.05,0) -- (-1.5,1.5);
	\draw [black!50!red,snake it,->] (0.05,0) -- (-1.41,1.58);
	\draw [black!50!red,fill=black!50!red] (0.1,0) -- (0.25,0.1) -- (0.1,0.1) -- (0.07,0.24) -- (-0.02,0.1) -- (-0.18,0.15) -- (-0.1,0) -- (-0.16,-0.16) -- (-0.01,-0.07) -- (0.1,-0.2) -- (0.1,0);
	\draw [black!10!red,fill=black!10!red] (0.08,0) -- (0.19,0.085) -- (0.08,0.08) -- (0.06,0.20) -- (-0.01,0.08) -- (-0.15,0.12) -- (-0.08,0) -- (-0.12,-0.12) -- (-0.00,-0.05) -- (0.085,-0.15) -- (0.08,0);
	\draw [black!5!orange,fill=black!5!orange] (0.06,0) -- (0.14,0.065) -- (0.06,0.06) -- (0.05,0.15) -- (-0.0,0.06) -- (-0.10,0.08) -- (-0.06,0) -- (-0.08,-0.08) -- (-0.00,-0.03) -- (0.06,-0.10) -- (0.06,0);
	\draw [black!5!yellow,fill=black!5!yellow] (0.03,0) -- (0.08,0.045) -- (0.04,0.04) -- (0.04,0.1) -- (-0.0,0.03) -- (-0.06,0.05) -- (-0.04,0) -- (-0.05,-0.043) -- (-0.00,-0.01) -- (0.043,-0.055) -- (0.03,0);
	\draw [thick,black!10!green!20!blue,opacity=0.02] (-0.61,-0.508) -- (0,3);
	\draw [thick,black!10!green!20!blue,opacity=0.02] (-1.39,-0.48) -- (0,3);
	\draw [thick,black!10!green!20!blue,opacity=0.04] (-0.64,-0.508) -- (0,3);
	\draw [thick,black!10!green!20!blue,opacity=0.04] (-1.36,-0.48) -- (0,3);
	\draw [thick,black!10!green!20!blue,opacity=0.08] (-0.67,-0.508) -- (0,3);
	\draw [thick,black!10!green!20!blue,opacity=0.08] (-1.33,-0.48) -- (0,3);
	\draw [thick,black!10!green!20!blue,opacity=0.12] (-1.3,-0.48) -- (0,3);
	\draw [thick,black!10!green!20!blue,opacity=0.12] (-0.70,-0.508) -- (0,3);
	\draw [thick,black!10!green!20!blue,opacity=0.16] (-1.27,-0.48) -- (0,3);
	\draw [thick,black!10!green!20!blue,opacity=0.16] (-0.73,-0.508) -- (0,3);
	\draw [thick,black!10!green!20!blue,opacity=0.23] (-1.24,-0.482) -- (0,3);
	\draw [thick,black!10!green!20!blue,opacity=0.23] (-0.76,-0.506) -- (0,3);
	\draw [thick,black!10!green!20!blue,opacity=0.3] (-1.21,-0.484) -- (0,3);
	\draw [thick,black!10!green!20!blue,opacity=0.3] (-0.79,-0.504) -- (0,3);
	\draw [thick,black!10!green!20!blue,opacity=0.4] (-1.18,-0.486) -- (0,3);
	\draw [thick,black!10!green!20!blue,opacity=0.4] (-0.82,-0.502) -- (0,3);
	\draw [thick,black!10!green!20!blue,opacity=0.5] (-1.15,-0.488) -- (0,3);
	\draw [thick,black!10!green!20!blue,opacity=0.5] (-0.85,-0.5) -- (0,3);
	\draw [thick,black!10!green!20!blue,opacity=0.6] (-1.12,-0.49) -- (0,3);
	\draw [thick,black!10!green!20!blue,opacity=0.6] (-0.88,-0.5) -- (0,3);
	\draw [thick,black!10!green!20!blue,opacity=0.7] (-1.09,-0.49) -- (0,3);
	\draw [thick,black!10!green!20!blue,opacity=0.7] (-0.91,-0.5) -- (0,3);
	\draw [thick,black!10!green!20!blue,opacity=0.8] (-1.06,-0.49) -- (0,3);
	\draw [thick,black!10!green!20!blue,opacity=0.8] (-0.94,-0.5) -- (0,3);
	\draw [thick,black!10!green!20!blue,opacity=0.9] (-1.04,-0.5) -- (0,3);
	\draw [thick,black!10!green!20!blue,opacity=0.9] (-0.97,-0.5) -- (0,3);
	\draw [thick,black!10!green!20!blue] (-1.02,-0.5) -- (0,3);
	\draw [thick,black!10!green!20!blue] (-1.0,-0.5) -- (0,3);
	\draw [] (-3,0) -- (0,3) -- (3,0) -- (0,-3) -- cycle;
	\draw [] (-3,0) to[out=-45,in=-135,distance=1.0cm] (3,0);
	\node [above] at (1.8,1.65) {$\mathscr{I}^+$};
	\node [below] at (1.8,-1.55) {$\mathscr{I}^-$};
	\node [above] at (0,3) {$i^+$};
	\node [below] at (0,-3) {$i^-$};
	\node [right] at (3,0) {$i^0$};
\end{tikzpicture}
\caption{A detector (blue) is a translationally-invariant operator localized at future null infinity $\mathscr{I}^+$, capable of measuring properties of a state at late times. Some detectors, such as the average null energy operator $\cE_2(\vec n)$, are localized on a light-ray. Other detectors can measure nontrivial angular distributions on the celestial sphere, as indicated by ``fuzziness" in the angular directions in the figure.\label{fig:pictureofdetector}}
\end{figure}

In this work, we explore an analogous story involving ``detectors'' in collider experiments. A local operator is, roughly speaking, ``anything one can measure at a point."
A detector is, roughly speaking, ``anything one can measure in scattering cross-sections," 
see figure~\ref{fig:pictureofdetector}.
Just as the space of local operators is determined by the dynamics of a theory, the space of detectors is determined by the theory as well.

In perturbation theory, a generic bare detector suffers from infrared (IR) divergences. For example, consider a detector $\cE_J(\vec n)$ that counts particles propagating in the direction $\vec n\in S^{d-2}$ on the celestial sphere, weighted by a power of their energy $E^{J-1}$. This observable is not IR safe when $J\neq 2$, since soft and collinear radiation conserves energy but not powers of energy.
The lack of IR-safety manifests as IR/collinear divergences in perturbation theory.
After suitably renormalizing the detector to remove the divergences, we obtain a new ``good" observable,
but its anomalous dimension (suitably-defined) is theory-dependent.

\begin{figure}
\centering
\begin{subfigure}[t]{0.5\textwidth}
\includegraphics[width=\textwidth]{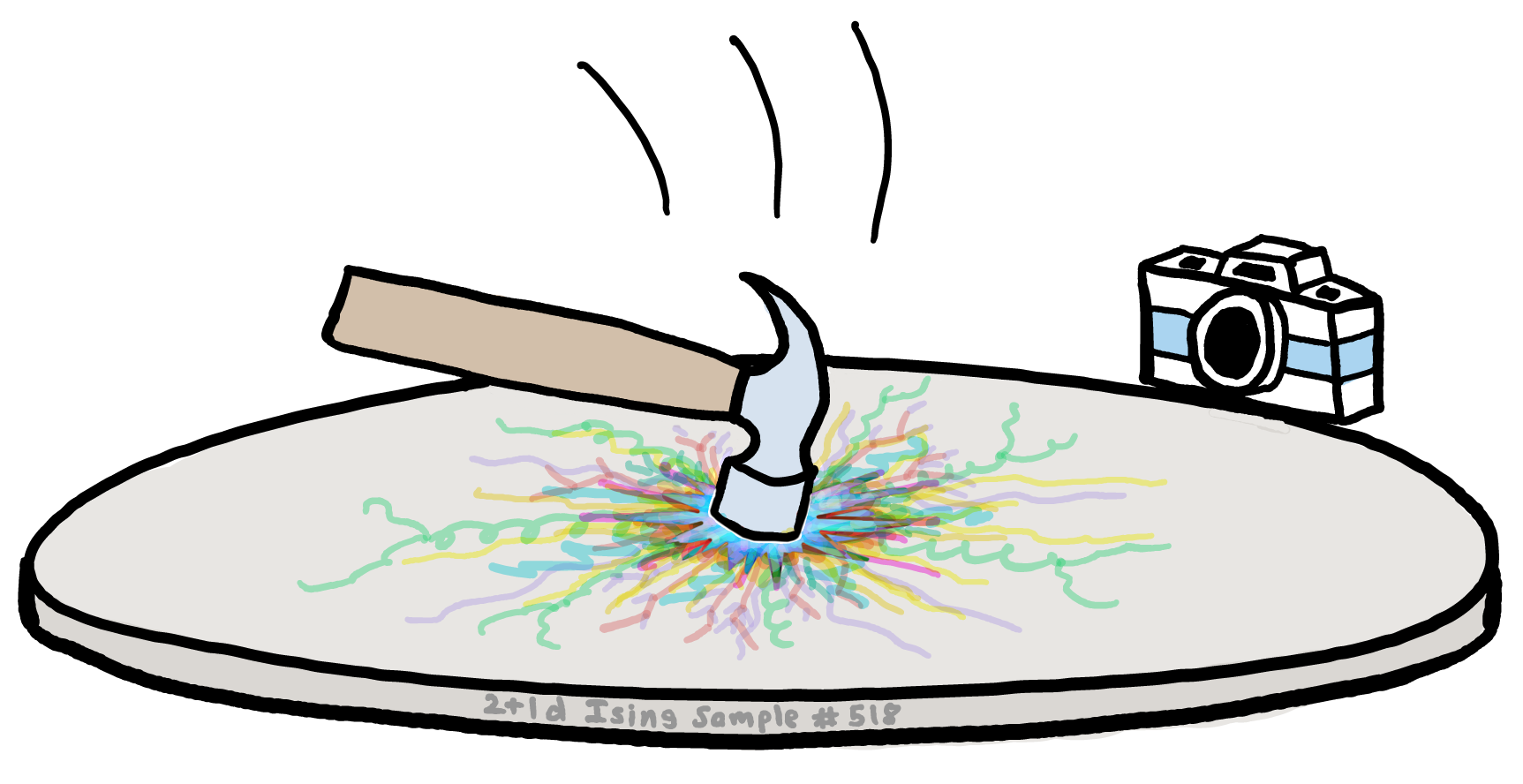}
\end{subfigure}
\begin{subfigure}{0.43\textwidth}
\vspace{-35mm}
\begin{align}
\raisebox{-4.5mm}{\includegraphics[width=0.3\textwidth]{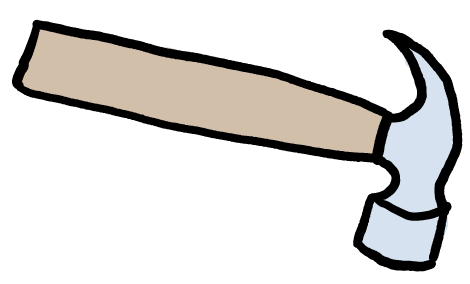}}\  &=\  \sum_i h_i \cO_i \label{eq:hammer}\\
\raisebox{-4mm}{\includegraphics[width=0.2\textwidth]{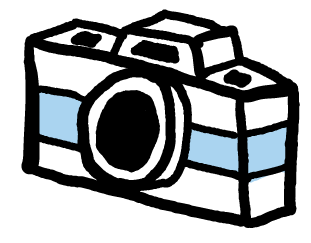}} \  &=\  \sum_j c_j \cD_j \label{eq:camera}
\end{align}
\end{subfigure}
\caption{In an experiment involving a QFT $\cQ$, the probe (hammer) can be expanded in operators $\cO_i$ that are intrinsic to $\cQ$, schematically eq.~(\ref{eq:hammer}). Similarly, a far-away measurement apparatus (camera) can be expanded in detectors $\cD_i$ that are intrinsic to $\cQ$, eq.~(\ref{eq:camera}). These twin expansions cleanly separate the details of the experiment (contained in the coefficients $h_i$ and $c_j$) from the dynamics of the theory (encoded in matrix elements $\<\cO_i|\cD_j|\cO_k\>$).\label{fig:eventshapecartoon}}
\end{figure}

Recall that the space of {\it local\/} operators has a simple nonperturbative definition via radial quantization in the UV CFT: it is its Hilbert space of states on $S^{d-1}$. Thus, local operators provide a basis of fundamental objects in which measurements at a point can be expanded. Similarly, detectors provide a basis of fundamental objects in which measurements near infinity can be expanded, see figure~\ref{fig:eventshapecartoon}.
However, we do not currently possess a similarly clean nonperturbative definition of the space of detectors.
They are less well-understood objects, and we seek to explore them in this work, focusing mostly on the case of conformal theories.
We summarize the analogy between detectors and local operators in table~\ref{table:local vs detector comparison}.

\begin{table}
\centering
\begin{tabular}{c|c}
local operator & detector \\
\hline
``measure at a point" & ``measure in cross-sections" \\
UV divergence & IR divergence \\
need to renormalize & need to renormalize \\
theory-dependent & theory-dependent \\
OPE & light-ray OPE \\
radial quantization & ?
\end{tabular}
\caption{A comparison between local operators and detectors. 
}
\label{table:local vs detector comparison}
\end{table}

The simplest kind of detector is the integral of a local operator along a light-ray at future null infinity $\mathscr{I}^+$.
In this case, the way renormalization works is easy to understand: the renormalized detector is the null integral of a renormalized local operator.
For example, in a free scalar theory, the operator $\cE_J$ just mentioned can be defined for even integer $J\geq 2$ as a null-integral of $\cO_J=\phi \ptl_{\mu_1}\cdots\ptl_{\mu_J}\phi$.
When interactions are turned on, $\cO_J$ gets an anomalous dimension, and thus so does $\cE_J$, leading to a nontrivial dependence on
infrared scales characterizing the measurement.

However, there exist detectors that are more general than simply the null integral of a local operator. For example, since energy is positive, we can consider $\cE_J$ for general complex $J$.
This cannot be written as the null integral of a local operator. Instead, it is a so-called ``light-ray operator'' $\mathbb{O}_J$ for the leading Regge trajectory \cite{Kravchuk:2018htv}.
In fact, such general light-ray operators appear naturally in the OPE of more conventional detectors like the average null energy operator $\cE_2$ \cite{Hofman:2008ar,Kologlu:2019mfz,Chang:2020qpj,Cordova:2018ygx}. 

Light-ray operators have a long history in quantum field theory.
Perhaps the most familiar example are parton distribution functions.
Their integer moments are equal to hadronic matrix elements of operators analogous to $\cO_J$ \cite{Christ:1972ms,Gross:1973ju,Georgi:1974wnj},
while parton distribution functions themselves are matrix elements of nonlocal light-ray operators \cite{Collins:1981uw,Balitsky:1987bk}.
In addition, detectors/light-ray operators appear in the Regge limit of correlation functions.
In gauge theories, null Wilson lines and their products realize the BFKL Pomeron \cite{Kuraev:1977fs,Balitsky:1978ic}
and control the leading behavior of correlators at large boost \cite{Mueller:1994jq,Balitsky:1995ub,Caron-Huot:2013fea}.

If our theory has a mass gap and trivial infrared dynamics, its detectors are simple to characterize:
they observe stable particles that propagate to infinity.
Without a mass gap, however, a typical event is accompanied with a burst of energy moving at the speed of light.
Focusing on conformal theories will allow us to better understand this signal when it is nontrivial.

A simplification in conformal theories is that there is nothing special about infinite distances:
future null infinity can be conformally mapped to a flat null sheet.
The space of detectors and the space of light-ray operators are thus equivalent.
This ``spacelike-timelike'' correspondence has found a number of applications, for example \cite{Hofman:2008ar,Hatta:2008st,Caron-Huot:2015bja,Vladimirov:2016dll,Mueller:2018llt,Dixon:2019uzg}.
In this paper we will thus use the terms ``detector" and ``light-ray operator" interchangeably.
However, note that some of our calculations, for example in section~\ref{sec:pomeron}, do not assume conformal symmetry.

One of our first tasks in this work will be to de-mystify light-ray operators/detectors like $\mathbb{O}_J$ by providing tools for defining them and computing with them directly in perturbation theory. We work with perhaps the simplest perturbative CFT: the Wilson-Fisher theory in $4-\e$ dimensions.

Light-ray operators can be classified according to their dimension $\De$ and spin $J$ \cite{Brower:2006ea,Costa:2012cb,Kravchuk:2018htv}.
The one that dominates in the Regge limit is the so-called ``Pomeron", which is the light-ray operator with the largest spin $J$ along the principal series $\De\in \frac d 2 + i \R$.
In this work, we will provide an explicit expression for the Pomeron of the Wilson-Fisher theory.
Other light-ray operators with smaller $J$ control subleading corrections in the Regge limit.
An important class of subleading corrections in the Wilson-Fisher theory come from novel ``horizontal trajectories" which we also explore.
These will be somewhat analogous to the null Wilson lines that appear in gauge theories.

This work is organized as follows. We begin in section~\ref{sec:chewfrautschi} with an extended introduction to detectors/light-ray operators and their appearance in the Wilson-Fisher theory. This section contextualizes and summarizes the rest of the paper. We also show that the known anomalous dimensions in Wilson-Fisher theory are sufficient to compute the Pomeron spin (the Regge intercept) up to and including $O(\e^4)$ terms, which corresponds to 5 loops. In section~\ref{sec:warmup twist2}, we describe the leading-twist Regge trajectory in the Wilson-Fisher theory in the language of detectors, recovering known results for its anomalous dimension from a different perspective. In section~\ref{sec:pomeron}, we apply this perspective to describe the mixing between the leading-twist trajectory and its so-called ``shadow," allowing us to identify the Pomeron of the Wilson-Fisher theory. In section~\ref{sec:WF horizontal}, we construct and explore some simple horizontal trajectories in the Wilson-Fisher theory. We conclude with discussion and future directions in section~\ref{sec:discussion}.

\section{The space of detectors in the Wilson-Fisher theory}
\label{sec:chewfrautschi}

In this section, we work out the main features of the space of detectors in the Wilson-Fisher theory.
We start by studying detectors for a free massless scalar $\phi$ in $d$ spacetime dimensions,
and then subsequently add a $\phi^4$ interaction and tune to the critical point. We will find hints about the space of detectors by studying singularities of anomalous dimensions as a function of $J$.
Surprisingly, it will turn out that light-ray operators are continuously connected with operators supported on full lightcones.

\subsection{$E^{J-1}$ flux in the free scalar theory}

Let us first define an operator $\mathcal{E}_2(\bn)$ that measures the energy flux in a spatial direction $\bn$ for a free massless scalar.
We start by expanding $\phi$ in creation and annihilation operators,
\be
	\f(x)=\int_{p^0>0} \frac{d^dp}{(2\pi)^{d-1}}\de(p^2)\p{a^\dagger(p) e^{-ipx}+a(p)e^{ipx}},
\ee
where we use a relativistic normalization where $a(p)$ is a Lorentz scalar.\footnote{Compared with the canonical normalization,
$a(p)=\sqrt{2p^0}a_{\text{canonical}}(\bp)$. The commutation relation is $[a(p),a^\dagger(p')]= 2p^0 (2\pi)^{d-1}\delta^{d-1}(\bp-\bp')$.}
The Hamiltonian in this convention is
\be
	H=\half\int d^{d-1}x:\!\p{(\ptl_0\f)^2+(\ptl_i\f)^2}\!:\,\propto \int d^{d-1}\bp\ a^\dagger(\bp)a(\bp),
\ee
where $a(\bp)\equiv a(p=(|\bp|,\bp))$.  We will not be concerned with the overall normalization of operators in this section, so we use a proportionality sign. The energy flux $\cE_2(\bn)$ should integrate to the Hamiltonian
\be
	\int_{|\bn|=1} d^{d-2}\bn\, \cE_2(\bn)=H.
\ee
Furthermore, $\cE_2(\bn)$ should only involve  creation and annihilation operators with momentum $\mathbf{p}$~in the direction $\bn$. Hence, we find
\be
\label{eq:etwodefinition}
	\cE_2(\bn)\propto\int_0^\oo dE\, E^{d-2} a^\dagger(E\bn)a(E\bn).
\ee

In the free theory, we can define more general operators $\cE_J(\bn)$ that measure the flux of powers of energy $E^{J-1}$ by inserting an additional factor of $E^{J-2}$ into the integral:
\be\label{eq:EJnon-cov}
	\cE_J(\bn)\propto\int_0^\oo dE\, E^{J+d-4} a^\dagger(E\bn)a(E\bn).
\ee
A nice property of $\cE_J$ is that it transforms in a simple way under Lorentz transformations. To see this, let us covariantize the expression (\ref{eq:EJnon-cov}). For a future-pointing null vector $z$, we set
\be\label{eq:EfromA}
	\cE_J(z)\propto\int_0^\oo d\b\, \b^{J+d-4} a^\dagger(\b z)a(\b z).
\ee
The expression (\ref{eq:EfromA}) is now Lorentz-invariant, and~\eqref{eq:EJnon-cov} can be recovered by setting $z=(1,\bn)$. In the form (\ref{eq:EfromA}), $\cE_J(z)$ becomes a homogeneous function of $z$ of degree $3-d-J$:
\be
	\cE_J(\l z)=\l^{3-d-J}\cE_J(z)\qquad (\l>0).
\ee
Interpreted as an operator in index-free notation, this means that $\cE_J(z)$ has Lorentz spin $3-d-J$.\footnote{See e.g.~\cite{Costa:2011mg} for an introduction to index-free notation.} Note that the mass dimension of $\cE_J(z)$ is $J-1$.

For even integer $J$,  the ``$E^{J-1}$ flux'' operator $\cE_J(z)$ can alternatively be defined in terms of the light-transform of a local operator:
\be\label{eq:EfromL}
	\cE_J(z)=2\wL[\cO_J](\infty,z).
\ee
Let us unpack this notation. Here, $\cO_J(x,z)$ is the leading-twist spin-$J$ primary operator built out of two $\phi$'s,
\be\label{eq:OJdefn}
	\cO_J(x,z)=N_J:\!\f(x)(z\.\ptl)^J \f(x)\!:+(z\.\ptl)(\cdots).
\ee
We use an index-free notation, where the indices of $\cO_J^{\mu_1\cdots\mu_J}(x)$ are contracted with a null polarization vector $z^\mu$.
The total derivative terms $(z\.\ptl)(\cdots)$ ensure that $\cO_J$ is primary (i.e.\ it is annihilated by the special conformal generators $K_\mu$), and $N_J$ is an inessential normalization factor. For example, we have $\cO_2=T$, where $T$ is the stress-tensor. 

The operation $\wL$ is a conformally-invariant integral transform called the light transform~\cite{Kravchuk:2018htv}. We will not need its complete definition here --- just some basic properties. Firstly, $\wL[\cO](x,z)$ is an integral of $\cO$ along the null direction $z$ starting at the point $x$. Secondly, when applied to a primary operator $\cO$ with scaling dimension $\De$ and Lorentz spin $J$, the light transform produces a primary operator $\wL[\cO]$ at $x$ with scaling dimension $\De_L=1-J$ and Lorentz spin $J_L=1-\De$.\footnote{In general, the spin $J_L$ is non-integer. This is not a problem since the Lorentz group $\SO(d-1,1)$ admits (infinite-dimensional) non-integer spin representations.}

The right-hand side of (\ref{eq:EfromL}) is a light transform evaluated at $x=\oo$ (spatial infinity), and is thus an integral along future null infinity $\mathscr{I}^+$.
Note that in a CFT, neither $x=\oo$ nor $\mathscr{I}^+$ are special --- CFTs live on the Lorentzian cylinder, and $x=\oo$ is an ordinary point there \cite{Luscher:1974ez} (see \cite{Kravchuk:2018htv} for a recent discussion). 
The right-hand side of (\ref{eq:EfromL}) is then clearly well-defined.

Let us check that the quantum numbers agree on both sides of~\eqref{eq:EfromL}. The operator $\cO_J$ has spin $J$ and scaling dimension $\De=\De(J)=J+d-2$. The light-transformed operator $\wL[\cO_J]$ has scaling dimension $\De_L=1-J$ and Lorentz spin $J_L=1-\De(J)=3-d-J$. These precisely match the quantum numbers of $\cE_J(z)$.\footnote{The attentive reader may be puzzled that the mass dimension of $\wL[\cO_J]$ is $1-J$ while that of $\cE_J$ is $J-1$. The resolution is that $\wL[\cO_J]$ is inserted at spatial infinity, which flips the sign of its mass dimension. Indeed, the definition of $\cO(\oo)$ for any $\cO$ is $\lim_{x\to \oo}|x|^{2\De_\cO}\cO(x)$, where $|x|^{2\De_\cO}$ is needed to obtain a non-zero result, which adds $-2\De_\cO$ units of mass dimension to $\De_\cO$ of $\cO$.} Furthermore, both sides are primary operators. Indeed, for operators inserted at spatial infinity, primariness is just translation-invariance. This is true for $\wL[\cO_J]$ by construction, and is also obviously true for the detector $\cE_J(x)$ as translations do not change the momenta of the particles. 

Thus,~\eqref{eq:EfromL} makes sense from the point of view of symmetries, and agreement with \eqref{eq:EJnon-cov} can be verified in free theory using the explicit expression~\eqref{eq:OJdefn} for $\cO_J$. 
Similar detectors with gravity have been recently discussed in \cite{Gonzo:2020xza}.

\subsection{Turning on interactions}

Let us now turn on the $\phi^4$ interaction and tune to the Wilson-Fisher fixed point. Our theory now does not have well-defined scattering states. In this setting, the notion of $E^{J-1}$ flux is ill-defined (except when $J=2$), since we cannot simply count particles weighted by powers of their energy. However, we can still build detectors from light-transforms of local operators $\wL[\cO_J]$, where now $\cO_J$  denotes a spin-$J$ operator in the {\it interacting} Wilson-Fisher theory. This gives a set of well-defined detectors with integer dimensions $\De_L=1-J$ in the interacting theory. However, their Lorentz spins are different from those in
the free theory: $J_L=1-\De(J)=3-d-J-\delta(J)$.

In the free theory, the ``$E^{J-1}$ flux" operators $\cE_J(z)$ made sense not just for integer $J$, but for any complex $J\in \C$. The reason is that energy is positive, so there is no ambiguity in defining $E^{J-1}$. So far, in the interacting theory, we have identified a set of detectors $\wL[\cO_J](\oo,z)$ that provide analogs of $\cE_J(z)$ for integer $J$. What about non-integer $J$? This leads to the question of defining an analytic continuation of $\wL[\cO_J]$ away from even $J\geq 0$.

We will follow \cite{Kravchuk:2018htv}, who considered precisely this problem and showed that one can define non-local light-ray operators $\mathbb{O}^+_J(x,z)$ such that the dependence on $J$ is analytic,
and for even integer $J\geq J_0$ ($J_0\leq 1$ is the Regge intercept~\cite{Costa:2012cb,Caron-Huot:2017vep}) we have
\be
	\mathbb{O}^+_J(x,z)=\wL[\cO_J](x,z).
\ee
This means that for general complex $J$ we can define $\cE_J(z)$ in an interacting CFT by
\be
	\cE_J(z)\propto\mathbb{O}^+_J(\infty,z).
\ee
In~\cite{Kravchuk:2018htv} it was argued that the operators $\mathbb{O}^+_J$ should exist non-perturbatively. To what extent this is true remains an open question. In this work, we will explore the construction of $\mathbb{O}^+_J$ and other light-ray operators in perturbation theory. We will see that they can indeed be straightforwardly defined and multiplicatively renormalized, at least to the lowest nontrivial order in perturbation theory.

These detectors can be naturally interpreted by thinking about measurements of ``$E^{J-1}$ flux''.
This phrase must be used with care  in an interacting theory, since no operator has exactly the desired quantum numbers (both dimension and spin).
Let us illustrate this in the case $J=3$. We know that energy flux $\cE_2(z)$ is well-defined, so we can construct transparent detectors that measure it.
More precisely, we can measure the flux through a finite angular region $\O$,
\be
	E_2(\O)\equiv \int_\O d^{d-2}\bn\, \cE_2(\bn).
\ee
Since our detectors are transparent, we can stack two of them to obtain an observable $\hat E_3(\O)$ defined by
\be
	\hat E_3(\O)=(E_2(\O))^2.
\ee
Intuitively, we may try to define the operator ``which measures the flux of $E^{2}$'' as the limit of $\hat E_3(\O)$ as the region $\O$ shrinks to a point around $\bn$.
This requires studying the OPE $\cE_2(\bn_1)\cE_2(\bn_2)$ as $\bn_1\to\bn_2$.  What happens is that this OPE has non-trivial scaling with respect to the angular size of $\O$,
$\theta\sim |\bn_1-\bn_2|$, which leads to \cite{Hofman:2008ar,Kologlu:2019mfz}
\be\label{eq:Esquaredscaling}
	\hat E_3(\O)\propto \theta^{d-2+\delta(3)}\cE_3(\bn)+\cdots.
\ee
Thus we defined an operator with manifestly the desired mass dimension $\Delta_L=2$ (it measures energy squared),
but it does not transform under Lorentz boosts like a density of energy squared on the celestial sphere, since $J_L={-}d-\delta(3)\neq -d$.
This ``anomalous spin'' is related, by the light transform, to the anomalous dimension $\delta(3)$ of local operators analytically continued to spin $J=3$.
Multiple energy correlators at the LHC were discussed recently in \cite{Lee:2022ige}.

Of course, this is just one particular way of defining ``$E^2$ flux,'' and one could imagine other measurements which for instance would have the correct Lorentz spin $J_L=-d$.
(Possibly by weighting by a suitable power of the momentum perpendicular to the axis $\bn$, or exploiting suitable time windows~\cite{Korchemsky:2021okt}.)
By solving $3{-}d{-}J{-}\delta(J)=-d$ for $J$, one would be able to predict the mass dimension of such a
measurement, as further discussed in section \ref{sec:reciprocity}.
Generally, a specific experiment may best be described by a linear combination of operators, which depend on fine details of the experiment and the theory.  This should not surprise us, since it also happens with local measurements.\footnote{The above discussion applies equally well to the case $J=1$. In the free theory, the operator $\cE_1(z)$ simply counts particle number in the $z$ direction. In the interacting theory, particle number is no longer well-defined, due to splitting as particles propagate to null infinity.  The rate of splitting can be quantified in different ways: the anomalous dimension $\de(1)$ is relevant for counting particles in a given angular region,
while its timelike counterpart $\delta\big|_{J_L=2-d}$ (see eq.~\eqref{gL from gT}) captures the dependence of the multiplicity of a jet on its invariant mass $Q^2$ \cite{Mueller:1981ex,Bolzoni:2012ii}. We thank Juan Maldacena for discussions on this point.
}

\subsection{The Chew-Frautschi plot of the Wilson-Fisher theory}

Before proceeding with explicit calculations, let us examine more closely the quantum numbers of the light-ray operators $\mathbb{O}_J^+$ in the Wilson-Fisher theory.
Recall that the leading-twist operators $\cO_J$ have spin $J$ and scaling dimension
\be
\label{eq:leadingreggetrajectoryabstract}
\De(J) = 2\De_\phi + J + \gamma(J),
\ee
where $\gamma(J)$ is known in perturbation theory and is well-defined for even $J\geq 0$.
(It differs from $\delta(J)$ in the preceding subsection by a simple shift: $\delta(J)=\gamma(J)+2\De_\phi+2-d$.)
The Lorentzian inversion formula provides a canonical analytic continuation of $\g(J)$ to $J\in \C$, which gives the quantum numbers of the light-ray operators $\mathbb{O}_J^+$ via $(J_L,\De_L)=(1-\De(J),1-J)$.

In perturbation theory in $d=4-\e$ dimensions, we have the following expansions~\cite{BREZIN1973227,Derkachov:1997pf, Henriksson:2022rnm}
\be
\De_\phi &= 1-\frac{1 }{2}\epsilon+\frac{1}{108}\epsilon^2+\frac{109}{11664}\epsilon^3+\left(\frac{7217}{1259712}-\frac{2 \zeta (3)}{243}\right) \epsilon^4 + O(\e^5),
 \\ \gamma(J)&= -\frac{1}{9 J (J+1)} \epsilon^2 +\left(\frac{22 J^2-32 J-27}{486 J^2 (J+1)^2} -\frac{2 H(J)}{27 J (J+1)}\right) \epsilon^3 + O(\e^4).\label{eq:gammaofJ}
\ee 
Here, $H(J)=\tfrac{\Gamma'(J+1)}{\Gamma(J+1)}+\gamma_E$ is the analytic continuation of the harmonic numbers, with $\g_E\approx 0.5772$ the Euler-Mascheroni constant.
The order $\e^4$ term in $\gamma(J)$ is known~\cite{Derkachov:1997pf},\footnote{This paper contains a typo in the result, see~\cite{Henriksson:2022rnm}.} but we do not reproduce it here for brevity.

\begin{figure}[htb]
\begin{center}
\includegraphics[scale=.6]{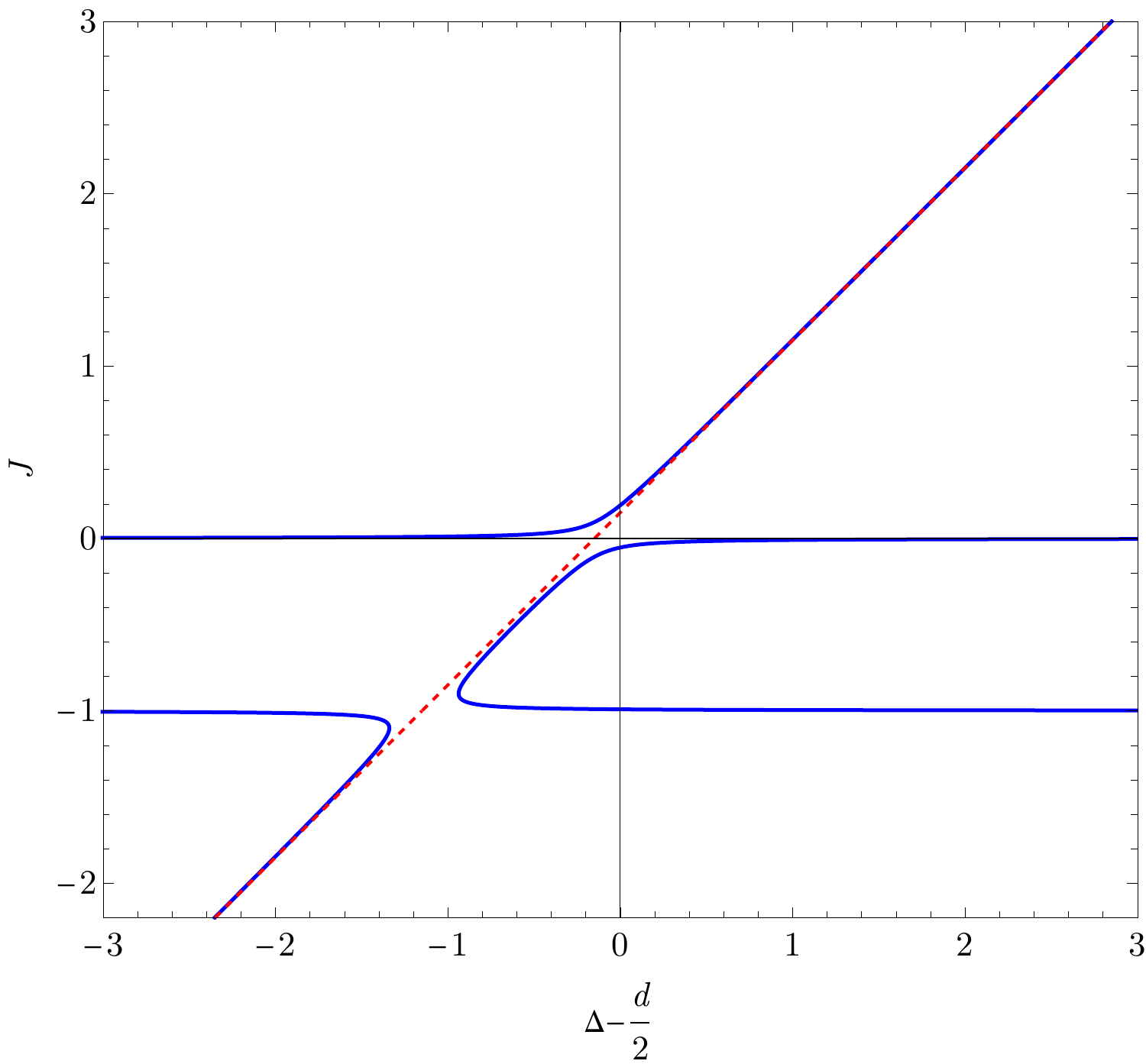}
\caption{Solid, blue: the naive leading Regge trajectory in Wilson-Fisher theory at $O(\e^2)$. Dashed, red: the same trajectory in free theory. The plot is made for $\e=0.3$.
The correct picture will be more complex, as described in the main text (see also figures~\ref{fig:intercept} and~\ref{fig:perturbativelines}).}
\label{fig:twist2poles}
\end{center}
\end{figure}

Focusing for now on the leading $\e^2$ contribution, we plot $\De(J)$ at $\e=0.3$ in figure~\ref{fig:twist2poles}. Following \cite{Brower:2006ea}, it will prove convenient to use coordinates $\De-\frac{d}{2}$ and $J$.
Given the interpretation of $\Delta$ in a conformal theory as energy in radial quantization, the resulting curve is often called a Regge trajectory, and we will refer to this type of plot as a Chew-Frautschi plot.
The above discussion associates to each point on the Regge trajectory a light-ray operator $\mathbb{O}_J^+$ and the corresponding detector $\cE_J(z)=\mathbb{O}_J(\oo,z)$.
We can immediately spot a problem with our Regge trajectory: the $\e^2$ result for $\g(J)$ has poles at $J=0$ and $J=-1$. At higher orders, poles appear at all non-positive integer $J$.

This na\"ively contradicts our claim that $\mathbb{O}_J^+$ and $\cE_J$ can be multiplicatively renormalized in perturbation theory for all $J$. However, as we will now explain, the poles are there for a good reason and simply need to be interpreted correctly.

\subsection{Shadow trajectories}
\label{sec:shadows}

The reason for the poles in $\g(J)$ is that there are other Regge trajectories missing from our Chew-Frautschi plot in figure~\ref{fig:twist2poles}. 
Let us first discuss the situation in the free theory, where the trajectory of $\mathbb{O}^+_J$ is a straight line. 
First, there are subleading, higher-twist trajectories that provide analytic continuations of $\wL[\cO]$, where $\cO$ is a local operator built out of more than two $\f$'s, potentially with extra contracted derivatives.\footnote{The degeneracy of these higher-twist operators grows with their spin since we have more ways to distribute the derivatives among the $\phi$'s as the spin increases. Therefore, since Regge trajectories are analytic in spin, the Regge trajectories for these higher-twist operators must be infinitely degenerate. By contrast, for the leading-twist trajectories there is a unique operator for each spin, see \eqref{eq:OJdefn}, and the leading trajectory is non-degenerate.}
We restrict to $\Z_{2}$-even Regge trajectories, so $\cO$ will be composed of an even number of $\f$'s.
In the free theory, these give lines parallel to the leading twist trajectory that we must add to our plot. Adding these, we obtain a set of Regge trajectories representing ``sensible lines drawn through local operators.'' 

We have introduced the light-ray operators $\mathbb{O}_J$ as devices that analytically continue light transforms of local operators, $\wL[\cO_J]$, in spin $J$. From this point of view, it seems reasonable to stop at the above set of trajectories. However, this does not solve the problem with poles in $\g(J)$ in the interacting theory.

One missing ingredient is shadow symmetry. There is a natural Lorentz-invariant integral transform acting on the space of light-ray operators: the ``spin shadow"~\cite{Kravchuk:2018htv} given by 
\be \label{eq:spin shadow definition}
	\wS_J[\mathbb{O}](x,z)=\int D^{d-2}z' (-2z\.z')^{2-d-J_L}\mathbb{O}(x,z').
\ee
Here, the measure is given by $D^{d-2}z=2 d^dz \de(z^2)\theta(z^0)/\vol \R$, so the $z'$-integral ranges over the forward null-cone.
(If we interpret $z$ as an embedding space coordinate on the celestial sphere $S^{d-2}$, then $\wS_J$ is the Euclidean shadow transform on the celestial sphere. This integral will be discussed in more detail in section~\ref{sec:shadowtransform}.)
The resulting operator $\wS_J[\mathbb{O}](x,z)$ has the same scaling dimension $\De_L$ as $\mathbb{O}$ but a different Lorentz spin $2-d-J_L$. Remembering that $\De_L=1-J$ and $J_L=1-\De(J)$, this corresponds precisely to the standard shadow transformation $\De\to d-\De$. Thus, the space of light-ray operators has a symmetry under $\De\to d-\De$, which should be reflected in the Chew-Frautschi plot.\footnote{Note that if $\mathbb{O}(x,z)$ is somehow localized along the null direction $z$, then its spin shadow $\wS_J[\mathbb{O}(x,z)]$ is delocalized over the entire future null cone. Consequently, the name ``light-ray operator" is perhaps a misnomer --- although a light-ray operator is always {\it labeled} by a null ray, it is not always localized on that null ray. It might be tempting to make a distinction between ``light-ray" operators localized on a light ray and ``light-cone" operators localized on a light cone. However, surprisingly, all such operators are continuously connected on the Chew-Frautschi plot, so such a distinction should be drawn with care.}

\begin{figure}[tb]
	\begin{center}
		\includegraphics[scale=.6]{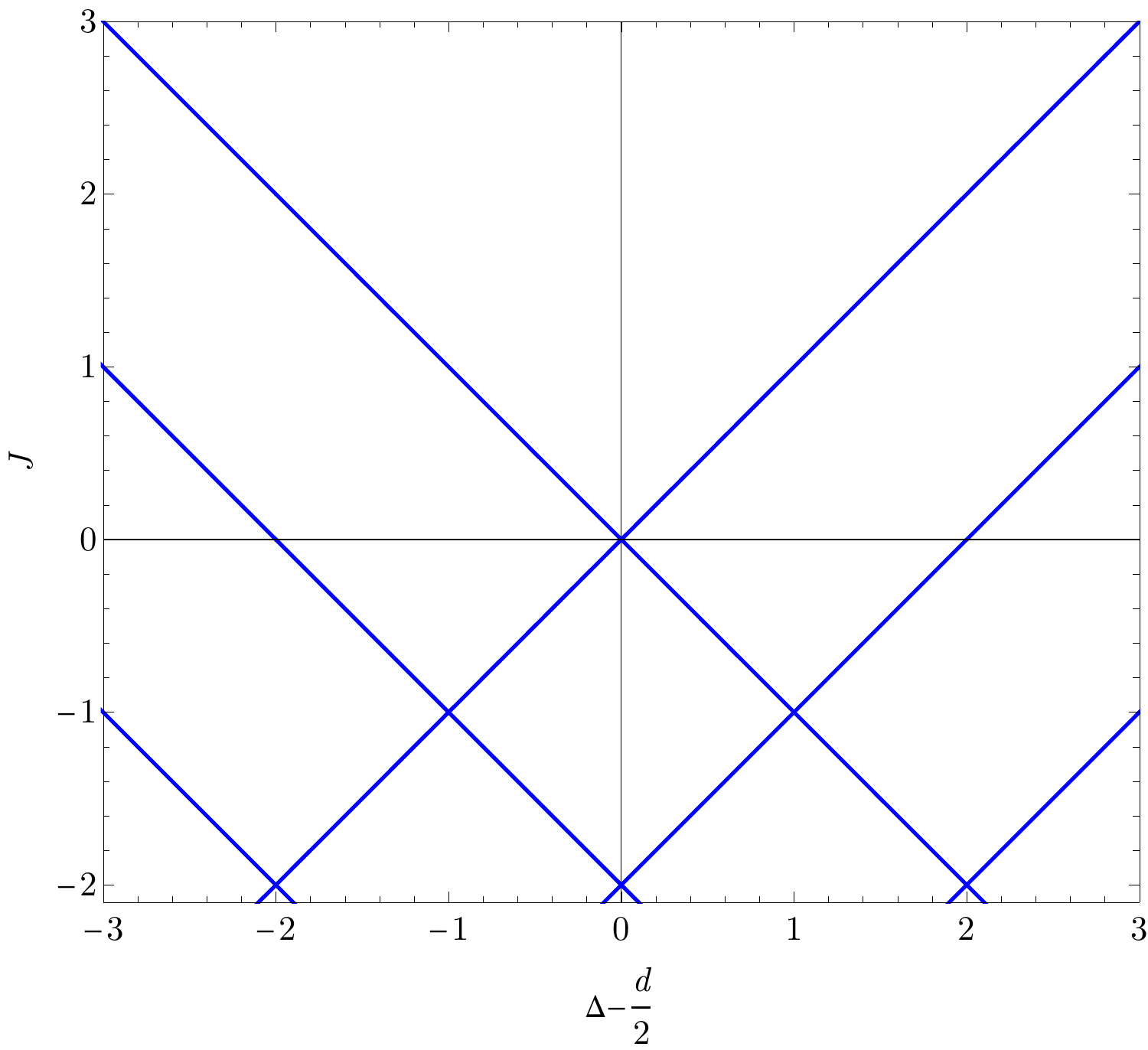}
		\caption{Chew-Frautschi plot of free theory in $d=4$ after accounting for higher-twist trajectories and their shadows. This is a less naive version of figure~\ref{fig:twist2poles}. The full picture is expected to be more complex and is described in the main text (see also figures~\ref{fig:intercept} and~\ref{fig:perturbativelines}).}
		\label{fig:freediagonals}
	\end{center}
\end{figure}

Let us then add to our Chew-Frautschi plot the shadows of everything we have discussed so far. The resulting plot in $d=4$ (free theory) is shown in figure~\ref{fig:freediagonals}. As we will explain, it is still missing some trajectories, but figure~\ref{fig:freediagonals} already makes one point clear: free-theory light-ray trajectories can intersect, and thus there can be degeneracies among its light-ray operators. In particular, the poles that we observed in $\g(J)$ at order $\e^2$ appear precisely at the intersections of the leading trajectory with shadow trajectories. In fact, this is true for the poles appearing at all known orders in $\e$. Our failure to renormalize the light-ray operators at these points, evidenced by these poles, can be attributed to a non-trivial mixing problem that must be solved at these intersections.

In this paper we will study the mixing problem that appears at the simplest intersection: that of the leading-twist trajectory with its shadow. Besides its simplicity, another reason for focusing on this intersection is that it coincides with the ``Regge intercept,'' or the largest value of $J$ among all trajectories at $\De(J)=\tfrac{d}{2}$. The Regge intercept determines the leading behavior of correlators in the Regge limit and the light-ray operator that sits at this point is known as the Pomeron~\cite{Costa:2012cb}.
One of the goals of this work is to answer the question, ``what is the Pomeron of the Wilson-Fisher theory?" 
After a warm-up in section~\ref{sec:warmup twist2}, we will study this question in detail in section~\ref{sec:pomeron}, where we will explicitly construct the necessary light-ray operators and solve the mixing problem at leading order.

\subsection{Mixing at the Regge intercept}
\label{sec:summary_mixing}

\begin{figure}[htb]
	\begin{center}
			\includegraphics[scale=.6]{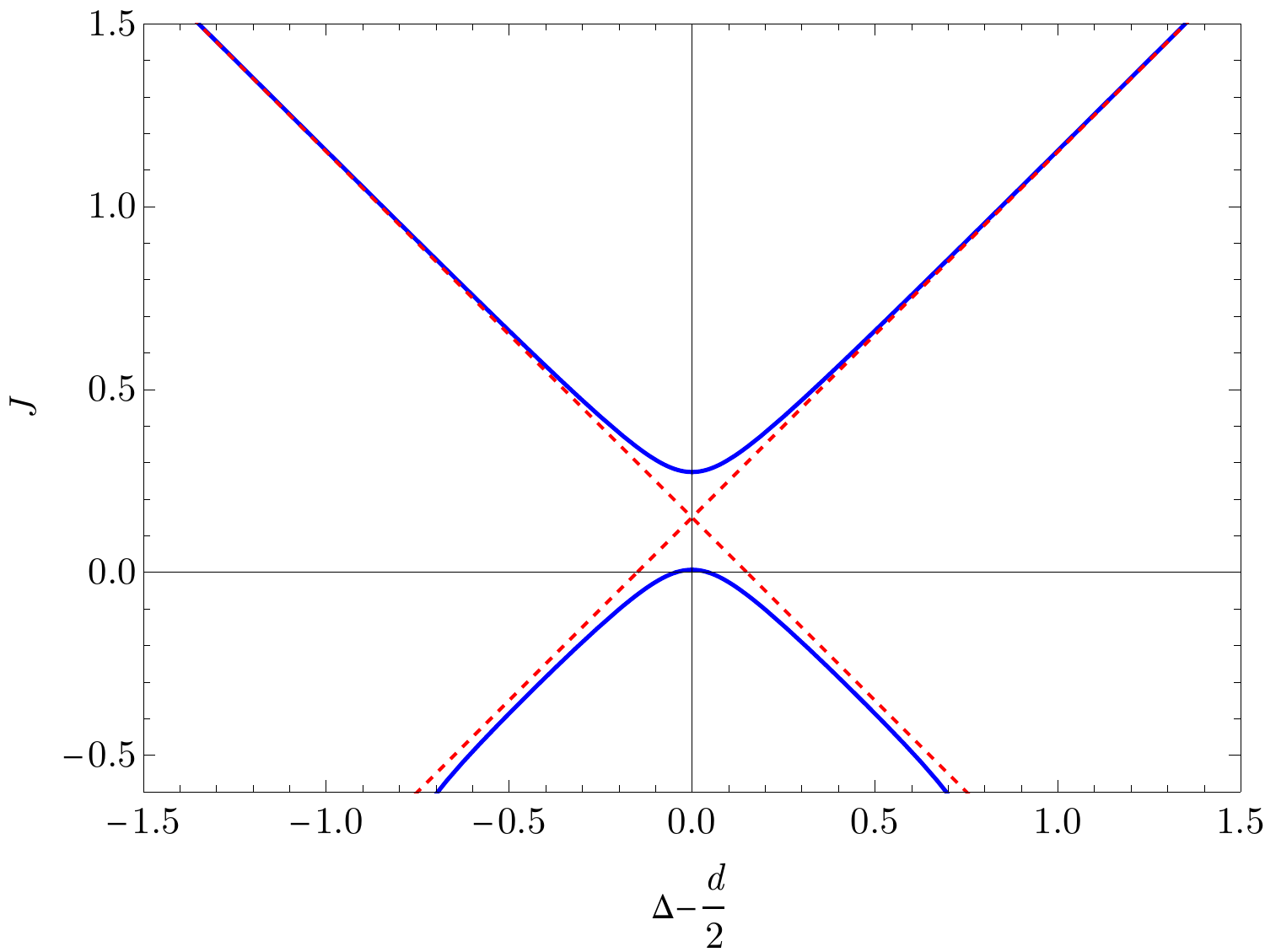}
		\caption{Chew-Frautschi plot of the leading Regge trajectory in Wilson-Fisher theory near the intercept at $O(\e^4)$ (solid, blue). Free theory trajectories are shown in dashed red. The plot is made at $\e=0.3$. See also figure~\ref{fig:3dintercept}.}
		\label{fig:intercept}
	\end{center}
\end{figure}

It turns out we can predict the quantum numbers of the solution to the mixing problem at the Regge intercept by correctly interpreting the expansion~\eqref{eq:gammaofJ}.
We simply conjecture that in an interacting theory, Regge trajectories can't diverge.  This implies that all perturbative singularities must get resolved in a way similar to figure~\ref{fig:intercept}.
Specifically, we will assume that we obtain a complex two-sheeted surface in a neighborhood of the intercept. We can describe such surfaces by\footnote{We thank Nikolay Gromov for a discussion about such parametrizations. A similar parametrization of  BFKL and twist-2 trajectory in $\cN=4$ SYM has been considered in~\cite{Brower:2006ea}, and a non-perturbative picture in the planar limit has been studied in~\cite{Gromov:2015wca}.}${}^{,}$\footnote{Here is an argument why this is always possible. We assume that we have a complex surface $\Sigma$ (which in this context is a neighborhood of the Regge intercept on the smooth Regge trajectory), and two holomoprhic functions $\nu_0:\Sigma\to \C$ and $J_0:\Sigma\to \C$ which embed it into $\nu,J$ space and have bounded images. We further assume that we have $n$ solutions $x_k$ of $J_0(x)=J$ for all $J$ in the image of $J_0$, except possibly for a discrete set of branch points. We denote these solutions by $x_k(J)$ ($k=1,\cdots,n)$, which are multi-valued functions of $J$. Under monodromies around the branch points they are permuted in some way. We then consider $F(\nu,J)=\prod_{k=1}^n(\nu-\nu_0(x_k(J)))$. It is a holomorphic function of $J$ which is single-valued, and can only have singularities at the branch points. However, since it is bounded, the singularities are removable and we get a function which is holomorphic in $J$ on the image of $J_0$. It is also obviously holomorphic in $\nu$, and thus is a holomorphic function of both variables. The set $F(\nu,J)=0$ is precisely the image of $\Sigma$.}
\be
	\nu^2 + f(J)\nu + g(J)=0,
\ee
where $f$ and $g$ are functions of $J$ that are analytic in the neighborhood of the intercept, and $\nu=\De-\tfrac{d}{2}$.\footnote{Our definition of $\nu$ in this section differs from the usual one $i\nu=\De-\frac d 2$ in e.g.~\cite{Costa:2012cb} by a factor of $i$.} Away from the intercept, this equation should have two roots, given by
\be
	\nu_\pm(J)=\pm\p{2\De_\f+J+\g(J)-\tfrac{d}{2}},
\ee
corresponding to the leading twist trajectory and its shadow. This allows us to compute $f(J)$ and $g(J)$ using Vieta's formulas,
\be
	f(J)&=-\nu_+(J)-\nu_-(J)=0\\
	g(J)&=\nu_+(J)\nu_-(J)=-\p{2\De_\f+J+\g(J)-\tfrac{d}{2}}^2.
\ee
The fact that $f(J)$ vanishes is forced by shadow symmetry, $\nu\to-\nu$, but near other intersections we might have $f(J)\neq 0$ (and higher-degree polynomials if more than two trajectories are mixing). The resulting equation is
\be \label{eq:physical state condition}
\nu^2 = (2\De_\phi-d/2+J+\gamma(J))^2.
\ee 
The nontrivial statement here is that the right-hand-side should be free of poles near $J=0$.  This is effectively a constraint on $\gamma(J)$.

Before proceeding, let us see in a toy model how $1/J$ poles can arise from expanding equations of the form~\eqref{eq:physical state condition}. Consider the equation
\be\label{eq:toy}
	\nu^2-J^2+\e^2=0.
\ee
At $\e=0$ this describes an intersection of straight lines, which gets resolved as in figure~\ref{fig:intercept} for $\e>0$. If we solve for $\De$ in this toy model, we get
\be
	\De=\tfrac{d}{2}\pm \sqrt{J^2-\e^2}.
\ee
Expanding in $\e$ at a generic $J$ we find
\be
	\De=\tfrac{d}{2}\pm \p{J-\frac{\e^2}{2J}}+\cdots,
\ee
which has a pole in $J$ even though the curve~\eqref{eq:toy} is perfectly smooth. Our conjecture is that the poles at $J=0$ in~\eqref{eq:gammaofJ} appear for a similar reason.

\begin{figure}[htb]
	\begin{center}
		\includegraphics[scale=.5]{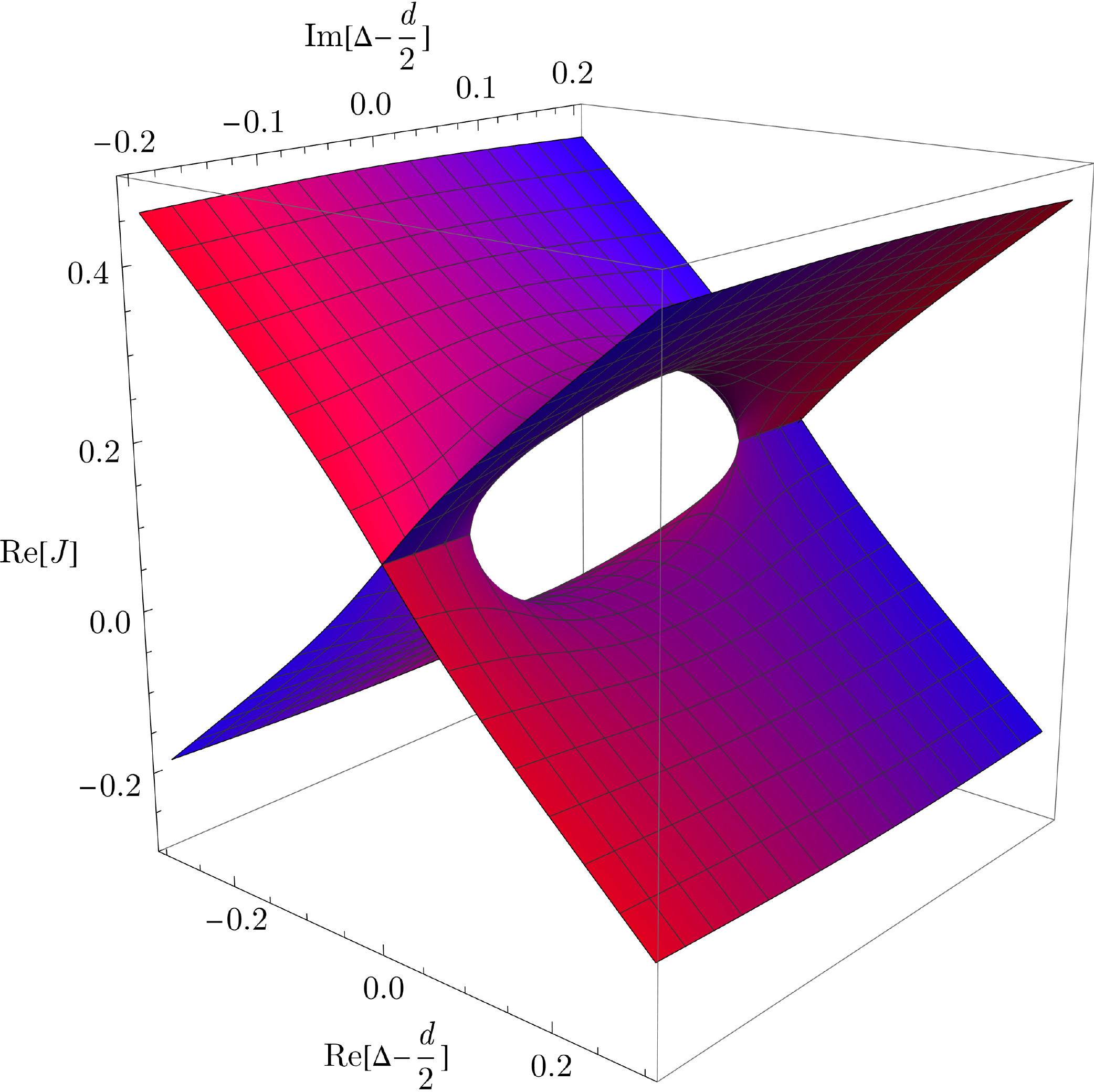}
		\caption{An $\R^3$ projection of the $\C^2$ Chew-Frautschi plot of the leading Regge trajectory in Wilson-Fisher theory near the intercept at $O(\e^4)$.  The imaginary part of $J$ is shown by color, with negative values in blue and positive values in red. Even though the two branches appear to intersect, they do not -- in order to intersect in $\C^2$, they need to intersect in this $\R^3$ projection and also have the same color.  The plot is made at $\e=0.3$.}
		\label{fig:3dintercept}
	\end{center}
\end{figure}

To verify this, we can use the anomalous dimension~\eqref{eq:gammaofJ} to compute for~\eqref{eq:physical state condition}
\be
\nu^2 &= (2\De_\phi-d/2+J+\gamma(J))^2
\nn \\ &= J^2-J \epsilon+\left(\frac{J}{27}+\frac{1}{4}-\frac{2}{9 (J+1)}\right) \epsilon ^2
\nn \\ &\quad +\left(\frac{109 J^3+164 J^2+265 J-114}{2916 (J+1)^2} -\frac{4 H(J)}{27 (J+1)}\right)\epsilon^3 + O(\e^4), \label{eq:WF leading trajectory}
\ee
where once again the order $\e^4$ term can be computed~\cite{Derkachov:1997pf,Henriksson:2022rnm}, but is omitted here for brevity. We see that the $1/J$ poles nicely cancel at each order in $\e$,
leaving a curve that is perfectly analytic in $J$ near $J=0$. We plot the trajectory in the real $(\nu,J)$-plane at four-loop order in the $\e$ expansion in figure~\ref{fig:intercept}, where we set $\e=0.3$. In fact,~\eqref{eq:WF leading trajectory} defines a complex surface that is perfectly regular as long as we stay away from the pole at $J=-1$ (to which we will return below).
We plot a projection of this surface in figure~\ref{fig:3dintercept}, which makes it clear that the two branches of the trajectory are connected in the complex plane.

One may ask: what does the cancellation of poles mean in terms of the pole structure of $\g(J)$?
A simple analysis shows that in order for all the poles to cancel, at each order in $\e$, 
the leading pole in the Laurent expansion of $\g(J)$ at $J\to 0$ is of order $\sim J^{1-n}\e^n$, and the coefficients of all multiple poles $J^{-k}$ for $k>1$ are fixed in terms of lower-order data.\footnote{More generally, knowing the full form of  $\g(J)$ up to and including $O(\e^{n_0})$ allows to predict the leading poles $J^k\e^{n}$ with $n+k\leq n_0-1$.
For example, the $O(\e^2)$ result for $\g(J)$ predicts the leading singularity $J^{-n+1}\e^n$ for all $n$. In the context of BFKL/DGLAP  mixing  in QCD and $\cN=4$ SYM the analogous properties were previously noticed in~\cite{Jaroszewicz:1982gr, Lipatov:1996ts, Kotikov:2000pm, Kotikov:2002ab, Brower:2006ea, Kotikov:2007cy}.} Assuming this holds, we can predict the $J^{-4},\dots,J^{-2}$ terms at $O(\e^5)$ from the known data. In fact, using additional information explained below, we can determine the $J^{-1}$ term at $O(\e^5)$ as well, giving
\be\label{eq:gammasingular}
	\g(J)=\cdots+\e^5\Big(&-\frac{5}{216}\frac{1}{J^4}+\frac{269}{8748}\frac{1}{J^3}+\frac{-227-78\pi^2+216\z(3)}{17496}\frac{1}{J^2}\nn\\
	&+\frac{82620 \zeta (3)-583200 \zeta (5)+6561 \pi ^4+46575 \pi ^2+96275}{7085880}\frac{1}{J}+O(J^0)\Big)+O(\e^6).
\ee

We can now extract the Regge intercept of the Wilson-Fisher theory by solving~\eqref{eq:WF leading trajectory} for $J$ at $\nu=0$, order-by-order in $\e$. Since the curve \eqref{eq:WF leading trajectory} intersects the $J$-axis at two points, see figure \ref{fig:intercept}, we find two solutions. The larger solution gives the Regge intercept:
\be\label{eq:reggeintercept}
J_0(\e)=&\p{\frac{1}{2}+\frac{\sqrt 2}{3}}\e-\frac{11\sqrt{2}+21}{162}\e^2+\frac{465+421\sqrt 2+54(4+3\sqrt 2)\pi^2-648\sqrt 2\zeta(3)}{17496}\e^3\nn\\
&+\frac{1}{9447840}\Big(
-486\pi^2(65(4+3\sqrt 2)+(28+27\sqrt 2)\pi^2)\nn\\
&\quad-5(76227+57760\sqrt 2+648(150+109\sqrt 2)\zeta(3)-233280\sqrt 2\zeta(5))
\Big)\e^4\nn\\
&+O(\e^5) \nn\\
=&\ 0.971405\e-0.225656\e^2+0.248731\e^3-0.631547\e^4+O(\e^5).
\ee
Here, we used the four-loop result (i.e.\ including $\e^4$) for~\eqref{eq:gammaofJ}~\cite{Derkachov:1997pf,Henriksson:2022rnm}, as well as the $J^{-1}$ term at $O(\e^5)$ from~\eqref{eq:gammasingular}. Note that the expansion for $J_0$ starts at order $\e$, even though~\eqref{eq:gammaofJ} started at $\e^2$. The second solution for $J$ is obtained by replacing $\sqrt 2\to -\sqrt 2$ in the above expression.

\begin{figure}[tb]
	\begin{center}
		\includegraphics[scale=.6]{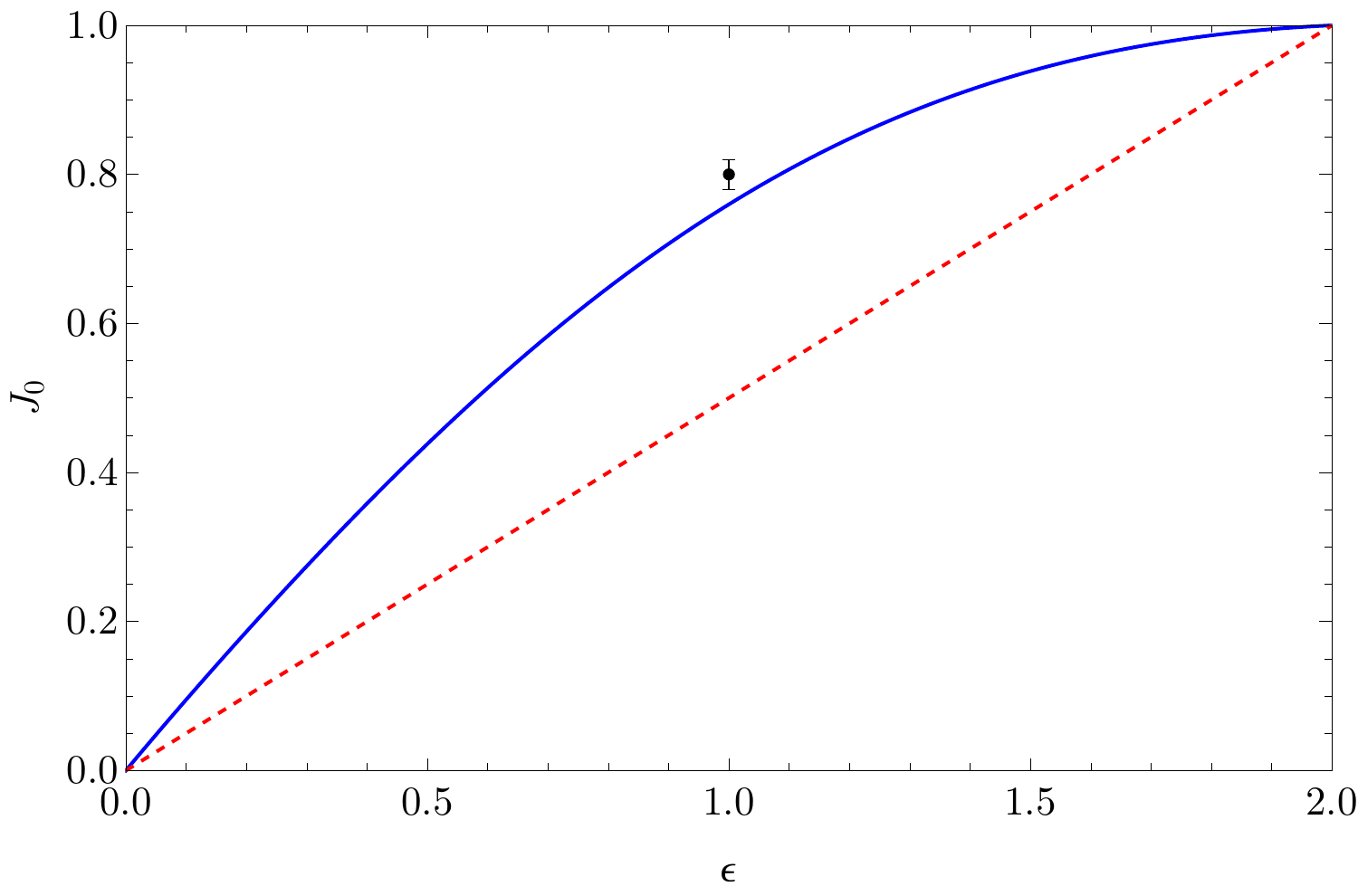}
		\caption{Solid, blue: the Pad\'e${}_{3,3}$ approximant to the intercept $J_0(\e)$ imposing the small-$\e$ expansion~\eqref{eq:reggeintercept} and known value in $d=2$ ($J_0(2)=1$). Dashed, red: the formal intercept value for free theory. The data point at $\e=1$ is the result from~\cite{Caron-Huot:2020ouj}, taken with the (non-rigorous) error bar estimated from their figure 12.}
		\label{fig:pade}
	\end{center}
\end{figure}

Note that the numerical coefficients in this expansion are not small. Therefore, as usual, setting $\e=1$ is not justified. As an amusing exercise, we can get a more reasonable estimate for $J_0(\epsilon)$ by assuming $J_0(2)=1$, which is where the leading twist trajectory intersects its shadow in the 2d Ising CFT, and constructing a Pad\'e${}_{3,3}$ approximant using this condition together with~\eqref{eq:reggeintercept}. This results in a monotonic curve interpolating between $J_0(0)=0$ and $J_0(2)=1$, as shown in figure~\ref{fig:pade}. With this approximation, we estimate the Regge intercept of the 3d Ising CFT to be $J_0(1)\approx 0.76$, which is close to the result $J_0(1)\approx 0.8$ from the analysis in~\cite{Caron-Huot:2020ouj}, see the caption to figure~\ref{fig:pade}. As pure speculation, we note that the plot  in figure~\ref{fig:pade} suggests the possibility that the slope of $J_0(\e)$ near $\e=2$ ($d=2$) might be 0, especially given the independent data point from~\cite{Caron-Huot:2020ouj}. This could be related to the constraints on a putative $(d=2+\e)$-expansion observed in~\cite{Li:2021uki}.

We see that by using the anomalous dimensions of the lightest spinning operators we can straightforwardly determine the spin of the Pomeron, which in turn determines how CFT correlators behave in the Regge limit. In fact, we get the full shape of the leading trajectory, which in principle allows one to analyze the Regge limit beyond the saddle-point approximation \cite{Caron-Huot:2020nem}.
 
\begin{figure}[tb]
	\begin{center}
		\includegraphics[scale=.6]{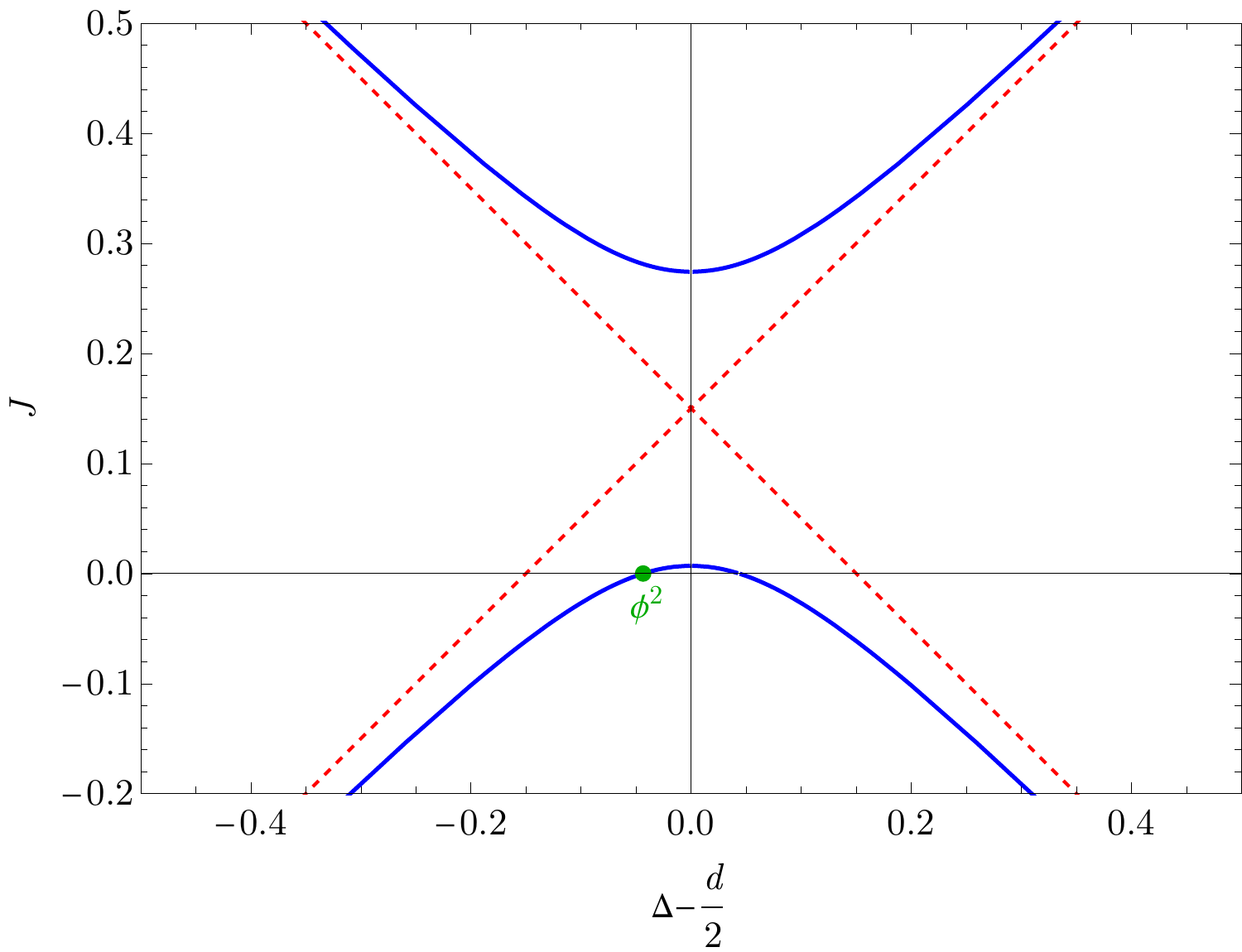}
		\caption{Chew-Frautschi plot of the leading Regge trajectory in Wilson-Fisher theory near the intercept at $O(\e^4)$. This is same as figure~\ref{fig:intercept}, but zoomed in to show the location of the $\f^2$ operator (more accurately, $\wL[\f^2]$).}
		\label{fig:phisquared}
	\end{center}
\end{figure}

Using~\eqref{eq:WF leading trajectory} we can also study where (light-transforms of) scalar operators sit on the Regge trajectories. 
Setting $J=0$ in~\eqref{eq:WF leading trajectory} and solving for $\nu$ yields:
\be\label{eq:epsilonnu}
	|\nu|=\frac{\e}{6}-\frac{19\e^2}{162}+\p{-\frac{937}{17496}+\frac{4\zeta(3)}{27}}\e^3+O(\e^4).
\ee
Here we only used $\gamma(J)$ up to order $\e^4$.\footnote{At $J=0$~\eqref{eq:WF leading trajectory} reads $\nu^2=\frac{1}{36}\e^2-\cdots$, and thus the $O(\e^4)$ term in~\eqref{eq:WF leading trajectory} only determines the $O(\e^3)$ term in~\eqref{eq:epsilonnu}.} Of the two corresponding values of $\De(0)=\tfrac{d}{2}\pm |\nu|$, the smaller one $\tfrac{d}{2}-|\nu|$ perfectly matches the known $\e$-expansion for the scaling dimension $\De_{\f^2}$ of the $\f^2$ operator. This strongly suggests that $\wL[\f^2]$ lives on the resummed leading-twist Regge trajectory~\eqref{eq:WF leading trajectory}, see figure~\ref{fig:phisquared}. This was first observed in the $\e$-expansion in~\cite{Alday:2017zzv}
 in a somewhat different form, and further discussed in~\cite{Caron-Huot:2020ouj}. 
In fact we can turn this argument around and use known results for the $\e$-expansion of $\De_{\f^2}$ \cite{Kazakov:1979ik,Henriksson:2022rnm} to determine the
$O(\e^5J^{-1})$ coefficient in $\g(J)$; this gives the prediction recorded in~\eqref{eq:gammasingular}.
If the $O(\e^5J^0)$ term were known in~\eqref{eq:gammasingular}, the same logic would fix all singular terms of $\g(J)$ at $J\to 0$ at $O(\e^6)$, and the $O(\e^5)$ term in $J_0$.

The above discussion is, however, somewhat unsatisfactory, because it is based on the assumption that the leading Regge trajectory is smooth at the intercept, and fundamentally is little more than a formal manipulation with known formulas.
In particular, it does not give insights into the question posed in section \ref{sec:shadows} on what is the explicit form of the Pomeron operator in Wilson-Fisher theory or explain how mixing between different trajectories is possible.
We will answer both questions in section~\ref{sec:pomeron} by directly renormalizing the light-ray operators for every point of the complex surface shown in figure~\ref{fig:3dintercept}.
As a consistency check, we will reproduce the leading correction term for the Regge intercept given in~\eqref{eq:reggeintercept}.

\subsection{Horizontal trajectories}
\label{sec:horizontalintroduction}

We have so far discussed Regge trajectories of local operators and their shadows. Is this a complete picture of the Chew-Frautschi plot?
It turns out it is not: we are missing more trajectories. While the Pomeron dominates in the Regge limit, these extra trajectories can give important subleading contributions.\footnote{In fact, in other symmetry sectors they can be the leading contributions -- see the  discussion in section~\ref{sec:discussion}.}

An important clue comes from gauge theories like $\cN=4$ SYM, which possess the well-known BFKL trajectory with $J=1$ (in the free theory) but arbitrary $\De$ --- i.e.\ a horizontal line on the Chew-Frautschi plot. It is unlikely that horizontal trajectories exist at $J>0$ in the Wilson-Fisher theory, since the anomalous dimension $\g(J)$ does not have poles for $J>0$ at any known order in perturbation theory. Furthermore, the construction of the BFKL trajectory uses gauge fields in an essential way.

However, it turns out that we can construct many horizontal trajectories in the Wilson-Fisher theory with $J\leq -1$. The first key idea is to consider products of light-transformed operators
\be
\label{eq:productoflighttransforms}
\wL[\cO_1](x,z_1) \wL[\cO_2](x,z_2),
\ee
where importantly we place $\wL[\cO_1]$ and $\wL[\cO_2]$ at the {\it same point\/} $x$. Formally, the product (\ref{eq:productoflighttransforms}) transforms like a primary light-ray operator with
\be
\De_L &= \De_{L,1}+\De_{L,2} \qquad\implies\qquad J=J_1+J_2-1.
\ee
In \cite{Kologlu:2019bco}, it was shown that the product (\ref{eq:productoflighttransforms}) is nonsingular if $J>J_0$, where $J_0$ is the Regge intercept of the theory. In this case, (\ref{eq:productoflighttransforms}) should {\it not} be considered a qualitatively new operator --- instead it can be expanded in an OPE in terms of other light-ray operators \cite{Kologlu:2019mfz,Chang:2020qpj}.\footnote{More precisely, the component of (\ref{eq:productoflighttransforms}) with ``transverse spin" $j$ on the celestial sphere is nonsingular and can be expanded in other operators when $J_1+J_2-1+j\geq J_0$.} However, if $J<J_0$, the product (\ref{eq:productoflighttransforms}) is singular and requires regularization. After such regularization, we may obtain a new operator.

Let us reiterate what will be a key working assumption: that we can identify independent operators by thinking about the regularization of products.
An analogy is the product of local operators like $\cO(x)\cO'(y)$ with $x\neq y$, which do not require regularization (beyond those of $\cO(x)$ and $\cO'(y)$) and do not add operators to the spectrum, while products like $\cO(x)\cO'(x)$ do.
A peculiarity of light-ray operators is that some products, even of operators labelled by the same base point, do not require regularization.
This feature is crucial to obtain a spin spectrum that is bounded above (for a fixed $\nu$).

\begin{figure}[tb]
	\begin{center}
		\includegraphics[scale=.6]{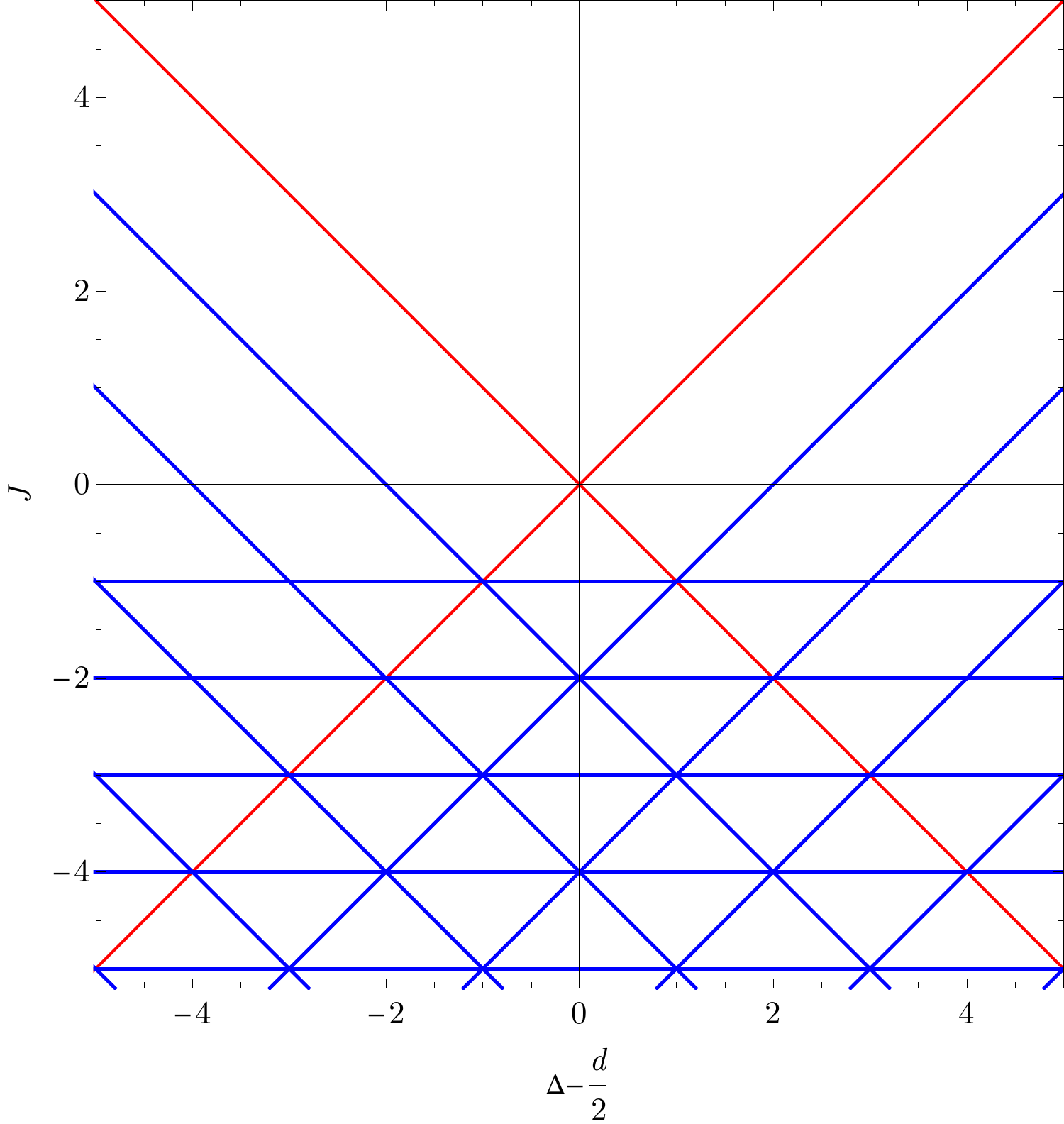}
		\caption{The expected structure of the perturbative ($\mathbb{Z}_2$-even, parity-even, traceless-symmetric) Regge trajectories in Wilson-Fisher theory. All the Regge trajectories (blue) except the leading trajectory (red) have degeneracies, which are expected to be broken at sufficiently high orders of perturbation theory.}
		\label{fig:perturbativelines}
	\end{center}
\end{figure}

Which operators can we choose for $\cO_1$ and $\cO_2$ in the Wilson-Fisher theory? The simplest choice may seem to be $\cO_1=\cO_2=\f$. However, as we discuss in section~\ref{sec:WF horizontal}, the only sensible definition of $\wL[\f]$ vanishes in the free theory, and in the interacting theory is related to $\l\wL[\f^3]$, where $\l$ is the $\f^4$ coupling.
Intuitively, in the free theory, a detector cannot absorb exactly one $\phi$ quantum while conserving energy and momentum.
Therefore, the minimal option is to set $\cO_1=\cO_2=\f^2$ and consider the product
\be
\label{eq:productsoflighttransformedscalars}
	\wL[\f^2](x,z_1)\wL[\f^2](x,z_2).
\ee
The product (\ref{eq:productsoflighttransformedscalars}) requires regularization, since $J_1+J_2-1=-1<J_0$. In the free theory, we can simply normal-order and define
\be
	\cH(x,z_1,z_2)\equiv\,:\!\wL[\f^2](x,z_1)\wL[\f^2](x,z_2)\!:.
\ee
The operator $\cH(x,z_1,z_2)$ then transforms like a light-ray operator with $J=-1$. However, it does not transform irreducibly under the Lorentz group, since it depends on two null polarizations. 
To obtain something that transforms irreducibly with spin $J_L$, we must convolve with a Clebsch-Gordan coefficient $K_{J_L}(z_1,z_2,z)$ for the Lorentz group:
\be\label{eq:project}
	\cH_{J_L}(x,z)=\int D^{d-2}z_1D^{d-2}z_2 K_{J_L}(z_1,z_2;z) \cH(x,z_1,z_2).
\ee
Here, we are free to choose any $J_L$ (and thus $\De=1-J_L$) we like, so we obtain a family of light-ray operators $\cH_{J_L}(x,z)$ that fill out a horizontal line at $J=-1$ on the Chew-Frautschi plot, see figure~\ref{fig:perturbativelines}. This is essentially the same as the construction of the BFKL trajectory (see \cite{Caron-Huot:2013fea} and references therein),
with Wilson lines replaced by $\wL[\f^2]$.
(When one chooses $x$ to be null infinity, the light-cone emanating from $x$ becomes the flat light-sheet like $x^0-x^1=0$, and the Clebsch-Gordan coefficients
are the color-singlet eigenfunctions from \cite{Lipatov:1985uk}, where the quantum number $\De$ is Fourier conjugate to the logarithmic size of the color dipole.)

This construction can be vastly generalized. Firstly, we are free to choose operators other than $\f^2$ in the light-transforms. As long as they are scalars, this gives new horizontal trajectories at $J=-1$, and in fact infinitely many of them. We can also take products of $n>2$ light-transforms of scalar operators. This will yield horizontal trajectories at negative spins $J=1-n$, again infinitely many at each spin.\footnote{It is interesting to ask whether we can construct horizontal trajectories with $J=0$ using these techniques. We expect that the answer is ``no," since we have been able to resolve the pole in $\g(J)$ at $J=0$ to relatively high order in $\e$ without taking such trajectories into account.}

But we can go even further: why restrict to light transforms of local operators? We can consider the product $\mathbb{O}^+_{J_1}(x,z_1)\mathbb{O}^+_{J_2}(x,z_2)$. This formally corresponds to a horizontal trajectory at $J=J_1+J_2-1$, which we now can tune continuously by dialing $J_1$ and $J_2$. We will consider such detectors in section~\ref{sec:WF horizontal}. There, we find that, at leading nontrivial order in perturbation theory, these operators need renormalization only for special values of $J$, and furthermore their divergence is proportional to an operator with fixed values of $J_1,J_2$. In other words, most operators of this kind are only additively renormalized, and so their anomalous dimensions are zero and they do not appear in RG equations for other observables. 

To summarize, we believe that the structure of perturbative Regge trajectories in the Wilson-Fisher theory is as in figure~\ref{fig:perturbativelines}. Note that the qualification ``perturbative'' is important, and the non-perturbative structure may be different. However, we expect that the leading trajectory and its shadow are robust (including the Regge intercept) up to the first intersection with other trajectories.

Among the plethora of horizontal trajectories, we will restrict our attention to the operators~\eqref{eq:project} and renormalize them explicitly in section~\ref{sec:renormalizehorizontal}. In section~\ref{sec:AppInCorrFuncs} we will show by direct calculation that they indeed appear in the Regge limit of local correlation functions, and so should be included in the Chew-Frautschi plot.

\section{The twist-2 trajectory in the detector frame}
\label{sec:warmup twist2}

\subsection{The detector frame}

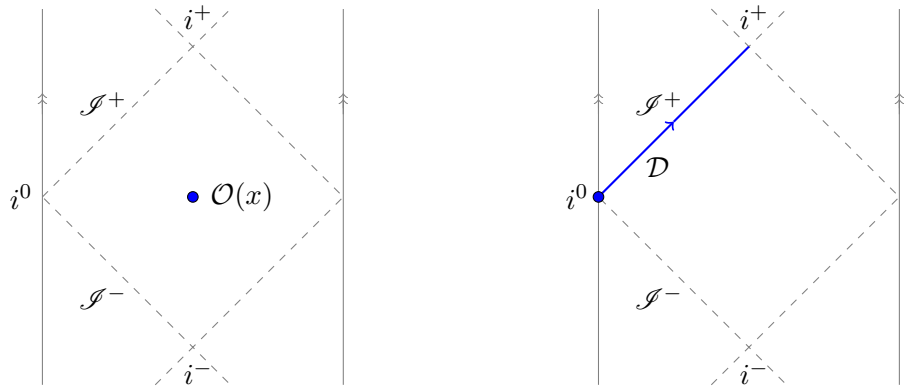
\begin{figure}
\centering
\begin{subfigure}[t]{0.5\textwidth}
\hspace{0.6in}
\begin{tikzpicture}
\draw [gray] (0,-0.5) -- (0,4.5);
\draw [gray] (4,-0.5) -- (4,4.5);
\draw [gray,->] (0,3.1) -- (0,3.3);
\draw [gray,->] (0,3.18) -- (0,3.38);
\draw [gray,->] (4,3.1) -- (4,3.3);
\draw [gray,->] (4,3.18) -- (4,3.38);
\draw [dashed,gray] (0,2) -- (2,4) -- (4,2) -- (2,0) -- (0,2);
\draw [dashed,gray] (1.5,-0.5) -- (2,0) -- (2.5,-0.5);
\draw [dashed,gray] (1.5,4.5) -- (2,4) -- (2.5,4.5);
\draw[black,fill=blue] (2,2) circle (2pt);
\node [right] at (2.1,2) {$\cO(x)$};
\node [above] at (0.8,2.95) {$\mathscr{I}^+$};
\node [below] at (0.8,0.95) {$\mathscr{I}^-$};
\node [left] at (0,2) {$i^0$};
\node [above] at (2.07,4.1) {$i^+$};
\node [below] at (2.07,-0.05) {$i^-$};
\end{tikzpicture}
\end{subfigure}
\begin{subfigure}[t]{0.45\textwidth}
\hspace{0.4in}
\begin{tikzpicture}
\draw [gray] (0,-0.5) -- (0,4.5);
\draw [gray] (4,-0.5) -- (4,4.5);
\draw [gray,->] (0,3.1) -- (0,3.3);
\draw [gray,->] (0,3.18) -- (0,3.38);
\draw [gray,->] (4,3.1) -- (4,3.3);
\draw [gray,->] (4,3.18) -- (4,3.38);
\draw [dashed,gray] (0,2) -- (2,4) -- (4,2) -- (2,0) -- (0,2);
\draw [dashed,gray] (1.5,-0.5) -- (2,0) -- (2.5,-0.5);
\draw [dashed,gray] (1.5,4.5) -- (2,4) -- (2.5,4.5);
\draw [->,thick,blue] (0,2) -- (1,3);
\draw [thick,blue] (1,3) -- (2,4);
\draw[black,fill=blue] (0,2) circle (2pt);
\node [right] at (0.5,2.4) {$\cD$};
\node [above] at (0.8,2.95) {$\mathscr{I}^+$};
\node [below] at (0.8,0.95) {$\mathscr{I}^-$};
\node [left] at (0,2) {$i^0$};
\node [above] at (2.07,4.1) {$i^+$};
\node [below] at (2.07,-0.05) {$i^-$};
\end{tikzpicture}
\end{subfigure}
\caption{Illustrations of the traditional conformal frame (left), with a local operator $\cO$ at a generic point inside Minkowski space, and the detector frame (right), where a detector $\cD$ lies along future null infinity $\mathscr{I}^+$. The detector $\cD$ transforms like a primary operator at spatial infinity $i^0$ (the blue point), which means that it is invariant under Minkowski translation generators, and this statement is exact in perturbation theory. In a CFT, future null infinity is not a special place, and can be reached by a simple shift on the Lorentzian cylinder. For simplicity of illustration, both figures show the 2d Lorentzian cylinder. For example, in the left figure, the two solid grey lines should be identified. Minkowski space is the interior of the diamond formed by dashed grey lines. The same is true for the figure on the right. \label{fig:differentframes}}
\end{figure}

Having described general features of the space of detectors in the Wilson-Fisher theory, let us now turn to  constructing detectors explicitly in perturbation theory.
Detectors live at future null infinity $\mathscr{I}^+$. This is not a special place in a CFT, since it can be mapped elsewhere by conformal transformations. However, perturbation theory does not respect conformal invariance in intermediate steps. Thus, defining detectors in perturbation theory is somewhat different from defining operators inside Minkowski space. The symmetries of $\mathscr{I}^+$ can provide powerful simplifications. (We will see examples where they trivialize some 2-loop integrals!) Furthermore, working with detectors leads to fruitful connections to scattering amplitudes and cross sections, as we explain shortly. We refer to the conformal frame with a detector at $\mathscr{I}^+$ as the ``detector frame.''

Before discussing the detector frame in detail, let us recall some facts about perturbation theory in the traditional conformal frame. Consider a spin-$J$ local operator $\cO(0)$ at the origin, see the left pane of figure~\ref{fig:differentframes}. To define $\cO(0)$ in perturbation theory, we start with a bare operator $\cO_0(0)$ with spin $J$ and dimension $\De_0$. Note that the spin $J$ of $\cO_0(0)$ is exact: it does not receive perturbative (or non-perturbative) corrections because the Lorentz generators are exact in perturbation theory. By contrast, the bare dimension $\De_0$ is not exact. The renormalized operator $\cO(0)$ develops an anomalous dimension $\g(J)$, and its full dimension $\De$ is given by
\be
\label{eq:deltaintermsofj}
\De &= \De_0+\g(J).
\ee
Unlike the Lorentz generators, the dilatation generator $D$ and special conformal generators $K^\mu$ do receive corrections. In particular the condition that $\cO(0)$ is a primary (i.e.\ that it is killed by $K^\mu$) gets corrections.

Now consider the detector frame, depicted on the right of figure~\ref{fig:differentframes}.
Detectors transform like primary operators at spatial infinity. Consequently, the condition of primariness for a detector is just translation invariance:
\be
\label{eq:primarinesscondition}
[P^\mu,\cD] &= 0.
\ee
(Recall that a primary operator at the origin is killed by $K^\mu$, while a primary at spatial infinity is killed by $P^\mu$.)
The statement of primariness for detectors is exact in perturbation theory because the translation generators are exact.

For example, consider a free scalar field $\f(x)$. We can define it at future null infinity via the limit\footnote{In the embedding formalism, this is equivalent to $\f(X)$ where $X=(X^+,X^-,X^\mu)=(0,-\a,z)$. In the following it will be useful to note that $\f(\a,z)$ has mass-dimension $0$ (dilatations still act on $\a$), and a homogeneity property $\f(\a,\l z)=\l^{-\De_\f}\f(\l^{-1} \a,z)$. These follow directly from the limit definition.}
\be
\f(\a;z) &= \lim_{L\to\oo} L^{\De_\f}\f(x+Lz),
\ee
where $\a=-2x\.z$ and $\De_\f=\frac{d-2}{2}$. Here, $z=(1,\vec n)$ is a future-pointing null vector with $\vec n\in S^{d-2}$ labeling a point on the celestial sphere, and $\a$ is twice the retarded time of $\f(\a,z)$. Translations simply shift retarded time:
\be
[P^\mu,\f(\a,z)] = -2z^\mu \ptl_\a \f(\a,z).
\ee
Thus, an example of a primary detector is 
\be
\label{eq:bareprimarydetector}
\cD_\psi(z) &= \int d\a_1\dots d\a_n\, \psi(\a_1,\dots,\a_n): \f(\a_1,z)\cdots \f(\a_n,z):\,,
\ee
where $\psi(\a_1,\dots,\a_n)$ is any 1-dimensional translationally-invariant kernel. When we turn on interactions, the operator $\cD_\psi(z)$ will remain primary to all orders in perturbation theory. More general detectors can be built from products of $\cD_\psi(z)$'s with different $\psi$'s and $z$'s, or from analogs of $\cD_\psi$ involving derivatives of $\f$.

We can also classify detectors into irreducible representations of the Lorentz group. We denote spin in the detector frame by $J_L$.  A traceless-symmetric-tensor detector $\cD(z)$ with spin $J_L$ is a homogeneous function of a null vector $z$ of degree $J_L$, which transforms under Lorentz transformations as
\be
U_\Lambda \cD(z) U_\Lambda^\dag &= \cD(\Lambda z).
\ee
(For detectors in more general Lorentz representations, $\cD(z)$ carries additional indices, and $J_L$ denotes the length of the first row of the Young diagram of the representation.) Note that $J_L$ need not be an integer. 
 For example, if the kernel $\psi(\a_1,\dots,\a_n)$ is homogeneous in the $\a$'s, then the operator $\cD_\psi(z)$ has spin
\be
J_{L}[\cD_\psi] &= n(1-\De_\f) + \deg_\a \psi,
\ee
where we've used that $d\a\, \f(\a,z)$ has degree $1-\De_\f$ in $z$. For instance, the following ``twist-2" detector has spin $J_L$:\footnote{We refer to (\ref{eq:twofieldoperator}) as ``twist-2" because in the free theory it has twist $2\De_\f=d-2$, which is 2 when $d=4$.}
\be
\label{eq:twofieldoperator}
\cD_{J_L}(z) &\equiv \frac 1 {C_{J_L}}\int d\a_1 d\a_2 |\a_1-\a_2|^{2(\De_\f-1)+J_L} : \f(\a_1,z)\f(\a_2,z):.
\ee
For future convenience, we choose the constant $C_L$ to be
\be
C_{J_L} &= 2^{J_L+d-1} \pi \sin\p{\pi \tfrac{J_L+2\De_\f}{2}} \G(2\De_\f+J_L-1).
\label{eq:constantfortwisttwo}
\ee
This choice ensures that formula (\ref{eq:twist two vertex}) below is as simple as possible.\footnote{As $J_L$ is varied, the constant $C_{J_L}$ may become 0 or singular. These features are somewhat of a red herring. Some of them are related to the fact that the factor $|\a_1-\a_2|^{2(\De_\f-1)+J_L}$ in~\eqref{eq:twofieldoperator} can become singular. For example, if $2\De_\f+J_L=1$, then $C_{J_L}$ has a pole. This pole cancels the pole coming from $|\a_1-\a_2|^{-1}$, turning it into a delta-function $\de(\a_1-\a_2)$. This ensures that the bare detector becomes the light transform of $\phi^2$. Other features can be explained by similar degenerations.} Because the Lorentz group is exact in perturbation theory, the quantum number $J_L$ does not receive corrections when we turn on interactions.

Finally, the dimension in the detector frame $\De_L$ is minus the eigenvalue of the dilatation generator $D$:
\be
\label{eq:minustheeigenvalue}
[D,\cD(z)] &= -\De_L \cD(z).
\ee
(The minus sign is because $\cD(z)$ transforms like a primary operator at infinity.) Of course $\De_L$ does receive perturbative corrections because $D$ does.

In summary, detectors are characterized by the following data
\be
&\begin{array}{l}
\textrm{primariness $[P^\mu,\cD(z)]=0$,} \\
\textrm{detector spin }J_L,
\end{array}
&&\Bigg\}
 \ \textrm{exact in perturbation theory,} \nn\\
&\,\textrm{detector dimension }\De_L. &&\Big\}
\  \textrm{corrected in perturbation theory.}
\ee
To define a detector in perturbation theory, we begin with a bare primary detector $\cD_0(z)$ with spin $J_L$ and tree-level dimension $\De_{L,0}(J_L)$. (For example, the tree-level dimension of the twist-2 operator (\ref{eq:twofieldoperator}) is $\De_{L,0}=J_L+d-2$.) Renormalizing $\cD_0$, we obtain an anomalous dimension $\g_L(J_L)$. The full dimension of the renormalized operator is then
\be
\label{eq:anomalousdetectordimension}
\De_L &= \De_{L,0}(J_L) + \g_L(J_L).
\ee

These considerations may seem elementary, but they give a surprising way to access Regge trajectories of light-ray operators that essentially flips the role of $\De$ and $J$! Consider the leading-twist detector $\mathbb{O}_J^+(\oo,z)$ in the interacting theory. To construct it using perturbation theory, we must start with a bare detector $\cD_{J_L}(z)$ with the same Lorentz spin, i.e.\ $J_L=1-\De(J)$, where $\De(J)$ corresponds to the full interacting theory. When we renormalize $\cD_{J_L}(z)$, it acquires a detector anomalous dimension $\g_L(J_L)$ and the renormalized operator $[\cD_{J_L}]_R$ will be $\mathbb{O}_J^+$:
\be
[\cD_{1-\De(J)}(z)]_R &= \mathbb{O}_J^+(\oo,z).
\ee
Using (\ref{eq:anomalousdetectordimension}) and the dictionary $(J_L,\De_L)=(1-\De,1-J)$, this gives a relation between $\De=\De(J)$ and $J$
\be
\label{eq:jintermsofdelta}
J &= J_0(1-\De)-\g_L(1-\De),
\ee
where we have defined the function $J_0(J_L)$ by $\De_{L,0}(J_L)=1-J_0(J_L)$.
Thus, the detector frame anomalous dimension $\g_L(J_L)$ naturally gives $J$ as a function of $\De$, instead of the more traditional $\De$ as a function of $J$. Due to (\ref{eq:jintermsofdelta}), we sometimes abuse terminology and refer to $-\g_L(1-\De)$ as an ``anomalous spin," since it is a correction to $J$.

{\bf To summarize:} In terms of the Chew-Frautschi plot, working in the traditional frame amounts to fixing the vertical position $J$ and computing corrections to the horizontal position $\De$. By contrast, working in the detector frame amounts to fixing the horizontal position $\De$ and computing corrections to the vertical position $J$. 

\subsubsection{Timelike anomalous dimensions and reciprocity}
\label{sec:reciprocity}

Thus, the traditional frame and the detector frame give us two ways to access the same Regge trajectory: (1) renormalize local operators in the traditional frame and compute $\g(J)$, or (2) renormalize detectors in the detector frame and compute $\g_L(1-\De)$. It turns out that $\g_L(1-\De)$ is a simple rewriting of a so-called ``timelike anomalous dimension" $\g_T(N)$, while $\g(J)$ is sometimes called a ``spacelike anomalous dimension." This leads to a simple explanation of the so-called ``reciprocity" relation between spacelike and timelike anomalous dimensions in CFT \cite{Basso:2006nk}.

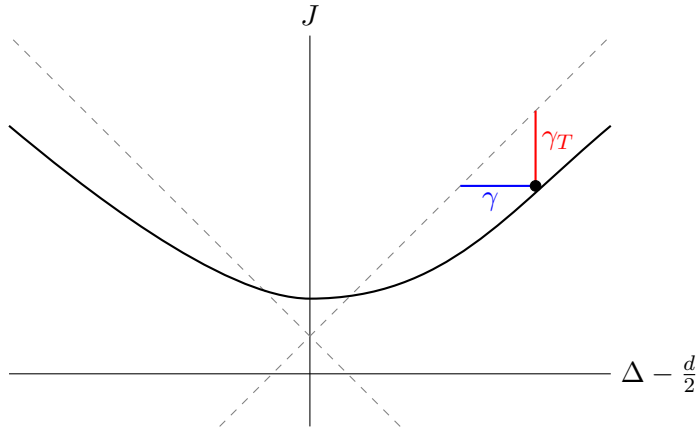
\begin{figure}
\centering
\begin{tikzpicture}
\draw[] (-4,-0.5) -- (4,-0.5);
\draw[] (0,-1.2) -- (0,4);
\node[right] at (4,-0.5) {$\De-\tfrac d 2$};
\node[above] at (0,4) {$J$};
\draw[gray,dashed] (-1.2,-1.2) -- (4,4);
\draw[gray,dashed] (1.2,-1.2) -- (-4,4);
\draw[thick] (-4,2.8) to[out=-40,in=180,looseness=0.65] (0,0.5) to[out=0,in=220] (4,2.8);
\draw[thick,red] (3,2) -- (3,3);
\draw[thick,blue] (3,2) -- (2,2);
\draw[black,fill=black] (3,2) circle (2pt);
\node[right] at (2.95,2.6) {\color{red}{$\g_T$}};
\node[below] at (2.4,2.05) {\color{blue}{$\g$}};
\end{tikzpicture}
\caption{The anomalous dimension $\g$ (shown in blue) measures the distance between a Regge trajectory and a 45$^\circ$ line, with fixed $J$, as appropriate for the traditional conformal frame. The timelike anomalous dimension $\g_T=\g_L$ (shown in red) measures the distance between a Regge trajectory and a 45$^\circ$ line, with fixed $\De$, as appropriate for the detector frame. They are equal because they form equal edges of an isosceles right triangle. This is the statement of reciprocity \cite{Basso:2006nk}.
\label{fig:reciprocity}}
\end{figure}

In more detail, consider a Regge trajectory with tree-level twist $\tau_0$. For example, the ``twist-2" operators (\ref{eq:twofieldoperator}) have $\tau_0=d-2$. 
In the traditional frame, we have
\be
\label{eq:spacelikedim}
\De=\tau_0+J+\g(J).
\ee
By contrast, in the detector frame, (\ref{eq:jintermsofdelta}) becomes
\be
\label{eq:timelikedim}
\De
&= \tau_0 + J + \g_L(1-\De).
\ee
Let us define 
the ``timelike anomalous dimension" $\g_T$ by the trivial redefinition
\be
\g_L(1-\De) &\equiv \g_T(\De-\tau_0). \label{gL from gT}
\ee
Together, (\ref{eq:spacelikedim}) and (\ref{eq:timelikedim}) imply the functional equation 
\be
\g_T(N)=\g(N-\g_T(N)).
\ee
Figure~\ref{fig:reciprocity} gives a geometric interpretation of this equality.
This is the statement of ``reciprocity" \cite{Basso:2006nk}.\footnote{This simple explanation of reciprocity clarifies its appearance in \cite{Dixon:2019uzg}, which studied the leading term in the OPE of energy detectors. In a CFT, this leading term is fixed by conformal symmetry to be a light-ray operator $\mathbb{O}_3$ with $J=3$ \cite{Hofman:2008ar}.  However, the work \cite{Dixon:2019uzg} studied the OPE in the detector frame, so they needed to access the $J=3$ operator by correctly tuning $\De$ and using reciprocity to relate timelike and spacelike anomalous dimensions.}

(Reciprocity also sometimes refers to a distinct phenomenon:
that the large-spin expansions of $\g$ proceeds in inverse powers of the conformal Casimir $h(h-1)$
where $h=\frac{\Delta(J)+J}{2}$, that is, that the function ${\cal P}(N)=\g(N-\tfrac12 {\cal P}(N)+\tfrac{\e}{2})$ admits an asymptotic series
in even powers of $1/(N+\frac12)$.  For conformal theories,
this is a consequence of the general structure of large-spin expansions as
manifested by the Lorentzian inversion formula \cite{Alday:2015eya,Simmons-Duffin:2016wlq,Caron-Huot:2017vep}.)

\subsubsection{The in-in formalism and weighted cross-sections}

Detectors annihilate the vacuum:
\be
\cD |\Omega\> &= 0.
\ee
In conformal field theory, this follows from the identification of detectors with light-ray operators at infinity $\mathbb{O}_i(\oo,z)$, which must annihilate the vacuum by representation-theoretic arguments \cite{Kravchuk:2018htv}. More generally, in a not-necessarily-conformal theory, the primariness condition (\ref{eq:primarinesscondition}) implies that $\cD|\Omega\>$ is a zero-energy state, and therefore proportional to the vacuum. If $\cD$ has nonzero Lorentz spin $J_L$ (in a general theory) and/or dimension $\De_L$ (in the case of a CFT), then the constant of proportionality must vanish.

Thus, the simplest non-vanishing matrix elements involving detectors are ``event shapes," i.e.\ matrix elements in a non-vacuum state $|\Psi\>$:
\be
\label{eq:eventshape}
\<\Psi|\cD|\Psi\>.
\ee
Such event shapes can be computed using the  Schwinger-Keldysh or ``in-in" formalism. The ket $|\Psi\>$ is described by a path integral with the usual Feynman rules, implementing forward time-evolution. The bra $\<\Psi|$ is described by a path integral with complex-conjugated Feynman rules, implementing backward time-evolution. The detector $\cD$ lives on a ``fold" connecting these two path integrals. In Feynman diagrams, we denote the fold pictorially by a gray line, with the region below the fold representing the ket $|\Psi\>$ and the region above the fold representing the bra $\<\Psi|$. See figures~\ref{fig:twist2vertex} and \ref{fig:twist2 2loops} below for examples.

Many interesting bare detectors are diagonalized on scattering states of the free theory:
\be
\label{eq:diagonalizedonscatteringstates}
\cD|p_1,\dots,p_k\>_\mathrm{out} = f_\cD(p_1,\dots,p_k) |p_1,\dots,p_k\>_\mathrm{out}.
\ee
For example, a product of average null energy (ANEC) operators has this property. When (\ref{eq:diagonalizedonscatteringstates}) holds, the event shape (\ref{eq:eventshape}) can be written as a sum over final states $|p_1,\dots,p_k\>_\mathrm{out}$
\be
\label{eq:finalstatesum}
\<\Psi|\cD|\Psi\> &= \sum_{k,p_1,\dots p_k} |{}_\mathrm{out}\<p_1,\dots,p_k|\Psi\>|^2 f_\cD(p_1\dots p_k),
\ee
where $\sum$ denotes a sum over particles and integral over phase space. In the in-in formalism, this representation comes about because propagators between the ket and bra sheets of the path integral are Wightman propagators, which are supported on-shell. Expressing them as sums over on-shell states, we obtain (\ref{eq:finalstatesum}).

We could choose $|\Psi\>$ itself to be a free theory scattering state $|\Psi\>=|q_1,\dots,q_l\>_\mathrm{in}$. This leads to an interpretation for matrix elements of $\cD$ in terms of weighted cross-sections
\be
\<\Psi|\cD|\Psi\> &= \sum_{k,p_1,\dots p_k} |{}_\mathrm{out}\<p_1,\dots,p_k|q_1,\dots,q_l\>_\mathrm{in}|^2  f_\cD(p_1\dots p_k) \nn\\
&= \sum_{k,p_1,\dots p_k} \s(q_1,\dots,q_l \to p_1,\dots,p_k) f_\cD(p_1,\dots,p_k).
\label{eq:weightedcrosssection}
\ee
While this interpretation is perhaps the most transparent one, it conflates two problems: the IR safety of the detector $\cD$ and the IR divergences associated with the initial state. For this reason, we will stick with states $|\Psi\>$ that are clearly well-defined in the interacting theory, such as $|\Psi\>=\cT\{\f(q_1)\cdots\f(q_l)\}|0\>$, where the momenta $q_i$ are generic and off-shell.

For well-defined $|\Psi\>$, the divergences in matrix elements of bare detectors $\cD_0$ are just the IR divergences in the weighted cross sections (\ref{eq:finalstatesum}). Traditionally, one focuses on IR-safe detectors with $f_\cD$ invariant under soft and collinear splittings. As we will see, there exist detectors that are not IR-safe and yet their associated IR divergence can be multiplicatively renormalized (as opposed to being absent altogether in the IR-safe case). 

\subsection{The leading Regge trajectory in the Wilson-Fisher theory}
\label{sec:leadingreggedetector}

Let us illustrate these ideas by renormalizing the twist-two detectors (\ref{eq:twofieldoperator}) in the Wilson-Fisher theory. This will provide our first example of fixing $\De$ and computing an ``anomalous spin" $-\g_L(1-\De)$. In the end, we will recover conventional results for anomalous dimensions of twist-two operators in the Wilson-Fisher theory, which serves as a useful consistency check on our methods. This computation will also serve as a warmup before tackling more exotic types of detectors in sections~\ref{sec:pomeron} and~\ref{sec:WF horizontal}.

Let us start by determining the Feynman rule for insertions of the bare operator $\cD_{J_L}(z)$. This can be read off from the tree-level matrix element
\be
\label{eq:vertex definition}
\<0|\f(-q)\cD_{J_L}(z)\f(p)|0\> = (2\pi)^d\delta^d(p-q) V_{J_L}(z;p).
\ee
We often abuse notation and write 
\be
\<0|\f(-p)\cD_{J_L}(z)\f(p)|0\> = V_{J_L}(z;p),
\label{eq:abusenotation}
\ee
where we implicitly strip off $(2\pi)^d$ times the momentum-conserving $\de$-function when the initial and final momenta of an event shape are equal.
A straightforward computation (see appendix~\ref{app:light transformed structures}) gives 
\be
\label{eq:treelevelvvertex}
\<0|\f(\a;z)\f(p)|0\> &= e^{-\frac{i\pi\De_\f}{2}}2^{\De_\f}\pi^{d/2}\int\limits_0^\oo d\b\, \de^d(p-\b z)\b^{\De_\f-1} e^{-\frac i 2\a\b}.
\ee
The $\delta$-function ensures that only particles moving in the direction $z$ contribute, and we see that their
energy $\beta$ is Fourier conjugate to the arrival time $\a$.
Plugging this in to (\ref{eq:twofieldoperator}), we find the following simple result for the vertex $V_{J_L}(z;p)$: 
\be
\label{eq:twist two vertex}
V_{J_L}(z;p)&=\int\limits_{0}^{\infty} d\b \b^{-J_L-1} \de^d(p-\b z).
\ee
In fact, this result for the vertex $V_{J_L}(z;p)$ is completely determined (up to normalization) by the symmetries, momentum conservation, and positivity of energy. 

\begin{figure}
\centering
\includegraphics[scale=.3]{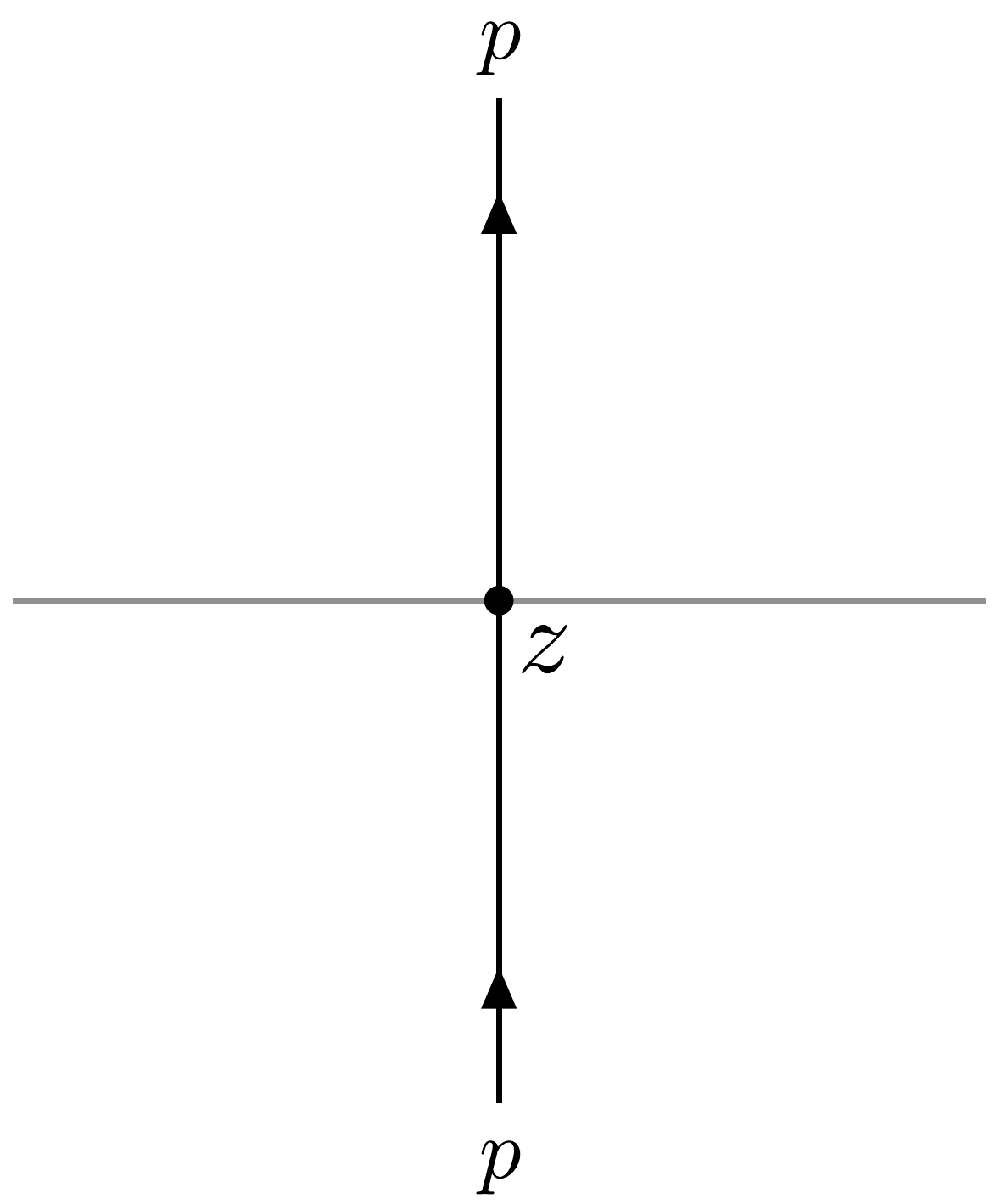}
\caption{Vertex for the twist-two light-ray operators. This represents a time-folded diagram where the gray horizontal line separates the lower and upper folds. The momentum runs from the lower to upper fold. The black dot on the gray horizontal line is the insertion of $\cD_{J}$ at null infinity.}
\label{fig:twist2vertex}
\end{figure}

A diagrammatic representation of $V_{J_L}(z;p)$ is shown in figure \ref{fig:twist2vertex}.
This is a time-folded diagram where the bottom and top edges correspond to $t=-\infty$, while the horizontal line in the middle corresponds to $t=\infty$. Time increases in the lower half of the diagram as we approach the horizontal line from below and decreases once we cross it.
The horizontal line represents the fold where we insert $\cD_{J_L}$ (indicated by a dot).
This picture comes from using a time-ordered path integral to create the ket state $\f(p)|0\>$ and an anti-time-ordered path integral to create the bra state $\<0|\f(-p)$,
and the fold, or ``cut'', separates the amplitude from its complex conjugate.

Overall, the Feynman rules for in-in calculations involving insertions of $\cD_{J_L}$ are: \begin{enumerate}
\item Each interaction vertex on the lower sheet gets a factor of $i\lambda \tilde{\mu}^{\epsilon}$.
\\
Each interaction vertex on the upper sheet gets a factor of $-i\lambda \tilde{\mu}^{\epsilon}$.
\item For propagators on the lower sheet use the time-ordered propagator $-i/(p^{2}-i0)$.
\\
For propagators on the upper sheet use the anti-time-ordered propagator $i/(p^{2}+i0)$.
\item For propagators between the lower and upper sheets use the Wightman propagator $2\pi\delta(p^{2})\th(p^0)\equiv (2\pi)\delta^+(p^2)$. Note that only positive momenta flow through the fold.
\item For a line with momentum $p$ passing through an insertion of $\cD_{J_L}(z)$, include a factor of $V_{J_{L}}(p;z)$. (Do not include extra propagators for the segments of the line on either side of the insertion, since these are already included in $V_{J_{L}}(p;z)$. Using cross-section interpretation, this corresponds to the usual fact that we should be computing amputated diagrams.)
\item For the time-ordered initial state $\cT\{\f(p_1)\cdots\f(p_n)\}|0\>$ add $n$ univalent vertices sourcing momenta $p_i$ on the lower sheet.
\item For the anti-time-ordered final state $\<0|\bar\cT\{\f(-q_1)\cdots\f(-q_m)\}$ add $m$ univalent vertices sinking momenta $q_i$ on the upper sheet.
\item Multiply by an overall momentum conserving $\delta$-function, $(2\pi)^d\delta(p_1+\ldots+p_n-q_1-\ldots-q_m)$.
\end{enumerate}

\begin{figure}
\centering
\includegraphics[scale=.3]{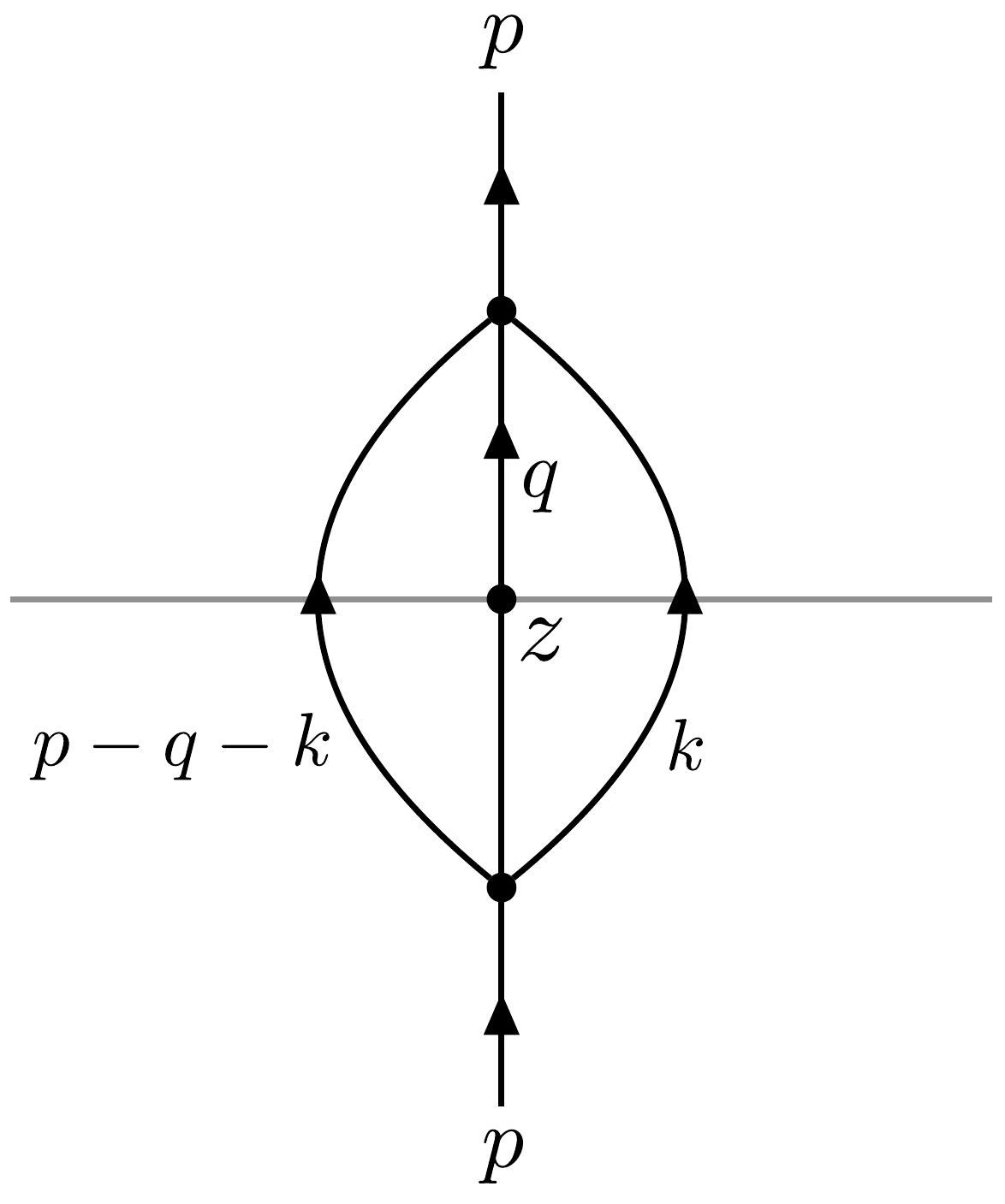}
\caption{Two loop correction to the vertex $V_{J_{L}}$. All momentum flows up from the lower to the upper fold.}
\label{fig:twist2 2loops}
\end{figure}

With these rules in hand, we are ready to study loop corrections to $\cD_{J_L}$. We focus on the event shape
\be
\label{eq:eventshapethatwestudy}
\<\Omega|\phi_R(-p) \cD_{J_L}(z) \phi_R(p)|\Omega\>,
\ee
where we implicitly strip off the momentum-conserving $\de$-function as in (\ref{eq:abusenotation}). To make the initial and final states well-defined, we have inserted renormalized operators
\be
\f_R(p) &= Z_\f^{-1/2} \f(p),
\ee
where the wavefunction renormalization factor for $\phi$ is given by
\be
Z_{\f}&=1-\frac{1}{\epsilon}\frac{\lambda^{2}}{12(4\pi)^{4}}+O(\lambda^{3}).
\ee
Any divergences in the event shape (\ref{eq:eventshapethatwestudy}) must be removed by multiplicative renormalization of the operator $\cD_{J_L}$, from which we can read off its anomalous dimension.

At one loop, there are no nontrivial contributions to (\ref{eq:eventshapethatwestudy}): the sole interaction vertex must lie either below or above the fold, and the resulting loop integral simply gives a mass correction to the propagator, which vanishes in dimensional regularization. At two loops, the only nontrivial diagrams are ``sunrise" diagrams. For these, we have three possibilities: (1) the interaction vertices lie on opposite sides of the fold, (2) both interaction vertices lie below the fold,  (3) both interaction vertices lie above the fold.

Case (1) is the most interesting, so we consider it first. By positivity of energy, the operator $\f(p)$ creating the initial state must connect to the interaction vertex below the fold, and $\f(-p)$ creating the final state must connect to the vertex above the fold. One of the lines between the vertices must pass through the $\cD_{J_L}$ insertion. The resulting diagram is depicted in figure~\ref{fig:twist2 2loops}, and is given by
\begin{align}
\mathcal{F}^{(2)}_{J_L}(z;p)=&\frac{(-i\l\tl\mu^\e)(+i\l\tl\mu^\e)}{2} \int \frac{d^dq}{(2\pi)^d}\frac{d^dk}{(2\pi)^d}\frac{i}{p^2+i0}\frac{-i}{p^2-i0}V_{J_{L}}(z;q)(2\pi)^2\de^+(k^2)\de^+((p-q-k)^2)
\nonumber
\\
=&(\l\tl\mu^\e)^2\frac{\vol S^{d-2}}{2^{d}(2\pi)^{2d-2}}\frac{\Gamma(\tfrac{d-2}{2})\G(-J_L)}{\G(-J_L+\tfrac{d-2}{2})}(-{2z\cdot p})^{J_L}(-p^2)^{\frac{d-4}{2}-J_L-2}\theta(-p^2). \label{eq:2looptwist2}
\end{align}
The symmetry factor  $\frac 1 2$ comes from swapping the lines that do not pass through $\cD_{J_L}$. To compute the integral, we used $\de^d(q-\b z)$ inside $V_{J_L}(q;z)$ to solve for $q$. The remaining integral over $k$ localizes to a sphere $S^{d-2}$, giving a factor $\vol S^{d-2}$. Finally, integrating over $\beta$ gives (\ref{eq:2looptwist2}).
Note that the $i0$'s play no role in this calculation, since kinematics force $p$ to be strictly timelike.

Na\"ively, the result (\ref{eq:2looptwist2}) appears to be finite.
However, there is a hidden divergence when we interpret it as a distribution in $p$.
 Setting $d=4-\epsilon$ and expanding around $\epsilon=0$ we find the pole
\be
(-{2z\cdot p})^{J_L}(-p^2)^{\frac{d-4}{2}-J_L-2}&=\frac{\pi}{2\e (J_L+1)} \int d\b\b^{-J_L-1} \de^{d}(p-\b z)+O(\e^0)
\nn \\ &=\frac{\pi}{2\e (J_L+1)}V_{J_L}(z;p)+O(\e^0). \label{eq:pole dist tw2}
\ee
The derivation of this identity can be found in Appendix \ref{app:poles dist}.
The divergence originates from the region where $q$, $k$ and $z$ are all collinear, whence the factor $V_{J_L}(z;p)$.
Plugging it into  (\ref{eq:2looptwist2}), this gives a divergence proportional to the tree-level vertex $V_{J_L}(z;p)$:
\be
\mathcal{F}^{(2)}_{J_L}(z;p) &= -\frac{1}{\epsilon}\frac{\lambda^{2}}{2(4\pi)^{4}}\frac{1}{J_L(J_L+1)} V_{J_L}(z;p) + O(\e^0).\label{eq:twoloopdiv}
\ee
This will make multiplicative renormalization possible.

Finally, let us consider cases (2) and (3), where the two-loop sunsets are entirely below or above the fold. In either case, the diagram is proportional to
\be
 \int d^dk\, d^d q \frac{1}{q^2} \frac{1}{k^2} \frac{1}{(p-q-k)^2} \frac{1}{p^2} V_{J_L}(z;p) 
&\ \propto\ p^{-2\e} V_{J_L}(z;p),
\label{eq:sunrisebelow}
\ee
where we have performed the integral by dimensional analysis and used $d=4-\e$. Note that the vertex $V_{J_L}(z;p)$ is supported on-shell, $p^2=0$. How then should we interpret the distribution $p^{-2\e} V_{J_L}(z;p)$? We claim that it vanishes. Indeed, if $\Re \e<0$, then the distribution clearly vanishes when paired with any smooth test function. In general, we define it by analytic continuation away from this region, so it vanishes identically. Alternatively, for $p^2=0$ the integral 
\be
	\int d^dk\, d^d q \frac{1}{q^2} \frac{1}{k^2} \frac{1}{(p-q-k)^2}
\ee
does not have a scale, and such integrals are known to vanish in dimensional regularization.

In summary, the bare detector event shape (\ref{eq:eventshapethatwestudy}) is given up to two-loop order by
\be
\<\Omega|\phi_R(-p) \cD_{J_L}(z) \phi_R(p)|\Omega\> &= Z_\f^{-1} \p{V_{J_L}(z;p) + \cF_{J_L}^{(2)}(z,p)} + O(\l^3),
\label{eq:twoloopbareeventshape}
\ee
where the $Z_\f^{-1}$ factor comes from the renormalized $\f_R$ operators.
There are divergences from $Z_\f^{-1}$ and also the $\e^{-1}$ pole in $\mathcal{F}^{(2)}_{J_L}(z;p)$. They can be cancelled up to $O(\l^3)$ by defining the renormalized operator
\be\label{eq:Zleading}
[\cD_{J_L}(z)]_R &= Z_{J_L}^{-1} \cD_{J_L}(z) \nn\\
Z_{J_L} &\equiv Z_\f^{-1} \p{1-\frac{1}{\epsilon}\frac{\lambda^{2}}{2(4\pi)^{4}}\frac{1}{J_L(J_L+1)}} + O(\l^3).
\ee
The anomalous dimension of $\cD_{J_L}$ in the detector frame is then 
\begin{align}
\label{eq:gammaltwisttwo}
-\gamma_{L}(J_L)=\frac{\partial \log Z_{J_{L}}}{\partial \lambda}\beta(\lambda)=\frac{\lambda^{2}}{(4\pi)^{4}}\left(\frac{1}{J_L(J_L+1)}-\frac{1}{6}\right)+O(\lambda^{3}),
\end{align}
where the $\beta$-function is 
\begin{align}
\beta(\lambda)=-\epsilon\lambda+3\frac{\lambda^{2}}{(4\pi)^{2}}+O(\lambda^3).
\end{align}
The minus sign on the left-hand side of (\ref{eq:gammaltwisttwo}) comes from the fact that we define the dimension of a detector as if it were a primary at infinity. (It is the same minus sign as in (\ref{eq:minustheeigenvalue}).)

The fixed-point value of the coupling is
\begin{align}
\lambda_*=\frac{(4\pi)^{2}}{3}\epsilon+\dots.
\end{align} 
Plugging this into $\g_L(J_L)$, the expression (\ref{eq:jintermsofdelta}) 
for the twist-2 Regge trajectory becomes
\be
J(\De) &= \De-(d-2) - \g_L(1-\De) \nn\\
&= \De - 2+\e+\frac{\e^2}{9} \p{\frac{1}{(\De-1)(\De-2)}-\frac 1 6} + O(\e^3).
\label{eq:jintermsofdeltaleadingregge}
\ee
Equation (\ref{eq:jintermsofdeltaleadingregge}) gives the promised perturbative expansion for $J$ in terms of $\De$. If we solve it for $\De$ in terms of $J$, the result agrees precisely with (\ref{eq:leadingreggetrajectoryabstract}) and (\ref{eq:gammaofJ}) at $O(\e^2)$.

One lesson from this exercise is that we can define and renormalize $\cD_{J_L}(z)$ using only its action on on-shell states.
Specifically, in the above computation, we needed only the vertex $V_{J_L}(z;p)$ --- never the explicit definition (\ref{eq:twofieldoperator}).
In fact, this lesson holds more generally. When defining and renormalizing detectors in perturbation theory, it is sufficient (and often more convenient) to specify their action on on-shell states.
This method is distinct from the usual renormalization of twist-2 operators (which can also exploit matrix elements between on-shell parton states, see e.g.~\cite{Moch:1999eb}),
since usually operators are inserted at finite positions rather than at infinity; the resulting Regge trajectory coincides at the critical point $\lambda=\lambda_*$.

\section{The leading intersection and the Pomeron}
\label{sec:pomeron}

In the previous section, we studied the renormalization of the leading-twist detectors $\cD_{J_L}$, which precisely
define ``$E^{J-1}$ flux'' for a certain $J=1-\Delta_L(J_L)$.
We found that a renormalized detector $[\cD_{J_L}]_R$ can be defined in~\eqref{eq:Zleading} so that divergences in its matrix elements cancel. However, the definition~\eqref{eq:Zleading} does not work for $J_L=0$ or $J_L=-1$, since the renormalization constant $Z_{J_L}$ becomes ill-defined. This also leads to the singularities at $\De=1$ and $\De=2$ in~\eqref{eq:jintermsofdeltaleadingregge}.

In section~\ref{sec:summary_mixing}, we anticipated that the problem at $J_L=-1$ ($\De=2$) is due to the intersection of the leading twist trajectory with its shadow, see figure~\ref{fig:freediagonals}. In this section, we explicitly confirm this expectation. We start with a discussion of the shadow trajectory in general. We then discuss its role in perturbation theory and see how the mixing happens.

\subsection{Shadow symmetry of Regge trajectories}
\label{sec:shadowtransform}

As discussed in section~\ref{sec:shadows}, the shadow of $\cD_{J_L}(z)$ takes the (schematic) form
\be\label{eq:shadowdetectorsketch}
	\wS_J[\cD_{J_L}](z)=\int D^{d-2}z' (-2z\.z')^{2-d-J_L}\cD_{J_L}(z').
\ee
The shadow detector $\wS_J[\cD_{J_L}]$ has the same scaling dimension as $\cD_{J_L}$, but its Lorentz spin is $2-d-J_L$ instead of $J_L$. The trajectories $\cD_{J_L}$ and $\wS_J[\cD_{J_L}]$ thus intersect at $J_L=\tfrac{2-d}{2}$.

The expression (\ref{eq:shadowdetectorsketch}) is schematic because we have been imprecise about the convergence of the integral. The variable $z'$ is integrated over the projective future null cone, which is compact, as it is just a parametrization of the celestial sphere. Assuming that $\cD_{J_L}(z')$ is well-defined, divergences can only come from the factor $(-2z\.z')^{2-d-J_L}$, which is singular when $z'\propto z$. To study the singularity, we parametrize $z=(z^+,z^-,z^i)=(1,y^2,y^i)$ using a $(d-2)$-dimensional coordinate $y$. The right-hand side of~\eqref{eq:shadowdetectorsketch} becomes
\be
\int d^{d-2}y' |y-y'|^{2(2-d-J_L)} \cD_{J_L}(y')
&=\int d^{d-2}x |x|^{2(2-d-J_L)} \cD_{J_L}(y+x) \nn\\
&=\int_{S^{d-3}} d\O_{d-3} \int_0^\oo dr r^{-1+(2-d-2J_L)}\cD_{J_L}(y+x),
\label{eq:integralinradialcoordinates}
\ee
where in the last line, we chose radial coordinates for $x\in \R^{d-2}$. Since nothing special happens to $\cD_{J_L}(y+x)$ at $x=0$, this integral is convergent near $r=0$ for $J_L<\frac{2-d}{2}$. (We don't have to worry about $r=\oo$ because it is merely a coordinate singularity and is a regular point on the celestial sphere.  Practically speaking, the $r\to \oo$ limit is regulated by the decay of $\cD_{J_L}(y+x)$ at large $x$, which in turn is due to the singular Weyl factor coming from the coordinate choice.)

More generally, as discussed in appendix~\ref{app:poles dist}, the integral (\ref{eq:integralinradialcoordinates}) should be defined by analytic continuation in $J_L$ away from the region where it converges. The result is that we can view the expression
\be
	|y-y'|^{2(2-d-J_L)}
\ee
as a well-defined distribution in $y'$ for all $J_L\in \C\setminus\{\tfrac{2-d}{2},\tfrac{2-d}{2}+1,\tfrac{2-d}{2}+2,\cdots\}$. Near $J_L=\tfrac{2-d}{2}+n$, we have simple poles
\be
	|y-y'|^{2(2-d-J_L)}\propto \frac{1}{J_L-\tfrac{2-d}{2}-n}\ptl^{2n}\de^{d-2}(y-y').
\ee

It is helpful to define the following rescaled version of $\wS_J$,
\be
	\wS'_J=\wS_J \frac{2}{\vol S^{d-3}\G(\tfrac{2-d}{2}-J_L)},
\ee
where $J_L$ reads off the spin of the object on which it acts. This cancels the above poles and also makes the coefficient of the delta-function $\de^{d-2}(y-y')$ at the $n=0$ pole equal to 1.
The transform $\wS'_J$ now has two key properties: it is well-defined for all $J_L\in \C$, and for $J_L=\tfrac{2-d}{2}$ it acts as the identity:
\be
	\wS'_{J}[\cD_{\frac{2-d}{2}}]=\cD_{\frac{2-d}{2}}. \label{wS' identity}
\ee

In perturbation theory, if we define a renormalized detector $[\cD_{J_L}]_R=Z_{J_L}^{-1}\cD_{J_L}$, the same factor $Z_{J_L}$ renormalizes $\wS'_{J_L}[\cD_{J_L}]$. This is because for any matrix element we can write
\be
	\<\Psi|Z_{J_L}^{-1}\wS'_{J_L}[\cD_{J_L}]|\Phi\>=\wS'_{J_L}[\<\Psi|Z_{J_L}^{-1}\cD_{J_L}|\Phi\>]=\wS'_{J_L}[\<\Psi|[\cD_{J_L}]_R|\Phi\>],
\ee
and the matrix elements $\<\Psi|[\cD_{J_L}]_R|\Phi\>$ are finite by construction. This means that the results of section~\ref{sec:leadingreggedetector} also renormalize the shadow of the leading twist trajectory (with the same caveats at $J_L\approx 0,-1$ as for the leading twist trajectory).

Our main motivation for discussing the shadow trajectory is to explain why we failed to define $[\cD_{J_L}]_R$ near $J_L=\tfrac{2-d}{2}\approx -1$ in perturbation theory.  We have claimed that this is due to mixing with the shadow trajectory at the intersection.
With this in mind, it will be convenient to normalize the shadow transform so that it squares to the identity for any $J_L$:
\be\label{eq:Shatdefn}
 \hat\wS_J=\wS_J \frac{\Gamma(-J_L)}{\pi^{\frac{d-2}{2}}\G(\tfrac{2-d}{2}-J_L)}\,.
\ee
This has the property that $\hat\wS_J^2=1$, which will simplify calculations below (and we still have 
$\hat\wS_{J}[\cD_{\frac{2-d}{2}}]=\cD_{\frac{2-d}{2}}$).
This comes at the expense of $\hat\wS_J$ having spurious poles for $J_L\in \Z_{\geq 0}$, but, importantly,
it is still regular for $J_L<0$. This includes the intercept point $J_L=-1$ and the points $J_L=1-\De$ with $\De$ being the scaling dimensions of the local operators, so the spurious poles will not cause problems in our calculations.

From what we have explained so far, it is not obvious how mixing can happen.
Operators with different tree-level dimensions don't mix in dimensional regularization, so we do not expect mixing away from the intersection of the trajectories. Furthermore, the tree-level trajectories intersect at $J_L=\tfrac{2-d}{2}$, but
$\hat\wS_{J}[\cD_{\frac{2-d}{2}}]=\cD_{\frac{2-d}{2}}$, so na\"ively there is only one detector at the intersection point!
(This property is important for the final picture in figure~\ref{fig:3dintercept} to be self-consistent:
otherwise the operators with $J_L=\tfrac{2-d}{2}$ would always be doubled by $\hat \wS_J$.)

As we will see, both of these problems have subtleties that resolve them. For the first problem, it is not true that there is no mixing in dimensional regularization: the coupling $\l$ always comes in the combination $\l\tl\mu^{\e}$, so mixing can occur between operators whose dimensions differ by multiples of $\e$. We can choose $J_L$ so that the scaling dimensions of the leading twist trajectory and its shadow differ by an integer multiple of $\e$, and this allows them to mix. For the second problem, even though there is only one operator at the intersection, there are two tangent spaces (as in figure~\ref{fig:freediagonals}),
and this non-analyticity will turn out to be sufficient to produce a new operator when interactions are turned on.

\subsection{The two-loop dilatation operator}

In this section we explain in detail how mixing happens between $\cD_{J_L}$ and its shadow. First, let us define the shadow detectors using the $\hat\wS_J$ version of the spin shadow transform,
\be
\tl\cD_{J_L}\equiv \hat\wS_J[\cD_{2-d-J_L}].
\ee
Note that $\tl\cD_{J_L}$ has Lorentz spin $J_L$. The property $\hat\wS_J^2=1$ ensures that 
\be
\cD_{J_L}=\hat\wS_J[\tl\cD_{2-d-J_L}],
\ee
which will simplify our calculations. Note that $\tl\cD_{J_L}$ is well-defined near $J_L=-1$.

Our goal now is to study the perturbative corrections to the matrix elements of $\cD_{J_L}$ and $\tl\cD_{J_L}$ and demonstrate that the divergences in these matrix elements can be canceled by a matrix of renormalization constants that mixes these two operators.

We start with a more careful analysis of the matrix elements
\be
	\<\cD_{J_L}\>\equiv \<0|\phi_R(-p)\cD_{J_L}\phi_R(p)|0\>\qquad\text{and}\qquad	\<\tl\cD_{J_L}\>\equiv\<0|\phi_R(-p)\tl\cD_{J_L}\phi_R(p)|0\>
\ee
near the intersection. In section~\ref{sec:leadingreggedetector} we showed that (eq.~\eqref{eq:twoloopbareeventshape})
\be
	\<\cD_{J_L}\>=Z_\f^{-1} \p{V_{J_L}(z;p) + \cF_{J_L}^{(2)}(z,p)} + O(\l^3),
\ee
where the two-loop correction is given exactly as (eq.~\eqref{eq:2looptwist2})
\be
	\mathcal{F}^{(2)}_{J_L}(z;p)=&(\l\tl\mu^\e)^2\frac{\vol S^{d-2}}{2^{d}(2\pi)^{2d-2}}\frac{\Gamma(\tfrac{d-2}{2})\G(-J_L)}{\G(-J_L+\tfrac{d-2}{2})}(-{2z\cdot p})^{J_L}(-p^2)^{\frac{d-4}{2}-J_L-2}\theta(-p^2). 
\ee
For generic $J_L$, this correction has a divergence at small $\e=4-d$ coming from the factor $(-{2z\cdot p})^{J_L}(-p^2)^{\frac{d-4}{2}-J_L-2}$, as given in~\eqref{eq:twoloopdiv}.
In terms of the matrix element $\<\cD_{J_L}\>$, it can be written as
\be\label{eq:DpoleE}
	\<\cD_{J_L}\>=\p{1+\frac{\lambda^2}{(4\pi)^4}\frac{1}{\e}\left[ \frac{1}{12} - \frac{1}{2J_L(J_L+1)}\right]}
\<\cD_{J_L}\>_\text{tree}+O(\l^2)+(\text{regular at $\e=0$}),
\ee
and the pole was removed in section \ref{sec:leadingreggedetector} by renormalization.

It turns out that the matrix element $\<\cD_{J_L}\>$ also has a divergence for any fixed $\e>0$ and a special value of $J_L$, which also comes from the factor $(-{2z\cdot p})^{J_L}(-p^2)^{\frac{d-4}{2}-J_L-2}$. Indeed, near $J_L=\Jo\equiv \frac{d-6}{2}$
($\Jo\approx -1$ for small $\e$; we will see the significance of its precise value shortly) we find that\footnote{See appendix~\ref{app:poles dist}. There are also other singularities in the two-loop correction to $\<\cD_{J_L}\>$ that can be classified using the results of appendix~\ref{app:poles dist}. They will not play a role in the present discussion.}
\be
	(-p^2)^{\frac{d-4}{2}-J_L-2}\theta(-p^2)\sim \frac{1}{\Jo-J_L}\de(p^2).\label{eq:pole2maintext}
\ee
On the pole, the angular factor $(-{2z\cdot p})^{J_L}$ coincides with that of the shadow transform of $V_{2-d-\Jo}$, so that we can
write
\be
\mathcal{F}^{(2)}_{J_L}(z;p) \sim \frac{1}{J_L-\Jo} \frac{\lambda^2\mu^{2\e}}{2(4\pi)^4}\cR(\e)\hat\wS_J[V_{2-d-\Jo}](z;p),
\ee
where, after a short calculation (setting $\mu=\sqrt{4\pi}e^{-\g/2}\tl\mu$ as usual), we find the coefficient
\be
	\cR(\e)=-\frac{(\tl\mu/\mu)^{2\e}2^{11-3d}\pi^{\frac{9}{2}-d}\G(d-4)\G(3-\tfrac{d}{2})\G(\tfrac{d-2}{2})}{\G(\tfrac{d-1}{2})\G(\tfrac{3d-10}{2})}=\frac{1}{\e}+1+O(\e).
\ee
In terms of the matrix element $\<\cD_{J_L}\>$ this means
\be\label{eq:DpoleJ}
	\<\cD_{J_L}\>&\sim
	\frac{1}{J_L-\Jo}\frac{\l^2\mu^{2\e}}{2(4\pi)^4}\cR(\e)\<\tl\cD_{\Jo}\>_\text{tree}+(\text{regular at $J_L=\Jo$}).
\ee

\begin{figure}[tb]
	\begin{center}
		\includegraphics[scale=.6]{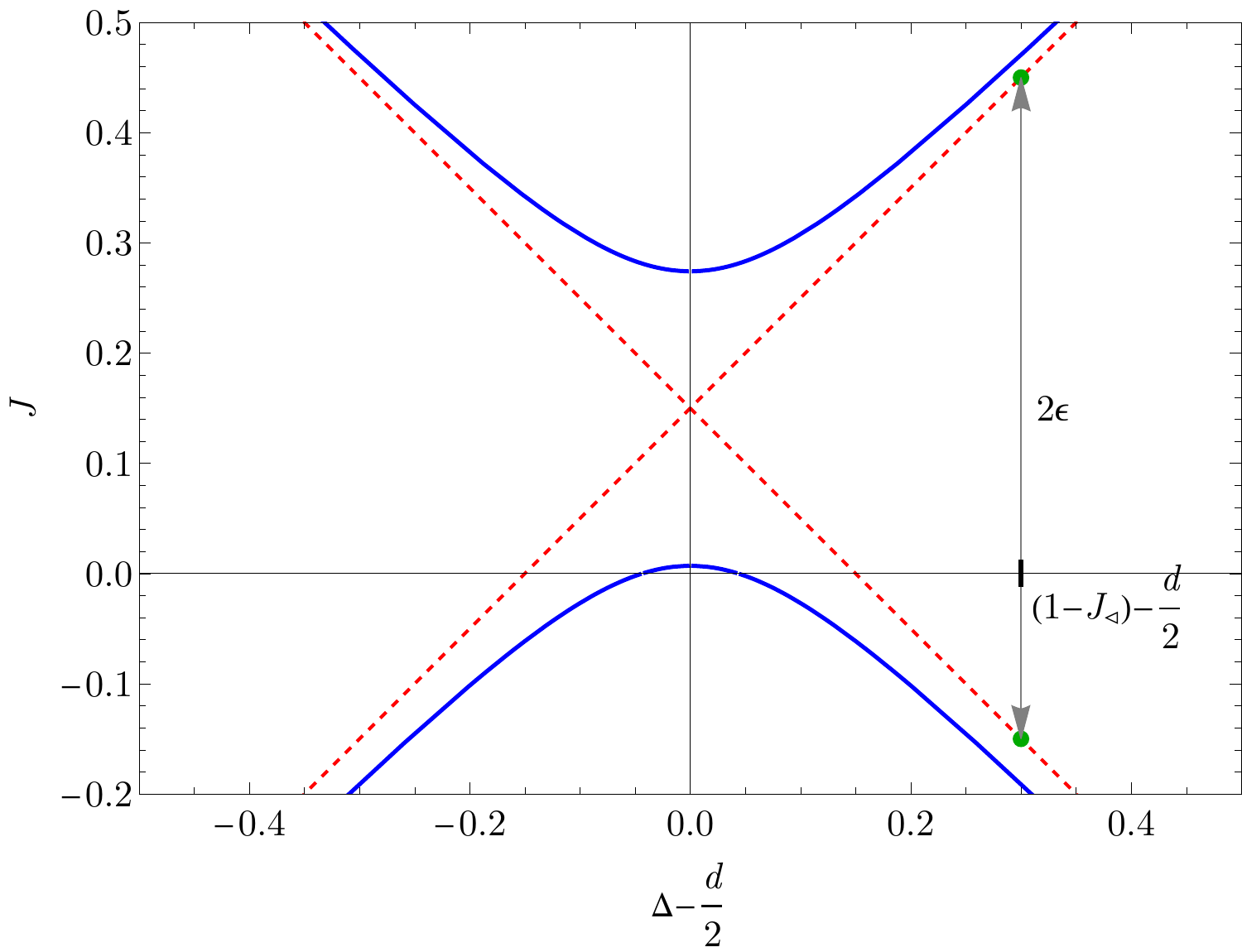}
		\caption{Free theory Regge trajectories (dashed, red), with the operators participating in the mixing at $J_L=\Jo$ shown as green dots. Their scaling dimensions $\Delta_L=1-J$ differ by $2\e$, which is the mass dimension of the coupling $\lambda^2\tl\mu^{2\e}$.
		The renormalized trajectory at $O(\e^2)$ is shown in solid blue. The plot is made at $\e=0.3$.}
		\label{fig:Jmixing}
	\end{center}
\end{figure}

Equation~\eqref{eq:DpoleJ} suggests why this divergence appears: the divergence in $\cD_{\Jo}$ is proportional to $\tl\cD_{\Jo}$.
We can thus say that $\cD_{\Jo}$ and $\tl \cD_{\Jo}$ mix. For this to be possible the mass dimensions must agree. Note that the (tree-level) mass dimension of $\cD_{\Jo}$ is $-\De_L(\Jo)=-\Jo-d+2$, while the mass dimension of $\tl\cD_{\Jo}$ is $-\De_L(2-d-\Jo)=\Jo$. Furthermore, two-loop contributions always appear with the factor $\tl\mu^{2\e}=\tl\mu^{2(4-d)}$, and so we have to solve
\be
	-\Jo-d+2=\Jo+2(4-d),
\ee
which gives $\Jo=\frac{d-6}{2}$, see figure~\ref{fig:Jmixing}. We use the subscript $\triangleleft$ for $\Jo$ due to the triangular shape formed by the lines in figure~\ref{fig:Jmixing}.

The divergence~\eqref{eq:DpoleJ} is worrisome. What does it mean? We expect our perturbative calculations to follow the following very simple logic: perturbation theory in $d=4$ is divergent; the divergences are regulated in $d=4-\e$ where they appear as $1/\e$ poles; we remove the $1/\e$ poles to get a sensible $\e$-expansion. Here we have a divergence that is not regulated in $d=4-\e$: the matrix elements of the bare detector $\cD_{\Jo}$ are infinite at two-loop order regardless of the value of $\e$.

When $d\approx 4$, this divergence happens at $J_L=\Jo\approx -1$.
This suggests to view it as the cause of the breakdown of perturbation theory near $J_L\approx-1$. This is similar to the breakdown of the $\e$-expansion near $\e=0$ that would happen due to $1/\e$ poles if we did not renormalize our operators.
We should thus try to improve perturbation theory by renormalizing our operators to remove the $1/(J_L-\tfrac{d-6}{2})$ poles.

As a first attempt, we may try the combination
\be
\cD_{J_L}-C\frac{\l^2}{J_L-\Jo}\tl\cD_{\Jo},
\ee
where $C$ is chosen using~\eqref{eq:DpoleJ} so that the $1/(J_L-\Jo)$ pole cancels in the matrix elements. While this combination can cancel the $1/(J_L-\Jo)$ pole, it is not consistent with dimensional analysis and the Lorentz spins of the two terms do not match. We may then consider an improved version
\be\label{eq:lincomb2}
	\cD_{J_L}-C\frac{\l^2 \mu^{2-d-2J_L}}{J_L-\Jo}\tl\cD_{J_L},
\ee
which now has matching dimensions and spins and still cancels the pole.
Note that even though the divergence occurs at one value of $J_L$, to cancel it we are forced to consider a mixed combination of the two trajectories for all values of $J_L$. In the above we could replace $C$ by any holomorphic function $C(J_L)$
that has the correct value at $\Jo$.
This is a scheme choice of the same kind we face when removing $1/\e$ divergences. Physical quantities will not depend on this choice.

The linear combination in~\eqref{eq:lincomb2} is sufficient to remove the $1/(J_L-\Jo)$ divergence, but it does not remove the standard $1/\e$ divergence. Simply adding the standard counter-term defined by~\eqref{eq:Zleading} will not work because it will reintroduce the $1/(J_L-\Jo)$ pole. This happens because, as one can see from~\eqref{eq:DpoleJ}, the constant $C$ in~\eqref{eq:lincomb2} itself contains a $1/\e$ pole: the divergences overlap.
The total divergence of the matrix element near $J_L=\Jo$ and $\e=0$ has the schematic form
\be\label{eq:divstructure}
	\<\cD_{J_L}\>\sim \frac{1}{(J-\Jo)\e}+\frac{1}{J-\Jo}+\frac{1}{\e}.
\ee
We seek a renormalization that removes all these divergences at the same time.

We can try to represent the divergence as
\be\label{eq:ansatz1}
 \<\cD_{J_L}\>=X(\e,J_L) \<\cD_{J_L}\>_\text{tree}+\mu^{2-d-2J_L}Y(\e,J_L)\<\tl\cD_{J_L}\>_\text{tree}+\text{finite},
\ee
and similarly for $\tl\cD$, so the renormalization factor becomes a $2\x2$ matrix
acting on the basis $(\cD_{J_L},\mu^{2-d-2J_L}\tl\cD_{J_L})$.
The coefficients are to be found by making all matrix elements in the renormalized basis free of the poles \eqref{eq:divstructure}.

A complication can be anticipated: because $\cD_{\frac{2-d}{2}}=\tl\cD_{\frac{2-d}{2}}$, the coefficients $X$ and $Y$ in eq.~\eqref{eq:ansatz1} will contain spurious poles at $J_L=\frac{2-d}{2}\approx -1$, due to the degeneracy of the basis.
In some sense this is really a problem with the tree-level basis.
Before proceeding with the two-loop renormalization,
we should first fix the tree-level problem.\footnote{Strictly speaking, we could proceed also in the degenerate basis. In fact, we will do so in section~\ref{sec:WF horizontal}. Here we will chose a non-degenerate basis in order to make the discussion of the intercept more transparent.} We thus define a basis that is non-degenerate near the intersection:
\be\label{eq:nice basis}
\mathbb{D}_{J_L} =
\begin{pmatrix}
\cD_{J_L} \\
\cD'_{J_L}
\end{pmatrix},
\qquad
\cD'_{J_L}\equiv\frac{\mu^{2-d-2J_L}\tl\cD_{J_L}-\cD_{J_L}}{J_L-\tfrac{2-d}{2}}.
\ee
Since the two entries have the same mass dimension and Lorentz spin, the renormalization
matrix in this basis will be dimensionless and Lorentz invariant.
The detectors $\cD'_{J_L}$ however do not have definite scaling dimension due to the explicit presence of $\mu$. Note in particular that $\cD_{\frac{2-d}{2}}$ and $\cD'_{\frac{2-d}{2}}$ are linearly-independent, but $\cD'_{\frac{2-d}{2}}$ does not scale in a standard way. Instead, $\cD'_{\frac{2-d}{2}}$ and $\cD_{\frac{2-d}{2}}$ form a log-multiplet in the free theory.
As we will see later, interactions break this log-multiplet into two independent operators.

The renormalization of $\cD_{J_L}$ can be worked out from the divergences in eqs.~\eqref{eq:DpoleE} and \eqref{eq:DpoleJ}, which give respectively:
\be
\<\cD_{J_L}\> &= \frac{\lambda^2}{(4\pi)^4}
 \frac{1}{\e} \left[\frac{1}{12}-\frac{1}{2J_L(J_L+1)} \right]\<\cD_{J_L}\>_{\rm tree}+\mbox{(regular at $\e\to 0$)},
\\
\<\cD_{J_L}\> &= \frac{\lambda^2}{(4\pi)^4}  \frac{1}{J_L-\Jo} \left[\frac{\cR(\e)}{2}
\<\cD_{J_L}\>_{\rm tree} -\frac{\e\cR(\e)}{2} \<\cD'_{J_L}\>_{\rm tree}\right] +\mbox{(regular at $J_L\to \Jo$)}.
\ee
As a sanity check, let us verify the compatibility of these equations.
Recalling that $\cR(\e)\sim \frac{1}{\e}$, the second line gives the double pole
\be
\<\cD_{J_L}\>  \sim  \frac{\lambda^2}{(4\pi)^4} \frac{\<\cD_{J_L}\>_{\rm tree}}{2\e(J_L+1)} +
\mbox{(less singular as $\e\to 0$, $J_L\to -1$)},
\ee
which is in perfect agreement with the first line.  This confirms that we identified all nearby  singularities,
and allows us to combine the divergences into a single expression:
\be\label{eq:twist2fulldiv}
  \<\cD_{J_L}\> = \frac{\lambda^2}{(4\pi)^4} \left(
  \left[\frac{1}{12\e} - \frac{1}{2\e J_L} + \frac{\cR(\e)}{2(J_L-\Jo)}\right] \<\cD_{J_L}\>_{\rm tree}
- \frac{\e \cR(\e)}{2(J_L-\Jo)} \<\cD'_{J_L}\>_{\rm tree}\right) + \mbox{(regular)}.
\ee
The divergences in $\<\tl\cD_{J_L}\>$ follow by the simple replacements $J_L\mapsto 2-d-J_L$ and
$\tilde{\cD}\leftrightarrow \cD$, which transforms $\cD'_{J_L}$ in a simple way.
Employing elementary algebra, we obtain the divergences in the matrix elements
of the basis \eqref{eq:nice basis}, or equivalently the renormalization factor which cancels them:
\be \label{bbDren}
	[\mathbb{D}_{J_L}]_R\equiv \cZ_{J_L}^{-1}\mathbb{D}_{J_L}
\ee
with
\be \label{Z nice basis}
\cZ_{J_L}= 1+\frac{\lambda^2}{(4\pi)^4}
 \begin{pmatrix}
  \frac{1}{12\e} - \frac{1}{2\e J_L} + \frac{\cR(\e)}{2(J_L-\Jo)}&  \frac{-\e \cR(\e)}{2(J_L-\Jo)}
\\[1mm] \frac{1}{\e J_L(J_L+d-2)} - \frac{\cR(\e)}{(J_L-\Jo)(J_L+d-2+\Jo)} \phantom{\quad} &
\frac{1}{12\e} - \frac{1}{2\e (2-d-J_L)} -\frac{\cR(\e)}{2(J_L-\Jo)} 
 \end{pmatrix}.
\ee
As expected, the renormalization factor has singularities at $\e\to 0$ and as $J_L\to \Jo\approx -1$ and its shadow $2-d-\Jo\approx -1$,
but no other singularities near the intersection point. We have thus succeeded at removing all the known singularities without introducing new ones!
The renormalized detectors $[\mathbb{D}_{J_L}]_R$ form a regular basis near the leading intersection.

From this result we can deduce how the dilatation operator acts as a $2\x2$ matrix in this basis,
\be
D[\mathbb{D}_{J_L}]_R=\mathscr{D}[\mathbb{D}_{J_L}]_R.
\ee
The dilatation operator $D$ captures the physical scale dependence of matrix elements, and is 
given on detectors at infinity by
\be
	D=D_\text{eng}-\frac{\ptl}{\ptl\log\mu},
\ee
where $D_\text{eng}$ is the operator counting the engineering mass dimensions.
We stress that this is not simply the $\mu$ dependence, because of
the explicit factors of $\mu$ in the basis \eqref{eq:nice basis}.  In fact, $D$ commutes with powers of $\mu$.
At tree-level it gives
\be
 D \cD_{J_L} = (2-d-J_L)\cD_{J_L} + O(\lambda^2),\qquad D \tl\cD_{J_L} =  J_L \tl\cD_{J_L} + O(\lambda^2),
\ee
and working through the basis change in eq.~\eqref{eq:nice basis} we find
\be
 \mathscr{D}_0 = \begin{pmatrix} 2-d-J_L \phantom{\quad}& 0 \\ 2 & J_L \end{pmatrix}.
\ee
It will be significant that this is non-diagonal already at tree-level.
At higher orders, acting on eq.~\eqref{bbDren} we have
\be
 \mathscr{D} = \cZ_{J_L}^{-1}\left( \mathscr{D}_0 +D_\l \right) \cZ_{J_L}
\ee
where $D_\lambda$ acts as renormalization group flow on the couplings in $\cZ$,
\be
 D_\l 
 = \b(\l)\frac{\ptl}{\ptl\l} =(-\e\l+O(\l^2))\frac{\ptl}{\ptl\l}.
\ee
Substituting in eq.~\eqref{Z nice basis} and commuting $\mathscr{D}_0$ across, we finally find
\be \label{eq:D}
\mathscr{D} = \mathscr{D}_0 + \frac{\lambda^2}{(4\pi)^4}
\begin{pmatrix}
\frac{1}{J_L}-\frac{1}{6} & \e \cR(\e) \\[1mm]
\frac{2}{J_L(2-d-J_L)}\phantom{\quad} & \frac{1}{2-d-J_L}-\frac{1}{6}\end{pmatrix} + O(\lambda^3).
\ee
This is the main result of this section.  Crucially, all entries are regular as $\e\to0$ and $J_L\approx -1$. 

\subsection{Interpreting the result}

What can we extract from the dilatation operator $\mathscr{D}$ in~\eqref{eq:D}?

Firstly, it should correctly reproduce the anomalous dimensions of $\cD_{J_L}$ away from the intercept $J_L=\tfrac{2-d}{2}$.
This is essentially guaranteed by construction, since it is just a change of basis away from the usual (diagonal)
dilatation operator that we discussed in the previous section.  It is still interesting to verify it directly from \eqref{eq:D}.
If we define $\cO=v [\mathbb{D}_{J_L}]_R$ for some row vector $v$, then
\be
	D\cO=v\mathscr{D} [\mathbb{D}_{J_L}]_R.
\ee
Thus, in order to have scaling detectors $\cO$ with the property $D\cO=-\De_L\cO$ we must choose $v$ to be a \textit{left} eigenvector of $\mathscr{D}$, $v\mathscr{D}=-\De_L v$.

At tree level, such eigenvectors are $v_1^{(0)}=(1,0)$ and $v_2^{(0)}=(\frac{1}{J_L-\frac{2-d}{2}},1)$ with eigenvalues
\be
	v_1^{(0)}\mathscr{D}=(2-d-J_L)v_1^{(0)},\qquad
	v_2^{(0)}\mathscr{D}=J_L v_2^{(0)}.
\ee
Unless $J_L=\tfrac{2-d}{2}$, these are non-degenerate, and thus the corrections to the eigenvalues are easily determined. The only subtlety is that $\mathscr{D}$ is not Hermitian, and so its perturbation theory is a bit more general than usually encountered in quantum mechanics. In practice this means that we should track both left eigenvectors $v_i$ and right eigenvectors $u_i$ for each eigenvalue.
Normalizing them so that $v_iu_j=\de_{ij}$ (which gives $u_1^{(0)} = (1,\tfrac{-1}{2-d-J_L})$ and $u_2^{(0)} = (0,1)$),
trivial modifications of the standard theory give for the $O(\l^2)$ eigenvalues
\be
	-(\De_L)_i=v_i^{(0)}\mathscr{D}u_i^{(0)}+O(\l^3),
\ee
and so in particular
\be
	(\De_L)_1=J_L+d-2+\frac{\l^2}{(4\pi)^2}\p{\frac{1}{6}-\frac{1}{J_L(J_L+1)}+O(\e)}+O(\l^3).
\ee
This is equivalent to~\eqref{eq:gammaltwisttwo} obtained in section~\ref{sec:leadingreggedetector}, and leads to the standard result for the anomalous dimension of the leading twist trajectory upon specializing to the fixed point $\l=\tfrac{(4\pi)^2}{3}\e+O(\e^2)$.

More generally, the characteristic equation for $\mathscr{D}$ gives:
\be
	0&=\det(\mathscr{D}+\De_L)\nn\\
	&=(\De_L+J_L)(\De_L+2-d-J_L)\nn\\
	&\quad+\frac{\lambda^{2}}{(4\pi)^{4}}\p{-\frac{2\De_L+2-d}{6}+\frac{(2-d)\De_L}{J_L(2-d-J_L)}+2(1-\e\cR(\e))}+O(\l^3),\label{eq:characteristic}
\ee
which generalizes the physical state condition (\ref{eq:WF leading trajectory})
to the theory away from the conformal fixed point $\lambda=\lambda_*$.
The singularities at $J_L\approx 0,-2$ will be discussed in the next section.

The advantage of the dilatation operator in the regular basis \eqref{eq:D} is that we are free to set $J_L=\tfrac{2-d}{2}$ directly,
since the matrix is perfectly regular. At tree level we find
\be \label{D0 Jordan form}
	\mathscr{D}_0=\begin{pmatrix}
		\tfrac{2-d}{2} & 0 \\ 2 & \tfrac{2-d}{2}
	\end{pmatrix}.
\ee
This does not only have degenerate eigenvalues, but also forms a non-trivial Jordan block. Thus, $\cD_{\frac{2-d}{2}}$ and $\cD'_{\frac{2-d}{2}}$ form a logarithmic multiplet at tree level.

This is surprising at first sight since $\mathscr{D}$ is expected to be self-adjoint
with respect to the inner product defined by the two-point functions of time-ordered operators. This is required by target-projectile duality in the Regge limit of correlators,
see section 2.3 of \cite{Caron-Huot:2013fea}.  Physically, correlation functions should depend only on the relative boost between a target and a projectile and not on individual boosts.
However, nothing ensures that this inner product is positive, nor nondegenerate at weak coupling in the regular basis \eqref{eq:nice basis}, which could explain eq.~\eqref{D0 Jordan form}.
It would be interesting to further study Regge factorization in this basis.

The Jordan form makes the determination of the eigenvalues harder than above, and the easiest way is to solve the characteristic equation in~\eqref{eq:characteristic}, restricted to $J_L=\tfrac{2-d}{2}$. It has solutions
\be
	(\De_L)_\pm=1-\frac{\e}{2}\mp\frac{\sqrt2\l}{(4\pi)^2}(1+O(\e))+O(\l^2).\label{eq:phi4intercept}
\ee
Writing this result in terms of $J=1-\De_L$ and evaluating at the fixed point $\l=\tfrac{(4\pi)^2}{3}\e+O(\e^2)$, we find 
\be
	J_\pm=\p{\frac{1}{2}\pm \frac{\sqrt2}{3}}\e+O(\e^2).
\ee
The Regge intercept is the larger of the two roots, which agrees with~\eqref{eq:reggeintercept} obtained from general analyticity assumptions. Here we observe two new features. Firstly, we can set $\e=0$ in~\eqref{eq:phi4intercept} to obtain the Regge intercept in the non-conformal massless 4d $\phi^4$ theory:
\be
	(\De_L)_\pm=1\mp \frac{\sqrt2\l}{(4\pi)^2}+O(\l^2). \label{De0}
\ee
Secondly, we can explicitly determine the (left) eigenvectors of $\mathscr{D}$,
\be
	v_\pm=\p{1,\pm\frac{\l}{\sqrt 2(4\pi)^2}(1+O(\e))+O(\l^2)}.
\ee
In particular, the detector $v_{+}[\mathbb{D}_{J_L}]_R$ is the Pomeron of the Wilson-fisher theory. 

In this calculation, we saw explicitly that the intersection of two trajectories gave rise to a logarithmic multiplet at tree level. We were able to construct a logarithmic partner for $\cD_{\frac{2-d}{2}}$ due to the existence of two tangent spaces at the intersection point, even though $\cD_{\frac{2-d}{2}}$ is the unique primary operator at that point. Turning on interactions broke the logarithmic multiplet into two conventional primary multiplets.  The Jordan form of the \emph{tree-level} dilatation operator was essential
to produce an $O(\lambda)$ splitting from a two-loop diagram.

In section \ref{sec:summary_mixing} we considered characteristic equations of the form
\be
	(\De_L-(\De_L)_1(J_L))(\De_L-(\De_L)_2(J_L))=0,
\ee
and observed that non-perturbative analyticity in spin implies an all-orders cancellation of poles near the intercept.
Here we realized such an equation from the determinant of a mixing matrix \eqref{eq:D}
which is guaranteed to be free of poles near the intercept, provided our renormalization procedure
(removing both $1/\e$ and $1/(J-\Jo)$-type poles) is self-consistent to all orders.
As in section~\ref{sec:summary_mixing}, this characteristic equation allows us to partially resum the perturbative expansion to obtain reliable results near the intercept, as we saw in the example of $J_L=\tfrac{2-d}{2}$.

\subsection{Can trajectories intersect?}

It is interesting to ask whether all intersections in the free spectrum (figure~\ref{fig:perturbativelines}) get resolved in a way similar to the resolution of the Regge intercept.
It is tempting to conjecture that any given intersection will always get resolved at a sufficiently high order in perturbation theory,
meaning that there will be no level crossing. Taking this conjecture to its extreme, one can conjecture that in the non-perturbative theory all  light-ray operators (with the same global symmetry) live on a single complex-analytic Regge trajectory.

Here we will simply make an observation of a statistical nature.
We saw that it is fruitful to view Regge trajectories as the solutions to a mixing problem,
where near intersections one diagonalizes a regular ``Hamiltonian'' representing $J=1-\Delta_L$.
In similar situations in quantum mechanics, one generically expects level crossings to be resolved, in the absence of symmetry or fine tuning.

Let us review this genericity argument, considering for example a $2\x 2$ $D(\nu)$,
\be
	D(\nu)=\begin{pmatrix}
		a(\nu) & b(\nu) \\ c(\nu) & d(\nu)
	\end{pmatrix},
\ee
where $\nu$ is a real parameter. In our situation $\nu$ is the horizontal coordinate in the Chew-Frautschi plot, $D$ represents the dilatation operator whose eigenvalues give $\De_L=1-J$. The discriminant of the characteristic equation for $D$ is
\be
	\mathrm{Discr}\,D(\nu)=\mathrm{Discr}_\l\p{\det(\l-D(\nu))}=(a(\nu)-d(\nu))^2+4b(\nu)c(\nu).
\ee 
The discriminant vanishes if and only if $D(\nu)$ has degenerate eigenvalues.
If $D(\nu)$ is Hermitian, then we would have that $a(\nu),d(\nu)\in \R$ and $c(\nu)=b(\nu)^*$. In this situation,
\be
	\mathrm{Discr}\,D(\nu)=4|b(\nu)|^2+|a(\nu)-d(\nu)|^2,
\ee
and $\mathrm{Discr}\,D(\nu)=0$ implies $a(\nu)=d(\nu)$ and $b(\nu)=0$. This gives 2 or 3 real conditions, depending on whether $b(\nu)$ is real or complex, for just one real variable $\nu$. Generically these cannot all be satisfied, making level-crossing non-generic in quantum mechanics. The discriminant is over-constraining because it is the sum of two non-negative terms.
If $n$ energy levels get close to crossing each other, one can extend this argument by considering an $n{\times}n$ sub-matrix capturing those states.

\begin{figure}[t]
	\centering
	\begin{tikzpicture}[scale=0.7]
		\draw[] (-0.5,-0.5) -- (4,-0.5);
		\draw[] (0,-1.2) -- (0,4);
		\node[right] at (4,-0.5) {$\nu$};
		\node[above] at (0,4) {$J$};
		\draw[thick,blue] (-0.5,0) -- (4,3);
		\draw[thick,blue] (-0.5,3) -- (4,0);
	\end{tikzpicture}
\raisebox{18mm}{\ $\longrightarrow$\quad}
	\begin{tikzpicture}[scale=0.7]
	\draw[] (-0.5,-0.5) -- (4,-0.5);
	\draw[] (0,-1.2) -- (0,4);
	\node[right] at (4,-0.5) {$\nu$};
	\node[above] at (0,4) {$J$};
	\draw[thick,blue] (-0.5,0) to[out=35,in=180,looseness=0.5] (1.75,1.25) to[out=0,in=145,looseness=0.5]  (4,0) ;
	\draw[thick,blue]  (-0.5,3) to[out=-35,in=180,looseness=0.5] (1.75,1.75) to[out=0,in=-145,looseness=0.5]   (4,3);
	\end{tikzpicture}
\raisebox{18mm}{\mbox{\ or\ \ }}
	\begin{tikzpicture}[scale=0.7]
	\draw[] (-0.5,-0.5) -- (4,-0.5);
	\draw[] (0,-1.2) -- (0,4);
	\node[right] at (4,-0.5) {$\nu$};
	\node[above] at (0,4) {$J$};
	\draw[thick,blue] (-0.5,0) to[out=35,in=-90,looseness=0.45] (1.3,1.5) to[out=90,in=-35,looseness=0.45]  (-0.5,3) ;
	\draw[thick,blue]  (4,3) to[out=215,in=90,looseness=0.45] (2.2,1.5) to[out=-90,in=145,looseness=0.45]   (4,0);
	\end{tikzpicture}
	\caption{Left: naive level-crossing. Right: two generic resolutions for a non-Hermitian Hamiltonian.\label{fig:level-crossing}}
\end{figure}
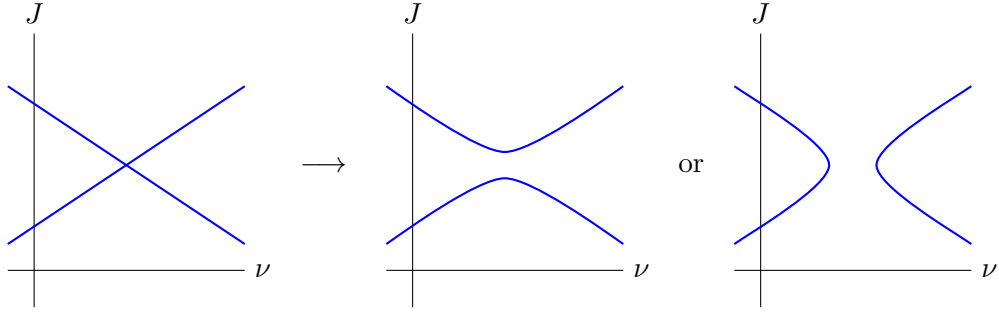

Now, in the case of Regge trajectories, we do not expect $D(\nu)$ to be Hermitian, at least not when written in a form that is regular in perturbation theory (witness the Jordan form in \eqref{D0 Jordan form}).

Naively, $\mathrm{Discr}\,D(\nu)=0$ is then a single real equation and one might generically expect a solution for $\nu\in \R$. 
However, such a solution, where $D(\nu)$ has a single zero, does not describe level crossing, rather it represents two real solutions colliding to become a complex-conjugate pair, as in the third of figure \ref{fig:level-crossing}.

In fact, if we define ``no level-crossing'' for Regge trajectories to mean that $\det(1-J-D(\nu))=0$ defines a non-singular complex surface in $\C^2$ (in particular it cannot look like the left of figure~\ref{fig:level-crossing} in any real section), then generically we do not expect level-crossing as long as $D(\nu)$ is (locally) holomorphic in $\nu$. Indeed, a singular point $(J,\nu)$ would need to satisfy three independent conditions,
\be	\det(1{-}J{-}D(\nu))=0,\quad \ptl_J\det(1{-}J{-}D(\nu))=0,\quad \ptl_\nu\det(1{-}J{-}D(\nu))=0,
\ee
and it can be checked that such a point exists only if $\mathrm{Discr}\,D(\nu)$ has a double-zero at some $\nu$, which is non-generic.
Moreover, the surface will be connected as long as $\mathrm{Discr}\,D(\nu)$ has a zero in the complex plane, which is generic. These arguments can be straightforwardly generalized to the $n\x n$ case.

Of course, none of this proves that level crossing is impossible, only that we do not expect to see it unless some yet unknown structure restricts the mixing between trajectories.

\section{Horizontal trajectories}
\label{sec:WF horizontal}

As explained in section~\ref{sec:horizontalintroduction}, we can build horizontal trajectories by taking a product of detectors $\mathbb{O}_1(\oo,z_1)\mathbb{O}_2(\oo,z_2)$ and convolving with a Clebsch-Gordan coefficient for the Lorentz group. Which Lorentz irreps can appear in this decomposition? Recall that $z_1$ and $z_2$ can be thought of as embedding-space coordinates for the celestial sphere $S^{d-2}$. Let us assume for simplicity that $\mathbb{O}_{1}(\oo,z_1)$ and $\mathbb{O}_2(\oo,z_2)$ transform like scalar operators on the celestial sphere. Then their product $\mathbb{O}_1(\oo,z_1)\mathbb{O}_2(\oo,z_2)$ can be decomposed into traceless symmetric tensor operators:
\be
\label{eq:convolvewithclebsch}
\int D^{d-2} z_1 D^{d-2} z_2 K_{J_L,j}(z_1,z_2;z) \mathbb{O}_1(\oo,z_1)\mathbb{O}_2(\oo,z_2).
\ee
 The corresponding representations are two-row Young diagrams for $\SO(d-1,1)$, with row lengths $(J_L,j)$. Here, $-J_L\in \C$ is a scaling dimension on the celestial sphere, and $j\in \Z_{\geq 0}$ is spin on the celestial sphere, which we call ``transverse spin". 
 
The Clebsch-Gordan coefficient that implements this decomposition is a conformal three-point function on the celestial sphere. For example, in the case $j=0$, we have
\be
K_{J_{L},0}(z_1,z_2;z) &= \<\cP_{-\tl J_{L1}}(z_1) \cP_{-\tl J_{L2}}(z_2) \cP_{-J_{L}}(z)\>,
\ee
where $\tl J_{Li}=2-d-J_{Li}$ and we have defined the standard celestial three-point structure
\be \label{standard 3pt}
\<\cP_{\de_1}(z_1) \cP_{\de_2}(z_2) \cP_{\de_3}(z_3)\> &= \frac{1}{z_{12}^{\frac{\de_1+\de_2-\de_3}{2}} z_{23}^{\frac{\de_2+\de_3-\de_1}{2}} z_{31}^{\frac{\de_3+\de_1-\de_2}{2}}},
\ee
where $z_{ij}\equiv -2 z_i\.z_j$.

In this section, we study horizontal trajectories built from products of the detectors $\cD_{J_L}(z)$. In other words, we will study renormalization of the composite detector 
\be
\label{eq:hdefinition}
\cH_{J_{L1},J_{L2}}(z_1,z_2) &\equiv \cD_{J_{L1}}(z_1) \cD_{J_{L2}}(z_2).
\ee
 A special case of (\ref{eq:hdefinition}) is the product of light-transformed operators $:\wL[\f^2](\oo,z_1) \wL[\f^2](\oo,z_2):$, which at tree level is proportional to $\cH_{3-d,3-d}(z_1,z_2)$. This operator will eventually be our main interest, since in section~\ref{sec:AppInCorrFuncs} we will find evidence that it appears in the Regge limit of a correlation function of local operators $\<\f^2\f^2\f^2\f^2\>$.

Before launching into a detailed discussion of $\cH_{J_{L1},J_{L2}}(z_1,z_2)$, let us first explain why we do not consider the apparently simpler expression
\be
:\wL[\f](\oo,z_1)\wL[\f](\oo,z_2):.
\ee
The reason is that in the free theory the operator $\wL[\f](\oo,z_1)$ vanishes.\footnote{There is an additional complication that the definition of the light-transform is not convergent on $\f$ for $d\leq 4$ (it converges on operators of dimension $\De$ and spin $J$ satisfying $\De+J>1$~\cite{Kravchuk:2018htv}). However, we can define the integral by analytic continuation in $\De$, similarly to what we did with the spin shadow transform in section~\ref{sec:shadowtransform}. In this sense, $(d-4)\wL[\f]$ is well-defined in a neighborhood of $d=4$.} Indeed, any matrix element of the schematic form
\be
	\<0|\f\cdots \f \wL[\f]\f\cdots \f|0\>
\ee
can be computed by Wick's theorem, and each term in the result will contain either $\<0|\f\wL[\f]|0\>$ or $\<0|\wL[\f]\f|0\>$. However, as shown in~\cite{Kravchuk:2018htv}, $\wL[\f]$ annihilates the vacuum on both the left and the right, so $\<0|\f\wL[\f]|0\>=\<0|\wL[\f]\f|0\>=0$.

One can also view this result as a consequence of equations of motion: $\wL \wS_J \wL[\f]$ is proportional to the usual shadow transform $\wS_\De[\f]$~\cite{Kravchuk:2018htv}, and (regularized) $\wS_\De$ acting on an operator $\f$ of dimension $\tfrac{d-2}{2}$ is just $\Box\f=0$. This suggests that as interactions are turned on, we have $\Box\f\propto\l \f^3$, and so $\wL[\f]\propto \l\wS_J\wL[\phi^3]$. In other words, $\wL[\f]$ lives on the shadow of whatever trajectory $\wL[\f^3]$ belongs to. 
So $\wL[\f^2]$ is indeed the simplest light-transform we can consider in the Wilson-Fisher theory.

\subsection{Renormalizing $\cH_{J_{L1},J_{L2}}$}
\label{sec:renormalizehorizontal}

\begin{figure}
	\centering
	\includegraphics[scale=.3]{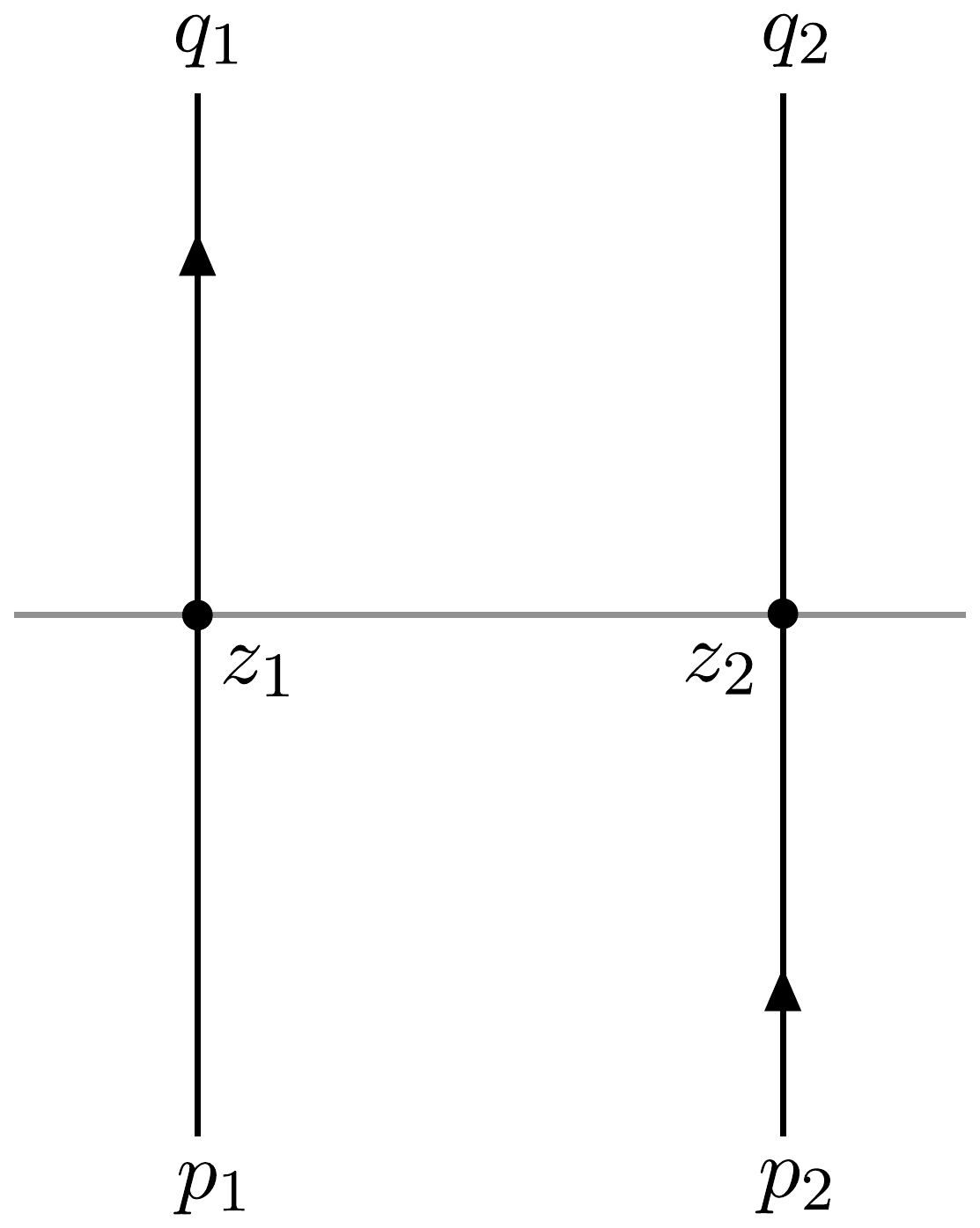}
	\caption{Tree-level contribution to \eqref{eq:eventshapewestudy} for $p_{1}\simeq q_1$ and $p_2\simeq q_2$.}
	\label{fig:treelevel horizontal}
\end{figure}

To renormalize $\cH_{J_{L1},J_{L2}}$, we study its event shape in a two-$\phi$ state:
\be
\label{eq:eventshapewestudy}
&(2\pi)^d\de(p_1+p_2-q_1-q_2)\<\cH_{J_{L1},J_{L2}}(z_1,z_2)\> \nn\\
&\quad\equiv\<0|\bar T\{\f_R(-q_1)\f_R(-q_2)\}\cH_{J_{L1},J_{L2}}(z_1,z_2) T\{\f_R(p_1)\f_R(p_2)\}|0\>,
\ee
Here, we continue to use $\<\cH_{J_{L1},J_{L2}}(z_1,z_2)\>$ as a convenient shorthand for the event shape with the momentum-conserving $\de$-function stripped off.

A tree-level contribution to (\ref{eq:eventshapewestudy}) is shown in figure~\ref{fig:treelevel horizontal}. Altogether there are four tree-level diagrams related to figure~\ref{fig:treelevel horizontal} by permuting $p_1\leftrightarrow p_2$ and $q_1\leftrightarrow q_2$. Their sum is:
\be
\<\cH_{J_{L1},J_{L2}}(z_1,z_2)\>_\mathrm{tree} &= (2\pi)^d \de(p_2-q_2) V_{J_{L1}}(z_1;p_1) V_{J_{L2}}(z_2;p_2)
 + (p_1\leftrightarrow p_2, q_1\leftrightarrow q_2),
\ee
where the vertex $V_{J_L}(z;p)$ is defined in (\ref{eq:twist two vertex}).
This event shape is simply the product of ``$E^{J_i-1}$ fluxes'' in two different directions.

There are no divergent one-loop diagrams contributing to the event shape (\ref{eq:eventshapewestudy}). At two-loop order, there are two types of diagrams we must distinguish. Firstly, there are diagrams where we decorate only one of the detectors with loops (such as those considered in section~\ref{sec:leadingreggedetector}), while the other detector simply contributes a tree-level vertex $V_{J_{Li}}(z_i,p_i)$. We refer to these as ``disconnected" contributions. Secondly, there are diagrams that connect both detectors $\cD_{J_{L1}}(z_1)$ and $\cD_{J_{L2}}(z_2)$ via loops. We refer to these as ``connected" contributions.

As explained in the previous section, disconnected contributions lead to subtle mixing between $\cD_{J_{Li}}(z_i)$ and its shadow $\tl \cD_{J_{Li}}(z_i)$. When analyzing operators close to $\wL[\f^2] \wL[\f^2]$, this mixing depends sensitively on the values $J_{L1},J_{L2}$, since they are very close to the Pomeron intercept. 

\subsubsection{Connected two-loop diagram}
\label{sec:Hconnected}

\begin{figure}
	\centering
	\begin{subfigure}{.46\textwidth}
	\centering
	\includegraphics[scale=.3]{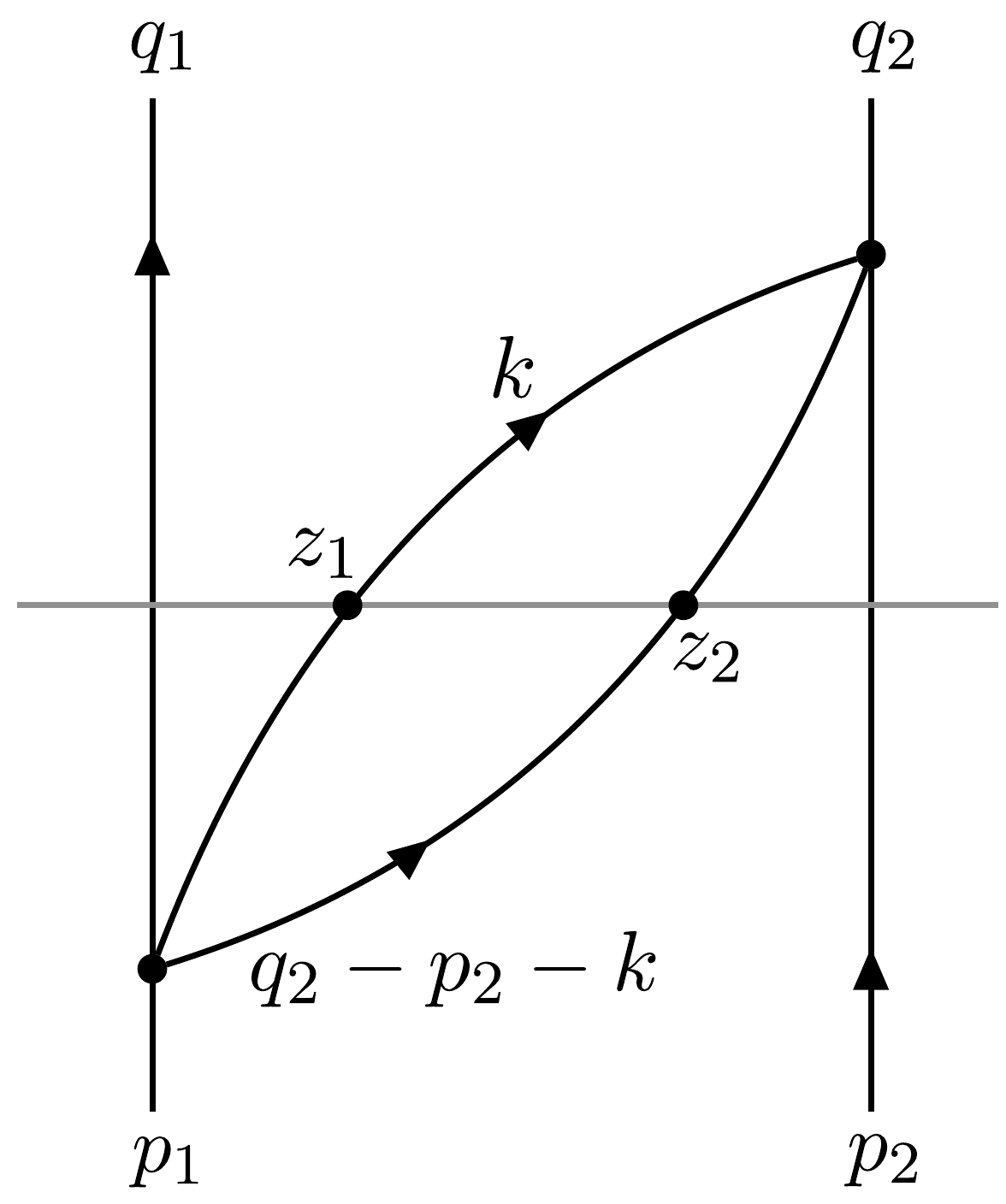}
	\caption{}
	\label{fig:horizontal loop}
	\end{subfigure}
	\begin{subfigure}{.46\textwidth}
	\centering
	\includegraphics[scale=.3]{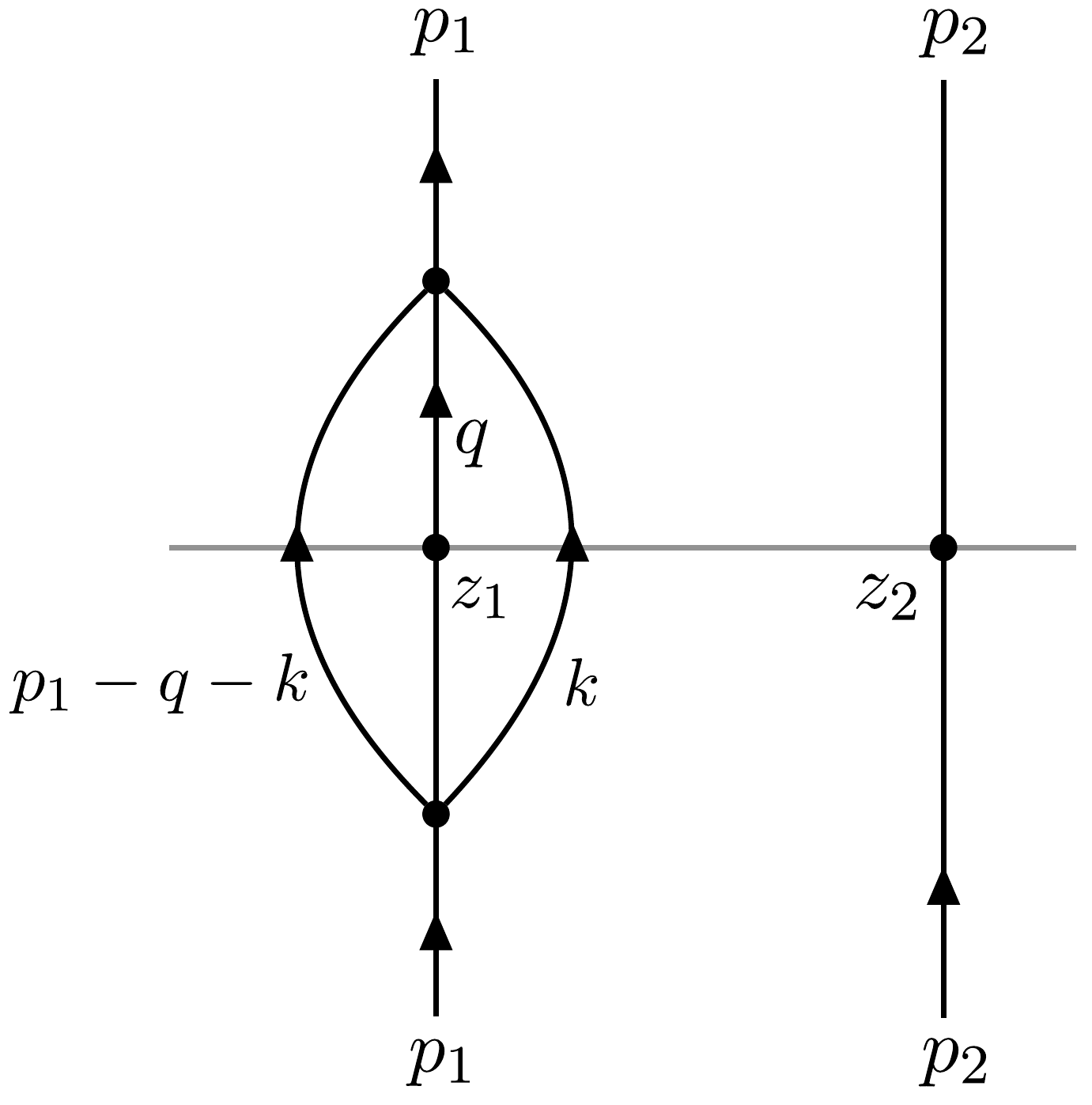}
	\caption{}
	\label{fig:disconnected horizontal loop}
	\end{subfigure}
	\caption{Loop-level contributions to the mixing of $\cH(z_1,z_2)$. (a) Connected diagram which induces ``smearing" on celestial sphere. (b) Disconnected diagram for the correction to an individual detector.
	There exist other diagrams at $O(\l)$ and $O(\l^2)$ but they do not contain IR divergences.
	}
\end{figure}

Up to two-loop order, the unique connected diagram (modulo permutations) that contributes a nontrivial divergence in our event shape is shown in figure \ref{fig:horizontal loop}. The physical interpretation is that the particle $p_1$ can split into other particles before arriving at the detector. The detector can then measure these other particles even if $p_1$ is not in the direction of $z_1$ or $z_2$. Hence, splitting due to interactions effectively ``smears" the detector over the celestial sphere.

The diagram is given by:
\be
\cG^{(2)}
&=(\l\tl\mu^\e)^2\,(2\pi\de(q_1^2))(2\pi\de(p_2^2))\int \frac{d^dk}{(2\pi)^{d}}V_{J_{L1}}(z_1;k)V_{J_{L2}}(z_2;q_2-p_2-k)\frac{1}{(p_1^2-i0)(q_2^2+i0)}\nn\\
	&= \frac{(\l\tl\mu^\e)^2}{(2\pi)^{d-2}}\de(q_1^2)\de(p_2^2)
	\int d\b_1d\b_2\, \b_1^{-J_{L1}-1}\, \b_2^{-J_{L2}-1}
\frac{\de^d(q_2-p_2-\b_1 z_1-\b_2 z_2)}{(p_1^2-i0)(q_2^2+i0)}.
\ee
Momentum conservation at the interaction vertices implies
\begin{align}
p_{1}=q_1+\beta_1z_1+\beta_2z_2,\label{eq:delta horizontal 1}
\\
q_{2}=p_2+\beta_1z_1+\beta_2z_2.\label{eq:delta horizontal 2}
\end{align}
Plugging these expressions into the propagators, we obtain
\be
\cG^{(2)}&=\frac{(\l\tl\mu^\e)^2}{(2\pi)^{d-2}}\de(q_1^2)\de(p_2^2)
 \int d\b_1d\b_2\b_1^{-J_{L1}-1} \b_2^{-J_{L2}-1}\nn\\
	&\quad\quad\quad \x \frac{\de^d(q_2-p_2-\b_1 z_1-\b_2 z_2)}{(2\b_1 q_1\.z_1+2\b_2 q_1\.z_2+2\b_1\b_2 z_1\.z_2)(2\b_1 p_2\.z_1+2\b_2 p_2\.z_2+2\b_1\b_2 z_1\.z_2)}.\label{eq:small beta G2}
\ee
All the angular integrals are gone because the detectors lie at definite angles $z_i$, we only have to integrate over the two energies $\beta_i$.

We have in mind $J_{Li} \approx -1$ (corresponding to $\wL[\f^2]$ detectors), so the $\beta_i\to 0$ limits separately converge.
However the integral has an IR divergence when both $\beta_i$ simultaneously go to zero if $J_{L1}+J_{L2}+2=0$.
This is the regime where particles formed by splitting become soft. To compute the divergence, we can ignore higher-order terms in $\beta_1,\beta_2$ in the denominator, and also ignore $\beta_1,\beta_2$ inside the $\de$-function. Making these approximations and changing variables to $\b_1=\b x^{1/2},\b_2=\b x^{-1/2}$, we obtain a pole $1/(J_{L1}+J_{L2}+2)$ from the integral over $\beta$:
\be
\cG^{(2)}&\sim
 \frac{(\l\tl\mu^\e)^2}{(2\pi)^{d-2}}\frac{-1}{J_{L1}+J_{L2}+2}\de(q_1^2)\de(p_2^2)\de^d(q_2-p_2)  K_\a(z_1,z_2;q_1,p_2),
 \label{eq:divergencewefind}
\ee
where $\a=\frac{J_{L1}-J_{L2}}{2}$ and we have defined the kernel
\be
\label{eq:kalphakernel}
K_{\a}(z_1,z_2;z_3,z_4) &\equiv \int_0^\oo \frac{dx \, x^{-\a}}{(x z_{13}+z_{23})(x z_{14}+z_{24})}
=
\frac{\pi}{\sin\pi \a} \frac{
\p{\frac{z_{13}}{z_{23}}}^{\a}-
\p{\frac{z_{14}}{z_{24}}}^{\a}
}{z_{24}z_{13}-z_{14}z_{23}},
\ee
where $z_{ij}\equiv -2 z_i\.z_j$ as before. 
Note that $K_\a$ in \eqref{eq:divergencewefind} has homogeneity degree $-1$ with respect to $q_1$ and $p_2$.

Since the divergence (\ref{eq:divergencewefind}) is proportional to the on-shell $\delta$-functions $\de(q_1^2)\de(p_2^2)$,
it can be written as an angular integral over tree-level event shapes.
The following identity applies to any function with the homogeneous scaling $f(\lambda z)=\lambda^{2-d-J_L}f(z)$:
\be
\label{eq:spheremeasure}
\int D^{d-2}z\, V_{J_L}(z;p) f(z)
&=2\int \frac{d\b}{\b}\b^{-d-J_L}\frac{\de(p^2/\b^2)}{\vol \R} f(p/\b) =2\de(p^2)f(p),
\ee
where we used the definition of the measure $D^{d-2}z$ given below (\ref{eq:spin shadow definition}).
Equation~(\ref{eq:divergencewefind}) can thus be written
\be
\cG^{(2)}&\sim
 \frac{(\l\tl\mu^\e)^2}{4(2\pi)^{2d-2}}\frac{-1}{J_{L1}+J_{L2}+2}(2\pi)^d \de^d(q_2-p_2)  
\nn\\
&\quad \x \int D^{d-2}z_3 D^{d-2}z_4  K_{\a}(z_1,z_2;z_3,z_4) V_{3-d}(z_3;p_1) V_{3-d}(z_4;p_2).
\ee

Other connected two-loop diagrams can be obtained by summing over $p_1\leftrightarrow p_2$ and $q_1\leftrightarrow q_2$. Overall, we find
\be\label{eq:Hconn}
\<\cH_{J_{L1},J_{L2}}(z_1,z_2)\>_{\substack{\textrm{2 loop,} \\ \textrm{conn.}}}
&\sim 
\frac{(\l\tl\mu^\e)^2}{4(2\pi)^{2d-2}} \frac{-1}{J_{L1}+J_{L2}+2}\nn\\
&\quad \x \int D^{d-2}z_3D^{d-2}z_4 K_{\a}(z_1,z_2;z_3,z_4) \<\cH_{3-d,3-d}(z_3,z_4)\>_{\textrm{tree}}
\ee
Note that the divergence in $\cH_{J_{L1},J_{L2}}$ is proportional to $\cH_{3-d,3-d}$, regardless of the values of $J_{L1}$ and $J_{L2}$.
Thus, only $\cH_{3-d,3-d}$ gets multiplicatively renormalized (with some integral kernel) --- other detectors get additively renormalized.
This happens because the homogeneity in $q_1$ and $p_2$ of \eqref{eq:divergencewefind}  determines the Lorentz spin of the resulting detector.

The multiplicative nature of the divergence can be made more explicit by exploiting the full power of Lorentz symmetry.
Indeed we have not yet diagonalized the \emph{total} spin of the operator on the left, which also includes an orbital component.
As anticipated in \eqref{eq:convolvewithclebsch}, eigenfunctions are simply labelled by three-point functions,
\be\label{eq:irreducibleH}
	\cH_{J_{L1},J_{L2};J_{L}}(z)&\equiv \int D^{d-2} z_1 D^{d-2} z_2 \<\cP_{-J_L}(z)\cP_{d-2+J_{L1}}(z_1)\cP_{d-2+J_{L2}}(z_2)\>\cH_{J_{L1},J_{L2}}(z_1,z_2).
\ee
By the uniqueness property of three-point functions, these automatically diagonalize the integral:
\be
&\int D^{d-2} z_1 D^{d-2} z_2 \<\cP_{-J_L}(z)\cP_{d-2+J_{L1}}(z_1)\cP_{d-2+J_{L2}}(z_2)\> K_\a(z_1,z_2;z_3,z_4) \nn\\
&\qquad= \pi^{d-2}\kappa_\a(J_L)\<\cP_{-J_L}(z)\cP_1(z_3)\cP_1(z_4)\>.
\ee
Thus the divergence in eq.~\eqref{eq:Hconn} can be written more simply as
\be\label{eq:Hconn1}
\<\cH_{J_{L1},J_{L2};J_{L}}(z)\>_{\substack{\textrm{2 loop,} \\ \textrm{conn.}}}
\sim 
\frac{(\l\tl\mu^\e)^2}{(4\pi)^d} \frac{-1}{J_{L1}+J_{L2}+2} \kappa_\a(J_L) \<\cH_{3-d,3-d;J_L}(z)\>_{\textrm{tree}}.
\ee
This is the main result of this subsection.  Here, as before $\a=\frac{J_{L1}-J_{L2}}{2}$.
The eigenvalue $\kappa_\a(J_L)$, computed in appendix~\ref{sec:diagonalizingkalpha}, takes the form
\be \label{kappa text}
\kappa_\a(J_L) &= \frac{2}{J_L+1} \cos(\tfrac{\pi J_L}{2}) \G(-\tfrac{J_L}{2})^2 \G(\tfrac{J_L+2}{2}-\a) \G(\tfrac{J_L+2}{2}+\a) +O(\e).
\ee
The eigenvalue is nonsingular for $\a\sim \e$ and generic $J_L$.

The full basis of eigenfunctions, according to the tensor product decomposition in eq.~\eqref{eq:convolvewithclebsch}, includes
states with nonzero transverse spin $j\neq 0$.  Remarkably, it turns out that the eigenvalue $\kappa_\a(J_L,j\neq 0)$ identically vanishes for all nonzero $j$!
This is shown in appendix~\ref{sec:diagonalizingkalpha}; for $\a=0$ this also follows from the results of~\cite{Chang:2020qpj} and the fact there there are no horizontal trajectories with $J=0,1,\cdots$ and $j=0$.
We will thus restrict our attention to $j=0$ states.
Note that this is distinct from what happens in the BFKL case, where all transverse spins evolve nontrivially
(although some may appear with vanishing OPE coefficient in specific examples due to selection rules, see \cite{Cornalba:2008qf}).

\subsubsection{Disconnected contributions and the complete renormalization}

In addition to the connected diagram in figure~\ref{fig:horizontal loop} there also exist disconnected diagrams which correct each of the two constituent detectors $\cD_{J_{Li}}$ inside of $\cH_{J_{L1},J_{L2}}$, see figure \ref{fig:disconnected horizontal loop}. For $J_{L1},J_{L2}\sim -1$ these diagrams lead to the mixing of $\cD_{J_{Li}}$ with $\tl\cD_{J_{Li}}$, as we discussed in section~\ref{sec:pomeron}. 

To interpret the results of section~\ref{sec:pomeron} in the context of $\cH_{J_{L1}, J_{L2}; J_L}$ we can just apply the conformal decomposition~\eqref{eq:irreducibleH} to the divergence~\eqref{eq:ansatz1}.
It is important however that after performing the angular integral in \eqref{eq:irreducibleH}, the shadow detectors $\tl\cD_{J_{Li}}$
do not define an independent family of detectors, rather they are just proportional to $\cD_{2-d-J_{Li}}$.
Thus,
\be\label{eq:Hdisonnected}
	\<\cH_{J_{L1}, J_{L2}; J_L}\>_{\substack{\textrm{2 loop,} \\ \textrm{disc.}}}=&\p{X(\e,J_{L1})+X(\e,J_{L2})}\<\cH_{J_{L1}, J_{L2}; J_L}\>_\text{tree}\nn\\
	&+\hat S(-J_L,d-2+J_{L2},[d-2+J_{L1}])Y(\e,J_{L1})\mu^{2-d-2J_{L1}}\<\cH_{2-d-J_{L1}, J_{L2}; J_L}\>_\text{tree}\nn\\
	&+\hat S(-J_L,d-2+J_{L1},[d-2+J_{L2}])Y(\e,J_{L2})\mu^{2-d-2J_{L2}}\<\cH_{J_{L1}, 2-d-J_{L2}; J_L}\>_\text{tree}\nn\\
	&+\text{finite}.
\ee
Here the shadow coefficient $\hat S(\de_1,\de_2,[\de_3])$ is defined by
\be
	\<\cP_{\de_1}(z_1)\cP_{\de_2}(z_2)\hat\wS_J[\cP_{\de_3}(z_3)]\>=\hat S(\de_1,\de_2,[\de_3])	\<\cP_{\de_1}(z_1)\cP_{\de_2}(z_2)\cP_{d-2-\de_3}(z_3)\>.
\ee
Explicitly,
\be\label{eq:Shatcoeff}
	\hat S(\de_1,\de_2,[\de_3])=&\frac{
		\pi^{\frac{d-2}{2}}\G(\de_3-\tfrac{d-2}{2})\G(\tfrac{d-2-\de_3+\de_1-\de_2}{2})\G(\tfrac{d-2-\de_3+\de_2-\de_1}{2})
	}{
		\G(d-2-\de_3)\G(\tfrac{\de_3+\de_1-\de_2}{2})\G(\tfrac{\de_3+\de_2-\de_1}{2})
	}\frac{\G(\de_3)}{\pi^{\frac{d-2}{2}}\G(\tfrac{2-d}{2}+\de_3)}\nn\\
	=&\frac{
		\G(\de_3)\G(\tfrac{d-2-\de_3+\de_1-\de_2}{2})\G(\tfrac{d-2-\de_3+\de_2-\de_1}{2})
	}{
		\G(d-2-\de_3)\G(\tfrac{\de_3+\de_1-\de_2}{2})\G(\tfrac{\de_3+\de_2-\de_1}{2})
	}.
\ee
In the first line, the first factor is the standard shadow coefficient in $d-2$ dimensions and the second factor comes from the definition~\eqref{eq:Shatdefn}. The coefficients $X(\e,J_L)$ and $Y(\e, J_L)$ can be deduced from~\eqref{eq:twist2fulldiv}, 
\be
	X(\e,J_L)=&\frac{\l^2}{(4\pi)^4}\p{    \frac{1}{12\e}  - \frac{1}{2\e J_L} +\frac{\cR(\e)}{2 (J_L-\Jo)} + \frac{\e\cR(\e)}{2(J_L-\Jo)(J_L-\frac{2-d}{2})}},\\
	Y(\e,J_L)=&-\frac{\l^2}{(4\pi)^4}{\frac{\e\cR(\e)}{2(J_L-\Jo)(J_L-\frac{2-d}{2})}}.
\ee
As discussed in section~\ref{sec:pomeron}, these have a spurious pole at $J_L=\frac{2-d}{2}$, but it cancels in the combination~\eqref{eq:ansatz1} due to $\cD_{\frac{2-d}{2}}=\tl\cD_{\frac{2-d}{2}}$. It therefore also cancels in the combination~\eqref{eq:Hdisonnected}.

Combining~\eqref{eq:Hdisonnected} with~\eqref{eq:Hconn1}, we can define the renormalized operator 
\be\label{eq:Hfinal}
	[\cH_{J_{L1}, J_{L2}; J_L}]_R=&(1-X(\e,J_{L1})-X(\e,J_{L2}))\cH_{J_{L1}, J_{L2}; J_L}\nn\\
	&-\hat S(-J_L,d-2+J_{L2},[d-2+J_{L1}])Y(\e,J_{L1})\mu^{2-d-2J_{L1}}\cH_{2-d-J_{L1}, J_{L2}; J_L}\nn\\
	&-\hat S(-J_L,d-2+J_{L1},[d-2+J_{L2}])Y(\e,J_{L2})\mu^{2-d-2J_{L2}}\cH_{J_{L1}, 2-d-J_{L2}; J_L}\nn\\
	&+\frac{(\l\tl\mu^\e)^2}{(4\pi)^d} \frac{\mu^{-J_{L1}-J_{L2}-2}}{J_{L1}+J_{L2}+2} \kappa_\a(J_L) \cH_{3-d,3-d;J_L}.
\ee
The operator $[\cH_{J_{L1}, J_{L2}; J_L}]_R$ has finite matrix elements at two-loop level.

\subsection{Dilatation operator and anomalous dimensions?} \label{ssec: anomalous dimension?}

The dilatation operator acting on $[\cH_{J_{L1}, J_{L2}; J_L}]_R$ can now be obtained straightforwardly. The structure of its action is evident from~\eqref{eq:Hfinal}:
\be
	D[\cH_{J_{L1}, J_{L2}; J_L}]_R\sim [\cH_{J_{L1}, J_{L2}; J_L}]_R+[\cH_{2-d-J_{L1}, J_{L2}; J_L}]_R+[\cH_{J_{L1}, 2-d-J_{L2}; J_L}]_R+[\cH_{3-d,3-d; J_L}]_R,
\ee
where we omit explicit coefficients. We can make two observations. Firstly, the subspace spanned by the three operators 
\be\label{eq:3md multiplet}
	\{[\cH_{3-d,3-d;J_L}]_R,[\cH_{3-d,-1;J_L}]_R,[\cH_{-1,-1;J_L}]_R\}
\ee
is closed under the action of $D$.
(There are three operators and not four because of Bose symmetry: $[\cH_{-1,3-d;J_L}]_R=[\cH_{3-d,-1;J_L}]_R$.)
Secondly, for all other $J_{L1},J_{L2}$ the disconnected part of $D$ acts within the subspace spanned by
\be\label{eq:othermultiplets}
	\{[\cH_{J_{L1}, J_{L2}; J_L}]_R,[\cH_{2-d-J_{L1}, J_{L2}; J_L}]_R,[\cH_{J_{L1}, 2-d-J_{L2}; J_L}]_R,[\cH_{2-d-J_{L1},2-d-J_{L2}; J_L}]_R\},
\ee
while the connected contribution adds an off-diagonal mixing with $[\cH_{3-d,3-d;J_L}]_R$.

As far as computing the eigenvalues of $D$ goes, this means that the connected diagrams only affect the eigenvalues on the subspace~\eqref{eq:3md multiplet}. On the subspace~\eqref{eq:othermultiplets} the connected contribution does not affect the eigenvalues, and only modifies the eigenvectors. In fact,  for most values of $J_{L1},J_{L2}$ there is technically no need to subtract $\cH_{3-d,3-d}$ in~\eqref{eq:Hfinal}, and the modification to the eigenvectors amounts to removing this subtraction.

As discussed in section~\ref{sec:horizontalintroduction}, this suggests that generically~\eqref{eq:othermultiplets} should not be treated as new operators, and only~\eqref{eq:3md multiplet} contains new dynamical information. We can then consider the problem of diagonalizing the dilatation operator on the subspace~\eqref{eq:3md multiplet}.

This turns out to be subtle for at least two reasons. Firstly, when we go to higher perturbative orders, it is possible that operators other than $\cH_{3-d,3-d,J_L}$ will appear from connnected divergences. Our preliminary analysis indicates that this happens already at $O(\l^3)$.\footnote{For example, if we replace one of the interaction vertices in the diagram in figure~\ref{fig:horizontal loop} with its two-loop correction (i.e.\ a sum of 3 subdiagrams and a counter-term), we will get an $O(\l^3)$ diagram which induces mixing of $\cH_{3-d,3-d}$ with $\cH_{3-d+\e/2,3-d}$. Replacing both vertices with their higher-order corrections we can get mixing with $\cH_{3-d+a\e/2,3-d+b\e/2}$ at $O(\l^{a+b+2})$.} These might include operators with other values of $J_{L1}$ and $J_{L2}$, or with a different number of fundamental fields, e.g.\ $\wL[\f^2]\wL[\f^4]$.  While it isn't necessarily a problem for diagonalizing the action of $D$ order by order, this infinite degeneracy raises the question of the meaning of its eigenvalues. Normally, when we have a finite degeneracy, we can write an RG equation which shows that computing anomalous dimensions $\g$ allows us to resum terms of the form $(\gamma \log x)^n$, where $x$ is some scale. Here, the structure of the resummation is unclear, and so the meaning of $D$ eigenvalues is also obscured. Specifically, it is not clear whether the solutions to the RG equation have the form of a discrete sum of powers $x^\gamma$. It would be interesting to study this question in more detail.

Secondly, even if we simply focus on computing the eigenvalues order-by-order, our dilatation operator $D$ has poles $\l^2/(J_{Li}-\frac{2-d}{2})\sim \frac{\l^2}{\e}$ due to degeneracy of the basis
\eqref{eq:3md multiplet} as $\e\to 0$.
This can be resolved as in section~\ref{sec:pomeron} by a change of basis. This results in the leading-order dilatation operator becoming a $3\x3$ Jordan block. This Jordan block structure complicates the perturbation theory for the eigenvalues. For example, in section~\ref{sec:pomeron} we saw that at the fixed point we needed $O(\e^2)$ dilatation operator to determine $O(\e)$ anomalous dimensions for a $2\x 2$ Jordan block.
For a $3\x 3$ Jordan block we need $D$ to $O(\e^3)$ to determine even just the $O(\e)$ anomalous dimensions.

To be more concrete, let us define the analog of the regular basis \eqref{eq:nice basis} for two detectors:
\be
	\cH_{1;J_L}&=[\cH_{3-d,3-d;J_L}]_R,\\
	\cH_{2;J_L}&=\frac{2}{\e}\p{\mu^{-\e}[\cH_{-1,3-d;J_L}]_R-[\cH_{3-d,3-d;J_L}]_R},\\
	\cH_{3;J_L}&=\frac{4}{\e^2}\p{\mu^{-2\e}[\cH_{-1,-1;J_L}]_R-2\mu^{-\e}[\cH_{-1,3-d;J_L}]_R+[\cH_{3-d,3-d;J_L}]_R}.
\ee
Setting $\mathbb{H}_{J_L}=\p{\cH_{1;J_L},\cH_{2;J_L},\cH_{3;J_L}}^T$,  we have
\be
	D\mathbb{H}_{J_L}=\mathscr{D}_H\mathbb{H}_{J_L}+O(\l^3),
\ee
where (treating $\l\sim \e$)
\be
	\mathscr{D}_{H}=&\begin{pmatrix}
			-2 & 0 & 0\\
			2 & -2{+}\e & 0\\
			0 & 4 & -2{+}2\e
		\end{pmatrix}+
\frac{\l^2}{(4\pi)^4}
		\begin{pmatrix}
	\frac{-7}{3}+s_{11} & 2& 0 \\
	2+s_{21}& \frac{-7}{3} & 1\\
	s_{31}& 4+s_{32}& \frac{-7}{3}+s_{33}
		\end{pmatrix}
		\nn\\&+
	\frac{\l^2}{(4\pi)^4}
		\begin{pmatrix}
			\kappa_0(J_L) & 0 & 0\\
			0 & 0 & 0\\
			-\ptl_\a^2\kappa_\a(J_L)\big\vert_{\a=0} & 0 & 0
		\end{pmatrix}+ O(\e^3). \label{dil H}
\ee
The first two terms come from the disconnected result \eqref{eq:D} at tree- and two-loop level respectively, restricted to $J_{Li}=3-d$,
while the last term comes from the connected divergence \eqref{eq:Hconn1}.
The latter has two columns that strictly vanish, because the divergence involves only $\cH_{3-d,3-d;J_L}$. The coefficients $s_{ij}$ originate from the shadow coefficient~\eqref{eq:Shatcoeff} and are given in appendix~\ref{app:Scoeffs}.

Naively, one would like to predict the scale dependence of matrix elements by exponentiating \eqref{dil H}:
\be \< \mathbb{H}_{J_L} \>_{\{\Lambda p_i\}} \overset{?}{=} \exp\left[\log(\Lambda) \mathscr{D}_{H}\right] \< \mathbb{H}_{J_L}\>_{\{p_i\}}
\ee
where the expectation value on the left is taken in a state with momenta rescaled by $\Lambda$.
The difficulty is that due to the $\sim \e^0$ terms below the diagonal, the unknown $\sim\e^3$ top-right entry $(\mathscr{D}_H)_{13}$ contributes to the exponential already
at leading logarithm order ($\sim \e^k \log^k\Lambda$). Thus, eq.~\eqref{dil H} does not even contain enough information to predict leading logarithms!
The same conclusion is reached by considering its characteristic equation: eq.~\eqref{dil H} is insufficient to predict $O(\e)$ eigenvalues.

One might try to make assumptions about $(\mathscr{D}_H)_{13}$ at $O(\l^3)$, but
representing $D$ as a $3\times 3$ matrix becomes questionable at that order since other operators start to mix.

A somewhat analogous situation exists in gauge theories, where evolution can produce an unbounded number of Wilson lines $U(z_i)$.
At weak fields these can be parametrized using a ``Reggeized gluon'' operator, $U(z_i)=e^{igW(z_i)}$.
While it is clear from the Wilson line picture that increasing the number of Reggeized gluons $W(z_i)$ costs powers of the coupling $g$,
it turns out that terms in the evolution that remove Reggeized gluons also cost the same powers of $g$.
This second property is crucial to approximately decouple sectors with different numbers of Reggeized gluons (each having a discrete spectrum \cite{Derkachov:2002wz}), but it is far from obvious in the Wilson line picture.
Rather, it follows from target-projectile duality of rapidity evolution, mentioned previously below eq.~\eqref{D0 Jordan form}.\footnote{The present detector-frame formulation corresponds to resumming so-called non-global logarithms, while target-projectile duality applies to rapidity evolution of correlators.
The evolution equations coincide in the critical dimension where QCD is conformal \cite{Hatta:2008st,Caron-Huot:2015bja}.}
This suggests that target-projectile duality should play an important role in scalar theories as well. 

The calculations here simply establish a general method for renormalizing individual diagrams, which applies even in the vicinity of singular points.
They do not answer the basic question of whether the evolution of general composite operators in the Wilson-Fisher theory
can be understood in terms of simple elementary excitations,
or even if there should be a single such ``Reggeon'' (corresponding perhaps to $\wL[\phi^2]$), a pair of them (corresponding say to the two eigenstates of eq.~\eqref{eq:D} at $J_L=3-d$), or more.
We leave these questions to future work.

\subsection{Appearance in correlation functions}
\label{sec:AppInCorrFuncs}

The construction and renormalization of the horizontal trajectories in the preceding subsections leaves open the question of whether they actually contribute to correlation functions. This question is further complicated by the subtleties discussed at the end of section~\ref{ssec: anomalous dimension?}. 

In this section, as a first step in addressing these issues, we consider correlators in a modified theory, or more simply a subset of the diagrams of the Wilson-Fisher theory, which corresponds to only the connected contribution in figure~\ref{fig:horizontal loop}.\footnote{The subset of diagrams that we consider here gives the full answer for two-magnon correlators in the large-$N$ limit of the conformal fishnet theory~\cite{Zamolodchikov:1980mb,Gurdogan:2015csr}. In fact, essentially the same calculation as we do here has been performed in~\cite{Gromov:2018hut}. The exact spectrum of two-magnon Regge trajectories is given by their equations (4.47) and (4.49) (in arXiv version 1 of~\cite{Gromov:2018hut}), which can be seen to contain a single $J=-1$ horizontal trajectory at weak coupling (the authors of~\cite{Gromov:2018hut} were only interested in $\mathrm{Re}\, J\geq 0$). In operator language, we expect that, due to the simplifications in the large-$N$ limit of the fishnet theory, no mixing is possible for the $\cH_{3-d,3-d}$ trajectory, which then gives the unique contribution at $J=-1$. Here, our goal is to describe the physical picture of how these operators appear in the correlation functions, as well as to explain that their contribution is non-zero in more general theories, such as the Wilson-Fisher theory.} 
The renormalization of \eqref{eq:Hconn1} then gives rise to a single near-horizontal trajectory with (setting $J_L=1-\Delta$)
\be \label{J connected}
 J(\Delta) = 1-\Delta_L(J_L) = -1+\frac{\lambda^2}{(4\pi)^4} \kappa_0(1-\Delta) +O(\lambda^3)\,.
\ee
Since this trajectory is constructed from a product of two $\mathbf{L}[\phi^{2}]$ operators, it is natural to expect it to be exchanged in the correlator $\<\phi^2\phi^2\phi^2\phi^2\>$. 
By examining a subset of Feynman diagrams and studying CFT inversion formulas, we will show that a Regge trajectory with (approximately) constant negative spin, $J\approx -1$,
indeed contributes to the Regge limit of this correlator.

First, as a brief reminder, in CFTs we can rewrite this position-space correlator as:
\be
\<\f^2(x_1)\f^2(x_2)\f^2(x_3)\f^2(x_4)\>=\frac{1}{x_{12}^{2\Delta_{\cO}}x_{34}^{2\Delta_{\cO}}}\mathcal{G}(z,\bar{z}), \label{eq:4phi2Convention}
\ee
with $\cO=\phi^2$ and where $(z,\bar{z})$ are the $d$-dimensional conformal cross-ratios,
\be
z\bar{z}=\frac{x_{12}^{2}x_{34}^{2}}{x_{13}^{2}x_{24}^{2}}, \qquad (1-z)(1-\bar{z})=\frac{x_{14}^{2}x_{23}^{2}}{x_{13}^{2}x_{24}^{2}}.
\ee
We can write the four-point function $\mathcal{G}(z,\bar{z})$ as an integral of conformal blocks over the principal series:
\begin{align}
\mathcal{G}(z,\bar{z})=\sum\limits_{J}\int\limits_{d/2-i\infty}^{d/2+i\infty}\frac{d\Delta}{2\pi i}C(\Delta,J)g_{\Delta,J}(z,\bar{z})
\end{align}
where $g_{\Delta,J}(z,\bar{z})$ are the conformal blocks and $C(\Delta,J)$ is the OPE function we want to study. 
To recover the standard conformal block expansion we close the contour to the right and pick up the poles in $C(\Delta,J)$.
The poles of $C(\Delta,J)$ in $\Delta$ at fixed, positive, integer $J$ then determine the spectrum of the exchanged operators.\footnote{The function $C(\De,J)$ and the conformal blocks also have kinematic poles in $\Delta$, but these will cancel out.}
The Lorentzian inversion formula \cite{Caron-Huot:2017vep} gives an analytic continuation of $C(\Delta,J)$ to generic $J\in \mathbb{C}$ and the poles of this analytic function are associated to light-ray operators \cite{Kravchuk:2018htv}.
The goal of this section is then to show that $C(\Delta,J)$ for $\<\phi^2\phi^2\phi^2\phi^2\>$ has a pole at $J\approx-1$ for arbitrary $\Delta$, which would correspond to a horizontal trajectory in a Chew-Frautschi plot.

In practice, we use two different methods.
In the first approach we compute the correlator $\<\phi^2\phi^2\phi^2\phi^2\>$ using standard Feynman rules, analytically continue to Lorentzian kinematics, and then evaluate the Lorentzian inversion formula in the Regge limit.
This approach lets us directly compute $C(\Delta,J)$ for $J$ near $-1$, but quickly becomes cumbersome at higher-loops. 
In the second approach, we use that many Feynman diagrams for $\<\phi^2\phi^2\phi^2\phi^2\>$ can be computed using harmonic analysis for the Euclidean conformal group. In this method we compute $C(\Delta,J)$ for positive, even integers using the Euclidean inversion formula and then analytically continue this expression to $J=-1$.
The fact these two methods agree gives a non-trivial consistency check for our results.
In this section we will mostly explain the first method, involving the Lorentzian inversion formula, and explain the second method in more detail in Appendix~\ref{app:diagonalizes}. 

For the remainder of this section we will work in $d=4$ dimensions.
We use the standard scalar propagator $\<\f(x_1)\f(x_2)\>=N_\f x_{12}^{-2}$, where $N_\f=\tfrac{1}{4\pi^2}$. At tree level, our correlator is given by
\be
\<\f^2(x_1)\f^2(x_2)\f^2(x_3)\f^2(x_4)\>_\mathrm{tree} &= G_\mathrm{tree}(x_1,x_2,x_3,x_4),
\ee
where $G_\mathrm{tree}$ is a sum of Wick contractions whose explicit form will not be needed here.

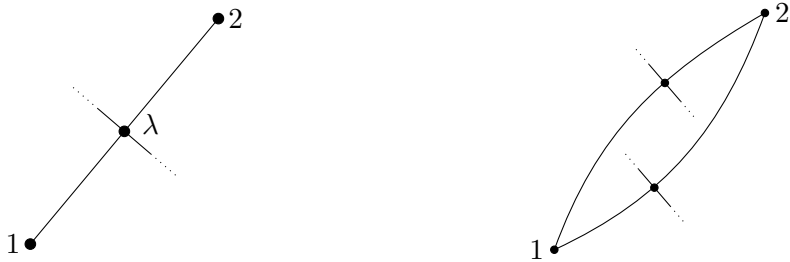
\begin{figure}[t]
\centering
\begin{subfigure}[c]{0.4\textwidth}
\centering\begin{tikzpicture}
\draw[] (0,0) -- (2.5,3);
\draw[black,fill=black] (0,0) circle (2pt);
\draw[black,fill=black] (2.5,3) circle (2pt);
\draw[black,fill=black] (1.25,1.5) circle (2pt);
\node[left] at (0,0) {$1$};
\node[right] at (2.5,3) {$2$};
\node[right] at (1.35,1.6) {$\l$};
\draw[] (1.25,1.5) -- (1.6,1.2);
\draw[] (1.25,1.5) -- (0.9,1.8);
\draw[dotted] (1.6,1.2) -- (1.95,0.9);
\draw[dotted] (0.9,1.8) -- (0.55,2.1);
\end{tikzpicture}
\end{subfigure}
\qquad
\begin{subfigure}[c]{0.40\textwidth}\centering
\begin{tikzpicture}[scale=0.7]
\draw[black,fill=black] (0,0) circle (2pt);
\draw[black,fill=black] (4,4.5) circle (2pt);
\draw[black,fill=black] (2.1,3.17) circle (2pt);
\draw[black,fill=black] (1.9,1.18) circle (2pt);
\draw[] (0,0) to[out=70,in=210] (4,4.5);
\draw[] (0,0) to[out=25,in=250] (4,4.5);
\draw[] (1.9-0.3,1.18+0.35) -- (1.9 + 0.3, 1.18 - 0.35);
\draw[dotted] (1.9 + 0.3, 1.18 - 0.35) -- (1.9 + 0.3 + 0.3, 1.18 - 0.35 - 0.35);
\draw[dotted] (1.9 - 0.3, 1.18 + 0.35) -- (1.9 - 0.3 - 0.3, 1.18 + 0.35 + 0.35);
\draw[] (2.1-0.3,3.17+0.35) -- (2.1 + 0.3, 3.17 - 0.35);
\draw[dotted] (2.1 + 0.3, 3.17 - 0.35) -- (2.1 + 0.3 + 0.3, 3.17 - 0.35 - 0.35);
\draw[dotted] (2.1 - 0.3, 3.17 + 0.35) -- (2.1 - 0.3 - 0.3, 3.17 + 0.35 + 0.35);
\node[left] at (0,0) {$1$};
\node[right] at (4,4.5) {$2$};
\end{tikzpicture}
\end{subfigure}
\caption{The operators $\f(x_1)$ and $\f(x_2)$ can create a $\f^2$ operator between them via the $\l \f^4$ interaction. When $\f(x_1)$ and $\f(x_2)$ are boosted apart, this $\f^2$ operator gets smeared over the null cone, creating the detector $\wL[\f^2]$. Right: when $\f^2(x_1)$ and $\f^2(x_2)$ are boosted, the same mechanism can create a pair of detectors $\wL[\f^2]\wL[\f^2]$.
\label{fig:creatingwlphisq}}
\end{figure}

To understand which diagrams are relevant for finding the horizontal trajectory, it is helpful to give a physical interpretation for how the product $\wL[\f^2] \wL[\f^2]$ appears in the Regge limit of the correlator. Let us begin with a pair of operators $\f^2(x_1)$ and $\f^2(x_2)$, and boost them in opposite directions. In the limit of large boost, the points $x_1$ and $x_2$ approach the tips of a null cone.
From each pair $\f(x_1)\f(x_2)$, we obtain an effective $\f^2$ operator using the $\l\f^4$ interaction, see the left panel of figure~\ref{fig:creatingwlphisq}.
This $\f^2$ operator is smeared near the null cone with a wavefunction that becomes independent of the null direction in the large boost limit --- in other words, at large boost, we end up with $\wL[\f^2]$.
Since we have two pairs $\f(x_1)\f(x_2)$, we obtain two $\wL[\f^2]$ operators, as shown on the right panel.
This argument shows that the detector $\wL[\f^2] \wL[\f^2]$ should appear in the Regge limit of $\<\f^2\f^2\f^2\f^2\>$ at order $O(\l^2)$.

\begin{figure}[t]
\centering
\begin{subfigure}[c]{0.4\textwidth}
\hspace{0.4in}
\begin{tikzpicture}[scale=0.8]
\draw[black,fill=black] (0,0) circle (2pt);
\draw[black,fill=black] (4,4.5) circle (2pt);
\draw[black,fill=black] (4,0) circle (2pt);
\draw[black,fill=black] (0,4.5) circle (2pt);
\draw[black,fill=black] (2.1,3.17) circle (2pt);
\draw[black,fill=black] (1.9,1.18) circle (2pt);
\draw[] (0,0) to[out=70,in=210] (4,4.5);
\draw[] (0,0) to[out=25,in=250] (4,4.5);
\draw[] (4,0) to[out=158,in=280] (0,4.5);
\draw[] (4,0) to[out=105,in=-26] (0,4.5);
\node[left] at (0,0) {$1$};
\node[right] at (4,4.5) {$2$};
\node[right] at (4,0) {$3$};
\node[left] at (0,4.5) {$4$};
\node[above] at (2.12,3.23) {$2'$};
\node[below] at (1.9,1.16) {$1'$};
\end{tikzpicture}
\end{subfigure}
\hfill
\begin{subfigure}[c]{0.43\textwidth}
\begin{tikzpicture}
\draw[black,fill=black] (0,0) circle (2pt);
\draw[black,fill=black] (2,0) circle (2pt);
\draw[black,fill=black] (4,0) circle (2pt);
\draw[black,fill=black] (0,2) circle (2pt);
\draw[black,fill=black] (2,2) circle (2pt);
\draw[black,fill=black] (4,2) circle (2pt);
\draw[] (0,0) -- (2,0);
\draw[] (0,0) -- (2,2);
\draw[] (0,2) -- (2,0);
\draw[] (0,2) -- (2,2);
\draw[] (2,0) -- (4,0);
\draw[] (2,0) -- (4,2);
\draw[] (2,2) -- (4,0);
\draw[] (2,2) -- (4,2);
\node[left] at (0,0) {$1$};
\node[left] at (0,2) {$2$};
\node[right] at (4,0) {$3$};
\node[right] at (4,2) {$4$};
\node[below] at (2,0) {$1'$};
\node[above] at (2,2) {$2'$};
\end{tikzpicture}
\end{subfigure}
\caption{On the left, a diagram showing an event shape for $\wL[\f^2] \wL[\f^2]$, embedded inside the four-point correlator $\<\f^2\f^2\f^2\f^2\>$. We imagine $\f^2(x_1) \f^2(x_2)$ creating the detector $\wL[\f^2] \wL[\f^2]$ via the mechanism in figure~\ref{fig:creatingwlphisq}. This detector is then measured in states created by $\f^2(x_3)$ and $\f^2(x_4)$. On the right, we re-draw the same diagram in a simpler way that makes clear that it is the square of the kernel $S$, which is one of the terms inside $SG_\mathrm{tree}$ in equation~(\ref{eq:sequenceofdiagrams}).
\label{fig:diagramshowingwphiincorrelator}}
\end{figure}
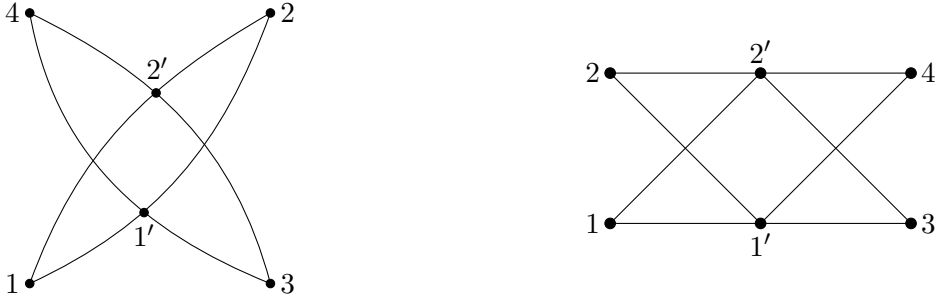

We can embed the second diagram of figure~\ref{fig:creatingwlphisq} into a standard Feynman diagram for $\<\f^2\f^2\f^2\f^2\>$ by evaluating its expectation value in a state created by $\f^2$. This leads to the diagram in figure~\ref{fig:diagramshowingwphiincorrelator}. We recognize this as the square of the standard conformal box diagram:
\begin{align}
\label{eq:squarebox}
	F^{(2)}(x_i)&=\frac{\lambda^{2}}{2}N_{\f}^{8}\int d^{d}x_1' d^{d}x_2'\frac{1}{x_{11'}^{2}x_{21'}^{2}x_{31'}^2x_{41'}^{2}}\frac{1}{x_{12'}^{2}x_{22'}^{2}x_{32'}^2x_{42'}^{2}}
	\nonumber
	\\
	&=\frac{\lambda^{2}}{2}N_{\f}^{8}\pi^{4}\frac{(z\bar{z})^{2}}{x_{12}^4x_{34}^4}\left(\Phi^{(1)}(z,\bar{z})\right)^{2},
\end{align}
where the factor of $1/2$ is a symmetry factor and
\begin{align}
	\Phi^{(1)}(z,\bar{z})&=\frac{1}{z-\bar{z}}\left(2\Li_2(z)-2\Li_2(\bar{z})+\log(z\bar{z})\log\left(\frac{1-z}{1-\bar{z}}\right)\right).
\end{align}

We next need to plug this result into the Lorentzian inversion formula.
The inversion formula for $\<\phi^2\phi^2\phi^2\phi^2\>$ in $4d$ is:\footnote{$\kappa_{\beta}$ is a kinematic factor and should not be confused with the eigenvalue $\kappa_{\alpha}(J_L)$ studied in the previous section.}
\begin{align}
C(\Delta,J)&=\frac{\kappa_{\Delta+J}}{2}\int\limits_{0}^{1}dz d\bar{z} \frac{(z-\bar{z})^{2}}{(z\bar{z})^4}g_{J+3,\Delta-3}(z,\bar{z})\text{dDisc}[\mathcal{G}(z,\bar{z})], \label{eq:inv4d}
\\
\kappa_{\beta}&=\frac{\Gamma(\frac{\beta}{2})^{4}}{2\pi ^{2}\Gamma(\beta-1)\Gamma(\beta)},
\end{align}
where here $g_{\Delta,J}$ is a $4d$ conformal block, 
\be
g_{\Delta,J}(z,\bar{z})&=\frac{z\bar{z}}{z-\bar{z}}\left(k_{\Delta+J}(z)k_{\Delta-J-2}(\bar{z})-(z\leftrightarrow \bar{z})\right) \label{eq:4dBlock}
\\
k_{2h}(z)&=z^h{}_2F_1(h,h,2h,z),
\ee
and the double-discontinuity is defined by:
\begin{align}
\text{dDisc}_{t}[\mathcal{G}(z,\bar{z})]&=\mathcal{G}(z,\bar{z})-\frac{1}{2}\left[\cG^{\circlearrowleft}(z,\bar{z})+\cG^{\circlearrowright}(z,\bar{z})\right].
\end{align}
The arrows in $\cG^{\circlearrowleft,\circlearrowright}$ indicate how we analytically continue around the branch cut that runs along $\bar{z}\in(1,\infty)$.\footnote{In general the inversion formula has distinct contributions from the $t$- and $u$- channel double-discontinuities, but $\<\f^2\f^2\f^2\f^2\>$ is invariant under $x_3\leftrightarrow x_4$ so they give identical results.}

To look for poles of $C(\De,J)$ in $J$, we should study the Regge limit
\begin{align}
	z=\sigma, \qquad \bar{z}=\eta \sigma,
\end{align}
with $\s\to 0$ and $\eta$ held fixed.\footnote{The pole at $J=-1$ will come from the $\s\ll1$ region of the integral so we can focus on this part of the integrand.}
In this limit, the $t$-channel $\dDisc$ of (\ref{eq:squarebox}) becomes
\begin{align}
	\dDisc_t\left[x_{12}^{4}x_{34}^{4}F^{(2)}(x_i)\right]\approx N_{\f}^{8} \lambda^2 2\pi^6 \frac{\eta^{2}\sigma^2\log^2(\eta)}{(1-\eta)^{2}}.
\end{align}
Plugging this into the Lorentzian inversion formula and evaluating the integral in the limit $\sigma\rightarrow 0$, we indeed find a pole at $J=-1$:
\begin{align}
	C^{(2)}(\Delta,J)\sim N_{\f}^{8}\lambda^{2}\frac{1}{J+1}\frac{\pi ^7 \sin (\pi  \Delta ) \Gamma \left(\frac{\Delta -1}{2}\right)^4}{\cos^4\left(\frac{\pi  \Delta }{2}\right) \Gamma (\Delta -2) \Gamma (\Delta -1)}\,. 
\label{eq:CtDirectInt}
\end{align}

This calculation shows that the correlator $\<\f^2\f^2\f^2\f^2\>$ does indeed contain a horizontal trajectory near $J=-1$. The result (\ref{eq:CtDirectInt}) encodes the appearance of this trajectory at $O(\l^2)$, as anticipated above. To find its anomalous spin, we must study the correlator at $O(\l^4)$. 
However, before doing so, it will be helpful to understand the appearance of the $1/(J+1)$ pole in (\ref{eq:CtDirectInt}) in a more efficient way using Euclidean harmonic analysis.

\begin{figure}[t]
\centering
\begin{tikzpicture}
\draw[black,fill=black] (0,0) circle (2pt);
\draw[black,fill=black] (2,0) circle (2pt);
\draw[black,fill=black] (0,2) circle (2pt);
\draw[black,fill=black] (2,2) circle (2pt);
\draw[] (0,0) -- (2,0);
\draw[] (0,0) -- (2,2);
\draw[] (0,2) -- (2,0);
\draw[] (0,2) -- (2,2);
\node[left] at (0,0) {$1$};
\node[left] at (0,2) {$2$};
\node[right] at (2,0) {$3$};
\node[right] at (2,2) {$4$};
\end{tikzpicture}
\caption{The kernel $S(x_1,x_2,x_3,x_4)$ in pictorial form.\label{fig:skernel}}
\end{figure}
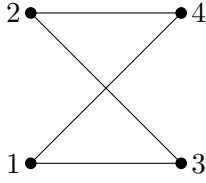

Staring at figure~\ref{fig:diagramshowingwphiincorrelator}, we recognize that it has the structure of a composition of kernels acting on pairs of spacetime points:
\be
(S_1S_2)(x_1,x_2,x_3,x_4) &= \int d^dx_1' d^dx_2' S_1(x_1,x_2,x_1',x_2') S_2(x_1',x_2',x_3,x_4).
\ee
Specifically, we can think of the diagram in figure~\ref{fig:diagramshowingwphiincorrelator} as part of the composition
\be
 \l^2 (SG_\mathrm{tree})(x_1,x_2,x_3,x_4) &\subset \<\f^2(x_1)\f^2(x_2)\f^2(x_3)\f^2(x_4)\>
\ee
where the kernel $S$ is given by
\be
\label{eq:thekernels}
S(x_1,x_2;x_3,x_4) &\equiv  \frac {N_\f^4} 2 \frac{1}{x_{13}^2 x_{14}^2 x_{23}^2 x_{24}^2},
\ee
see figure~\ref{fig:skernel}.

In fact, figure~\ref{fig:diagramshowingwphiincorrelator} is part of a series of diagrams obtained by iterating the kernel $S$:
\be
\label{eq:sequenceofdiagrams}
 \<\f^2(x_1)\f^2(x_2)\f^2(x_3)\f^2(x_4)\>&\supset\sum_{n=0}^\oo \l^{2n} (S^n G_\mathrm{tree})(x_1,x_2,x_3,x_4)
\nn\\
&= \p{\frac{1}{1-\l^2 S} G_\mathrm{tree}}(x_1,x_2,x_3,x_4).
\ee
Note that $S(x_1,x_2;x_3,x_4)$ contains a factor of $1/2$. The terms in (\ref{eq:sequenceofdiagrams}) thus contain factors of $1/2^n$ that account for the symmetry factors of diagrams obtained by composing multiple $S$'s. In the last line of (\ref{eq:sequenceofdiagrams}), we summed the geometric series.

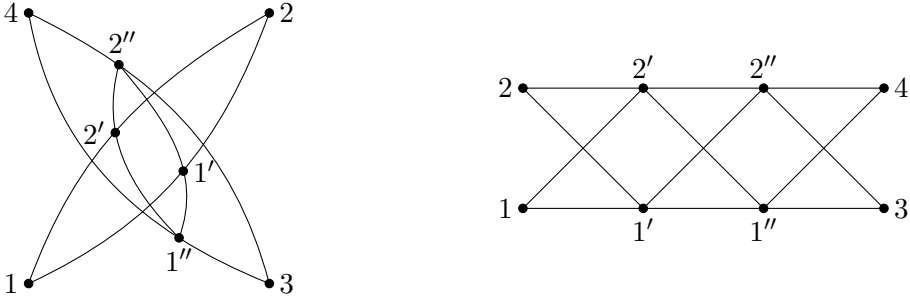
\begin{figure}[t]
\centering
\begin{subfigure}[c]{0.35\textwidth}
\hspace{0.3in}
\begin{tikzpicture}[scale=0.8]
\draw[black,fill=black] (0,0) circle (2pt);
\draw[black,fill=black] (4,4.5) circle (2pt);
\draw[black,fill=black] (4,0) circle (2pt);
\draw[black,fill=black] (0,4.5) circle (2pt);
\draw[black,fill=black] (1.5,3.64) circle (2pt);
\draw[black,fill=black] (2.5,0.76) circle (2pt);
\draw[black,fill=black] (2.57,1.87) circle (2pt);
\draw[black,fill=black] (1.44,2.51) circle (2pt);
\draw[] (0,0) to[out=70,in=210] (4,4.5);
\draw[] (0,0) to[out=25,in=250] (4,4.5);
\draw[] (4,0) to[out=158,in=280] (0,4.5);
\draw[] (4,0) to[out=105,in=-26] (0,4.5);
\draw[] (1.5,3.64) to[out=255,in=135] (2.5,0.76);
\draw[] (1.5,3.64) to[out=-50,in=70] (2.5,0.76);
\node[left] at (0,0) {$1$};
\node[right] at (4,4.5) {$2$};
\node[right] at (4,0) {$3$};
\node[left] at (0,4.5) {$4$};
\node[above] at (1.56,3.66) {$2''$};
\node[below] at (2.5,0.75) {$1''$};
\node[right] at (2.57,1.87) {$1'$};
\node[left] at (1.44,2.51) {$2'$};
\end{tikzpicture}
\end{subfigure}
\hfill
\begin{subfigure}[c]{0.52\textwidth}
\begin{tikzpicture}[scale=0.8]
\draw[black,fill=black] (0,0) circle (2pt);
\draw[black,fill=black] (2,0) circle (2pt);
\draw[black,fill=black] (4,0) circle (2pt);
\draw[black,fill=black] (6,0) circle (2pt);
\draw[black,fill=black] (0,2) circle (2pt);
\draw[black,fill=black] (2,2) circle (2pt);
\draw[black,fill=black] (4,2) circle (2pt);
\draw[black,fill=black] (6,2) circle (2pt);
\draw[] (0,0) -- (2,0);
\draw[] (0,0) -- (2,2);
\draw[] (0,2) -- (2,0);
\draw[] (0,2) -- (2,2);
\draw[] (2,0) -- (4,0);
\draw[] (2,0) -- (4,2);
\draw[] (2,2) -- (4,0);
\draw[] (2,2) -- (4,2);
\draw[] (4,0) -- (6,0);
\draw[] (4,0) -- (6,2);
\draw[] (4,2) -- (6,0);
\draw[] (4,2) -- (6,2);
\node[left] at (0,0) {$1$};
\node[left] at (0,2) {$2$};
\node[right] at (6,0) {$3$};
\node[right] at (6,2) {$4$};
\node[below] at (2,0) {$1'$};
\node[above] at (2,2) {$2'$};
\node[below] at (4,0) {$1''$};
\node[above] at (4,2) {$2''$};
\end{tikzpicture}
\end{subfigure}
\caption{On the left, an embedding of the anomalous dimension diagram from figure~\ref{fig:horizontal loop} inside the four-point correlator $\<\f^2\f^2\f^2\f^2\>$. We imagine $\f^2(x_1) \f^2(x_2)$ creating the detector $\wL[\f^2] \wL[\f^2]$ via the mechanism in figure~\ref{fig:creatingwlphisq}. This detector is measured by particles that propagate from $x_3,x_4$, split at $x_{1}''$ and $x_{2}''$, and are finally detected at $x_1'$ and $x_2'$.
On the right, we re-draw the same diagram in a simpler way that makes clear that it is the cube of the kernel $S$, which is one of the terms inside $S^2G_\mathrm{tree}$  in equation~(\ref{eq:sequenceofdiagrams}).\label{fig:scubed}}
\end{figure}

Nicely, the term $S^2 G_\mathrm{tree}$ in (\ref{eq:sequenceofdiagrams}) can be interpreted as an embedding of figure~\ref{fig:horizontal loop} into the four-point function, see figure~\ref{fig:scubed}. Thus, we expect this term to reveal the anomalous spin of the horizontal trajectory. The full series (\ref{eq:sequenceofdiagrams}) represents the exponentiation of this anomalous spin.

Because $S$ is conformally-invariant, it can be diagonalized via its conformal partial wave decomposition, with eigenvalue $s(\De,J)$. The key observation is that the eigenvalue $s(\De,J)$ exhibits a pole at $J=-1$, as we show in appendix~\ref{app:diagonalizes}:
\be
\label{eq:seigenvalue}
s(\De,J) &\sim -\frac{\sin(\tfrac{\pi \De}{2})}{128\pi^2 (\De-2)\cos(\tfrac{\pi\De}{2})^2} \frac{1}{J+1}.
\ee
This matches the $1/(J+1)$ pole in~(\ref{eq:CtDirectInt}). However, we can now proceed further and use the eigenvalues of $S$ in the re-summed series (\ref{eq:sequenceofdiagrams}), giving
\be
\label{eq:shiftedpole}
C(\De,J) &\supset \frac{1}{1-\l^2 s(\De,J)} C_\mathrm{tree}(\De,J).
\ee
Plugging in the eigenvalue (\ref{eq:seigenvalue}) near $J=-1$, we find a pole in $C(\De,J)$ at the shifted location
\be
J = -1 +\frac{\lambda^2}{(4\pi)^4} \frac{2\pi^2\sin(\tfrac{\pi \De}{2})}{(2-\De) \cos(\tfrac{\pi\De}{2})^2},
\ee
in precise agreement with eqs.~\eqref{J connected} and \eqref{kappa text}!

Thus, we have found a match between a subset of diagrams contributing to the renormalization of $\cH$ and a subset of diagrams contributing to the pole near $J=-1$ in the four-point function $\<\f^2 \f^2 \f^2 \f^2\>$.
This gives evidence that near-horizontal Regge trajectories do contribute to correlators of this theory, at a subleading power. 
We emphasize that we included only a subset of diagrams --- those that were naturally related to the connected graph in figure~\ref{fig:horizontal loop}.
It would be interesting to include all diagrams and understand how the corrections exponentiate, which is challenging as discussed in subsection \ref{ssec: anomalous dimension?}.
We leave this for the future.

\section{Discussion}
\label{sec:discussion}
In this paper we have made some initial progress towards understanding the perturbative structure of the space of asymptotic detectors (equivalently, Regge trajectories, or non-local light-ray operators) in CFTs,
and especially in the Wilson-Fisher CFT. 
Detectors, roughly, are anything one can measure at infinity in a scattering experiment.
We have shown that the detector point of view on light-ray operators not only gives them a physical interpretation, but is also a convenient technical tool for perturbative computations.

Operators involving products of fields along a light ray have long been studied in quantum field theory.
The simplest, leading-twist operators, involve products of two fields time-integrated against some kernel.
Perhaps surprisingly, we find that such ``light-ray'' operators continuously connect with ones spread over a full null plane:
quantum mechanics can prevent localization to a light ray.

As our first technical result, in section~\ref{sec:pomeron} we explained the precise perturbative mechanism behind
the resolution of the $J=0$ singularity of the leading-twist anomalous dimension in the Wilson-Fisher CFT.
We showed that this singularity is due to mixing between the leading-twist trajectory and its shadow at their unique intersection point:
to renormalize light-ray operators near this intersection (including the Pomeron operator), novel $1/J$-type subtractions are needed in addition to the standard $1/\e$-type subtractions.
Accounting for this, the dilatation operator becomes a regular $2\x2$ matrix (see eq.~\eqref{eq:D}), whose eigenstates smoothly connect leading-twist operators to their shadows,
which are delocalized.

Our second technical result is the construction and renormalization of novel horizontal Regge trajectories in the Wilson-Fisher theory in section~\ref{sec:WF horizontal},
which are qualitatively similar to null Wilson lines in gauge theories.
This result connects to previous work of some of us on the light-ray OPE~\cite{Kologlu:2019bco,Kologlu:2019mfz,Chang:2020qpj},
where it was shown that some pairs of light-ray operators can be inserted on the same null plane without encountering singularities. 
Here we showed how to renormalize the products when they are singular, producing new well-defined operators.
In section~\ref{sec:WF horizontal} we studied the simplest example of such trajectories, built from $\wL[\f^2]\wL[\f^2]$, which in the free theory sits at fixed $J=-1$ in the Chew-Frautschi plot.
We showed, by matching with a computation based on the Lorentzian inversion formula, that the connected part of the dilatation operator precisely reproduces the contribution of a class of diagrams to the four-point function of $\phi^2$.
Interestingly, as we discuss in section~\ref{sec:Hconnected}, the connected contributions to the dilatation operator vanish for non-zero transverse spin.

It is important to note that horizontal trajectories are not always subleading in the Regge limit (as was the case for the trajectories in section~\ref{sec:WF horizontal}). Even in the Wilson-Fisher theory, the leading Regge intercept in other symmetry sectors can receive corrections from mixing with horizontal trajectories. For example, the leading-twist local $\Z_2$-odd operators have twist 3 in $d=4$, and thus their trajectory intersects with its shadow at $J=-1$, where $\Z_2$-odd horizontal trajectories also exist (e.g.\ $\wL[\f^2]\wL[\f^3]$). Parity-odd local operators have even higher twists, and thus horizontal trajectories should be the only contributions to the Regge intercept in parity-odd sectors.

Intersections between perturbative Regge trajectories are ubiquitous (there are infinitely many already in the Wilson-Fisher theory), and it would be very interesting to apply our techniques to other examples.
One intersection in the Wilson-Fisher theory which might be accessible is the one related to the $J=-1$ pole in the two-loop leading twist anomalous dimension, where the mixing presumably involves both the horizontal and the higher-twist trajectories
(see figure \ref{fig:perturbativelines}).
In gauge theories, there is a prominent intersection between leading twist trajectories that control deep inelastic scattering (``DGLAP evolution''), and near-horizontal ones that control
forward scattering (``BFKL evolution'').  Localized light-ray operators are supposed to become delocalized dipoles \cite{Balitsky:2013npa} as they evolve through this intersection.
Our results suggest it may be natural to unite these formalisms into a smooth matrix,
that would effectively remove the small-$x$ singularities of the DGLAP kernel as well as the collinear singularities of the BFKL kernel, and yet avoid double counting, at the cost of introducing an off-diagonal term that mixes them.
This would be similar, if possible, to how the matrix \eqref{eq:D} (or its characteristic equation \eqref{eq:WF leading trajectory}) smoothens out the leading intersection in the Wilson-Fisher theory.

More generally, the language of asymptotic detectors appears to provide a new and potentially useful point of view on splitting functions and other timelike quantities.
For instance, as discussed in section~\ref{sec:reciprocity}, it gives a simple explanation of the reciprocity between timelike and spacelike anomalous dimensions -- the two anomalous dimensions describe the same curve but in different coordinates.
The curves are the same simply because there is nothing special about infinity in a conformal theory, so measurements at infinity (``timelike'') are conformally equivalent to other measurements.
This is related to a timelike-spacelike correspondence which has sometimes been exploited in non-conformal theories.
For example, in minimal subtraction schemes, anomalous dimensions do not depend on the spacetime dimension $d=4-\e$ while rapidity evolution equations do; choosing the spacetime dimension so that the $\b$-function vanishes
then yields interesting relations (see \cite{Basso:2006nk,Vladimirov:2016dll,Dixon:2018qgp}).
A technical advantage of the detector frame is that it does not require a rapidity regulator since these divergences are regulated by $\e$.

Our results leave us with many open questions.
Firstly, it would be interesting to understand how to resum the contributions of horizontal trajectories in the Wilson-Fisher theory.
The technical challenge, discussed in section \ref{ssec: anomalous dimension?}, is that while each diagram can be renormalized straightforwardly and a dilatation operator can be calculated order-by-order,
it is not clear how to exponentiate it: new operators appear at each loop order.
In gauge theories, the analogous problem is solved perturbatively by the Reggeized gluon,
and one can wonder if a similar effective degree of freedom exists in the Wilson-Fisher theory.

Secondly, this leads to the question of studying the more general operators $\cH_{J_{L1},J_{L2}}$. As we showed in section~\ref{sec:WF horizontal}, their (connected) 2-loop divergences arise only for $J_{L1}+J_{L2}\simeq -2$ and are proportional to $\cH_{3-d,3-d}$. A preliminary analysis suggests that the situation at higher loop orders is more involved and additional operators appear in the divergences.  It is an important open problem to develop a systematic understanding of what this means for the set of the Regge trajectories in Wilson-Fisher theory, both at a fixed loop level and non-perturbatively. Similarly, one should consider the operators of the type $\wL[\f^n]\wL[\f^m]$ which should also correspond to Regge trajectories at $J=-1$, as well as operators involving more light-rays which would live at $J=-n$ with $n>1$. It would also be interesting to understand whether the horizontal trajectories survive in some form down to $\e=2$ and make sense in the 2d Ising CFT.

Thirdly, we have not discussed the higher-twist trajectories that appear as diagonal lines on the Chew-Frautschi plot. Beyond the leading one, these are infinitely-degenerate and their renormalization is not  well understood for non-integer $J$
(see figure 5 of \cite{Braun:1999te} for examples of distinct analytic continuations).  It is an important problem to develop at least a perturbative picture of these trajectories.

Our results in sections~\ref{sec:pomeron} and~\ref{sec:WF horizontal} suggest that the picture of the Regge trajectories in the Wilson-Fisher is relatively simple above and in the neighborhood of the leading twist trajectory for $J>-1$, but infinite degeneracies in the free theory obscure the other regions of the Chew-Frautschi plot.  Nonperturbatively, we expect a set of discrete trajectories that characterize the physical measurements that can be made on a null plane.
It is striking how little we understand about this plot even in such a well-studied theory.

\section*{Acknowledgements}	

We thank Cyuan-Han Chang, Lance Dixon, Nikolay Gromov, Juan Maldacena, Ian Moult, Andy Strominger, Sasha Zhiboedov, and especially Shota Komatsu for valuable discussions. We additionally thank the organizers and participants of the workshops ``Analytic approaches to the bootstrap" in the Azores in 2018, and Bootstrap 2019 at the Perimeter Institute, where this work was initiated. PK and DSD also thank the organizers and participants of the 2019 conference ``From Scattering Amplitudes to the Conformal Bootstrap" at the Aspen Center for Physics (supported by National Science Foundation grant PHY-1607611). DSD is supported by Simons Foundation grant 488657 (Simons Collaboration on the Nonperturbative Bootstrap) and a DOE Early Career Award under grant No.\ DE-SC0019085. PK was supported by DOE grant No.\ DE-SC0009988, the Adler Family Fund, as well as the Corning Glass Works Foundation Fellowship Fund at the Institute for Advanced Study.
Work of SCH is supported by the Simons Collaboration on the Nonperturbative Bootstrap, the
Canada Research Chair program and the Sloan Foundation.
The work of MK is supported by funding from
the European Research Council (ERC) under the European Union's Horizon 2020 research and innovation programme (grant agreement No 787185), and by a Yale Mossman Prize Fellowship in Physics. DM was supported by the DOE under grant No.\ DE-SC0011632 and by the NSF grant PHY-2014071.

\pagebreak
	
\appendix

\section{The tree-level twist-2 vertex}
\label{app:light transformed structures}

In this appendix, we derive formula (\ref{eq:treelevelvvertex}) for the tree-level vertex of $\f(\alpha;z)$, which is needed as an intermediate step in computing the tree-level vertex $V_{J_L}$. In position space we have,
\be
	\<0|\f(\a;z)\f(y)|0\>&=\lim_{L\to +\oo} L^{\De_\f}\frac{N_{\f}}{(x+Lz-y)^{2\De_\f}} 
	=\frac{N_{\f}}{(-\a-2y\cdot z+i\e)^{\De_\f}},\label{eq:positionvertex}
\ee
where we restored the $i\epsilon$ prescription and made the identification $\alpha=-2x\cdot z$. The standard normalization in perturbation theory is,
\be
N_{\f}=\frac{\Gamma(\Delta_{\f})}{\pi^{d/2}2^{d-2\Delta_{\f}}},
\ee
where $\De_\f=\frac{d-2}{2}$, which corresponds to the time-ordered propagator
\be
	\<\f(p)\f(0)\>=\frac{-i}{p^2-i\e}.
\ee

We must now Fourier transform \eqref{eq:positionvertex} in $y$ to obtain $\<0|\f(\a;z)\f(p)|0\>$. Using Lorentz symmetry, we can take $z$ to be along $+$ direction, i.e.\ $z^+=1$, and other components to vanish. We then find
\be
\int d^{d}ye^{ip\cdot y}\frac{1}{(-\a-2y\cdot z+i\e)^{\De_\f}}&=\frac{1}{2}\int d^{d-2}\vec y dy^+dy^- e^{-i\half p^+y^--i\half p^-y^++i\vec y\vec p}\frac{1}{(-\a+y^-+i\e)^{\De_\f}}		
\nn \\ &=  \frac{2^{1-\De_\f}e^{-\frac{i\pi\De_\f}{2} }}{\Gamma(\De_\f)}e^{-i\half \a p^+}(2\pi)^{d}\de(p^-)\de^{d-2}(\vec p)\theta(p^+)(p^+)^{\De_\f-1}.   
\ee
We can compare the final answer with the ansatz
\be
\int\limits_{0}^{\infty} d\b\, \de^d(p-\b z)\b^a e^{ib\b}&=2 \int\limits_{0}^{\infty} d\b\, \de^{d-2}(\vec p)\de(p^-)\de(p^+-\b)\b^a e^{ib\b}\nn\\
&=2 \de^{d-2}(\vec p)\de(p^-)(p^+)^a e^{ibp^+},
\ee
From this we conclude that for general $z$,
\be
\<0|\f(\a;z)\f(p)|0\> &= N_{\f}\frac{2^{-\De_\f}e^{-\frac{i\pi\De_\f}{2} }}{\Gamma(\De_\f)}(2\pi)^{d}\int\limits_0^\oo d\b\, \de^d(p-\b z)\b^{\De_\f-1} e^{-i\half\a\b}\nn\\
&=\frac{e^{-\frac{i\pi\De_\f}{2}}}{2(2\pi)^{d/2}}\int\limits_0^\oo d\b\, (2\pi)^{d}\de^d(p-\b z)\b^{\De_\f-1} e^{-i\half\a\b}.
\ee

\section{Poles of distributions}
\label{app:poles dist}

In this appendix, we discuss the analytic continuation of distributions of a special form, following~\cite{Bernstein_Gelfand,doi:10.1002/cpa.3160230202}. The key example is
\be
f_{a,b,c}(x,y)=x^ay^b(y-x^2)^c\theta(x)\theta(y-x^2), \label{eq:f dist}
\ee
analytically continued in $a,b,c$, which appears when renormalizing twist-two operators. The methods discussed here apply more generally to distributions of the form $\prod_i f_i(x)^{a_i}$, although we will not discuss the generalization explicitly.

Let us start by considering a simpler example, the distribution $f_a(x)=x^a\theta(x)$ on $\R$~\cite{Gelfand:105396}.  For $\Re a>-1$ this is a locally-integrable function, and thus defines a distribution by
\be
	f_a(\f)=\int_{-\oo}^{+\oo}dx f_a(x)\f(x)=\int_{0}^{+\oo}dx x^a\f(x)\qquad(\Re a>-1).
\ee
In fact, the number $f_a(\f)$ depends on $a$ holomorphically, i.e.\ $a\mapsto f_a(\f)$ defines a holomorphic function for $\Re a>-1$. We can then ask if this holomorphic function can be analytically continued to other values of $a$. It is well-known~\cite{Gelfand:105396} that the answer is yes, and it can be continued to all of $\C\setminus\{-1,-2,\cdots\}$. Furthermore, for every $a$ in this set, the analytic continuation $f_a(\f)$ depends on the test function $\f$ in such a way that it defines a distribution. Therefore, we can speak of the analytic continuation of $f_a$ to $a\in \C\setminus\{-1,-2,\cdots\}$ without referring to $\f$. From now on, we will write simply $x^a\theta(x)$ instead of $f_a$. Finally, and crucially for us, at $a\in \{-1,-2,\cdots\}$ $x^a\theta(x)$ has simple poles with residues proportional to derivatives of a delta-function. Specifically,
\be
x^a\theta(x)=\frac{(-1)^{n-1}\de^{(n-1)}(x)}{(n-1)!}\frac{1}{a+n}+\cdots.
\ee
There are many ways to see this. For example one can relate $x^a\theta(x)$ to $(x\pm i0)^{a}$ which are well-defined distributions for all $a\in \C$ and depend on $a$ holomophically (this follows because they are boundary values of holomorphic functions).

Note that the step function $\theta(x)$ in $x^a\theta(x)$ is somewhat of a red herring -- we use it to define the simplest distribution that we can. Distributions supported for all $x$ can be defined by linear combinations. For example, we can consider
\be
	x^a\theta(x)-(-x)^a\theta(-x).
\ee
The pole at $a=-1$ cancels in this combination, and setting $a=-1$ we obtain the principal value distribution $\mathrm{p.v.}\frac{1}{x}$.

The above gives a complete description of the properties of $x^a\theta(x)$. We can now use it as a building block for more complicated distributions, e.g.
\be
	x^a y^b\theta(x)\theta(y).
\ee
Here we are multiplying two distributions together. In general, this is not allowed, but in this case they depend on two different variables, so this is fine and we simply get a distribution on $\R^2$ that has poles whenever $a$ or $b$ is a negative integer. A more general way of phrasing this is that whenever the singular loci of the two distributions (in this example, the lines $x=0$ and $y=0$) intersect (i.e.\ at $x=y=0$), they intersect transversely, and we can chose the local coordinates appropriately to justify the multiplication of the distributions (here $x,y$ already work).

This way of thinking allows us to consider, for example, the distribution
\be
	x^a y^b (1-x-y)^c\theta(x)\theta(y)\theta(1-x-y).
\ee
In this case we can multiply the distributions because when, say,\ $x=0$ intersects $1-x-y=0$ at $(x,y)=(0,1)$, we can use coordinates $s=x,t=1-x-y$ to justify multiplying $s^a\theta(s)$ by  $t^c\theta(t)$ (while $(1-s-t)^b\theta(1-s-t)$ is just a smooth function near this point). Therefore, we obtain a distribution with poles whenever one of $a,b,c$ is a negative integer, and we can easily compute the Laurent series near any pole.
 
We now want to repeat this analysis for the distribution $f_{a,b,c}(x,y)$ defined by~\eqref{eq:f dist}. Unfortunately, the above discussion isn't enough here. Indeed, in this case we have a product of three distributions which are singular along $x=0$, $y=0$, and $y=x^2$. All three curves intersect at $(x,y)=0$, and so this is not a transverse intersection. This means that there is no choice of local coordinates that could justify the multiplication of the three distributions. One could guess that perhaps we should treat $y-x^2\approx y$ near this point, but as we will see this leads to wrong results.

To bypass this problem one can follow the logic described in~\cite{Bernstein_Gelfand,doi:10.1002/cpa.3160230202}. The idea is to find a manifold $R$ and a smooth map $\tau:R\to \R^2$ such that the pullback of $f_{a,b,c}(x,y)dxdy$ along $\tau$ is better behaved than $f_{a,b,c}(x,y)dxdy$ itself. Concretely, we take $R=\R^2$ and
\be
	\tau(u,v)=(u,vu^2).
\ee
First of all, as functions,
\be\label{eq:xytouv}
	f_{a,b,c}(x,y)dxdy=u^{a+2b+2c+2}v^b(v-1)^c\theta(u)\theta(v-1)dudv.
\ee
This means that at least for $\Re a,\Re b,\Re c>0$
\be\label{eq:correspondence}
	\int dx dy f_{a,b,c}(x,y) \phi(x,y)=\int_R du dv u^{a+2b+2c+2}v^b(v-1)^c\theta(u)\theta(v-1)\phi(u,vu^2).
\ee
Note that as long  as $\phi(x,y)$ is a smooth function, so is $\phi(u,vu^2)$. Therefore, if we manage to analytically continue the right-hand side of~\eqref{eq:xytouv} in $a,b,c$ as a distribution, we will also find an analytic continuation of $f_{a,b,c}(x,y)$. But we already know how to perform the analytic continuation of~\eqref{eq:xytouv} since the singularities $v=1$ and $u=0$ intersect transversely ($v=0$ is not really a singularity since it is inaccessible due to $\theta(v-1)$). Morally speaking, what we have constructed is a resolution of the singularity at $x=y=0$, which turned a non-transverse intersection into several transverse ones.

We then conclude immediately that $f_{a,b,c}$ can be analytically continued as a distribution to all complex values $a,b,c$ except those where
\be
a+2+2b+2c=-n\quad\text{or}\quad c=-m, \qquad n,m\in\mathbb{Z}_{>0} \label{eq:abcpoles},
\ee
where we have simple poles.\footnote{Note that merely approximating $y-x^2\approx y$ would yield poles at $a=-m$ and $b+c=-n$, which is a completely different (and wrong) set of poles.} We will be interested in two poles in this sequence, the one at 
\be
	a+2+2b+2c=-1,
\ee
and the one at 
\be
	c=-1.
\ee

For the first pole, we have
\be
	u^{a+2b+2c+2}v^b(v-1)^c\theta(u)\theta(v-1)=\frac{\de(u)v^b (v-1)^{c} \theta(v-1)}{a+2b+2c+3}+\cdots
\ee
Using this in~\eqref{eq:correspondence} we find
\be
	\int dx dy f_{a,b,c}(x,y) \phi(x,y)&=\frac{\phi(0,0)}{a+2b+2c+3}\int_1^\oo dv v^b(v-1)^c +\cdots\nn\\
	&=\frac{\G(-1-b-c)\G(1+c)}{\G(-b)}\frac{\phi(0,0)}{a+2b+2c+3}+\cdots.
\ee
This implies, near this pole,
\be
	f_{a,b,c}(x,y)=\frac{\G(-1-b-c)\G(1+c)}{\G(-b)}\frac{\de(x)\de(y)}{a+2b+2c+3}+\cdots.\label{eq:pole1}
\ee
A similar procedure for the second pole yields
\be
	f_{a,b,c}(x,y)=\frac{x^{a+2b}\de(y-x^2)\theta(x)}{c+1}+\cdots.\label{eq:pole2}
\ee
We can now apply these results to the two-loop calculations from the main text.

There, we found the following density (see~\eqref{eq:2looptwist2}), 
\be
d^{d}p(-{2z\cdot p})^{J_L}(-p^2)^{\frac{d-4}{2}-J_L-2}\theta(-p^2).
\ee 
To find the map to $f_{a,b,c}(x,y)$ it is convenient to use lightcone coordinates and take $z=e_+$,
\be
d^{d}p(-{2z\cdot p})^{J_L}(-p^2)^{\frac{d-4}{2}-J_L-2}\bigg|_{z=e_{+}}=\frac{1}{2}dp^+dp^-d^{d-2}\vec{p}(p^-)^{J_L}(p^+p^--\vec{p}^{2})^{\frac{d-4}{2}-J_L-2}.
\ee
If we define $x=|\vec{p}|$ and $y=p^+p^-$ then we find,
\be
&d^{d}p(-{2z\cdot p})^{J_L}(-p^2)^{\frac{d-4}{2}-J_L-2}\theta(-p^2)
\nn \\ &=\half dx dy dp^+ d\O_{d-3} (p^+)^{-1 - J_{L}}y^{J_{L}}x^{d-3}(y-x^2)^{-J_{L}+\frac{d-4}{2}-2}\theta(y-x^2)\theta(x).
\ee
Comparing to the function $f_{a,b,c}(x,y)$ in \eqref{eq:f dist} we see,
\be
a=d-3, \quad b=J_{L}, \quad c=-J_{L}+\frac{d-4}{2}-2~. \label{eq:abc to dJL}
\ee
Setting $d=4-\epsilon$ and using \eqref{eq:abcpoles} we see there are poles when,
\be
-1-2\epsilon=-n,
\\
-2-J_L-\epsilon/2=-m,
\ee
for positive integer $n$ and $m$. We see that for generic $J_L$ there is a single pole when $\epsilon=0$ coming from $n=-1$.  This is the first pole that we analyzed above. Using~\eqref{eq:pole1} we find
\be
&y^{J_{L}}x^{d-3}(y-x^2)^{-J_{L}+\frac{d-4}{2}-2}\theta(y-x^2)\theta(x)
=	\frac{1}{1+J_L}\frac{\de(x)\de(y)}{4\e}+\cdots.
\ee
Taking into account $x=|\vec p|$ and thus
\be
d\Omega_{d-3}dx \delta(x)=\vol S^{d-3}d^{d-2}\vec{p}\delta^{d-2}(\vec{p}),
\ee
for generic $J_L$ we find,
\be
&d^{d}p(-{2z\cdot p})^{J_L}(-p^2)^{\frac{d-4}{2}-J_L-2}\theta(-p^2)\nn\\
&=\frac{1}{4\e(1+J_L)} (p^+)^{-1-J_L}d\Omega_{d-3}\delta(x)\delta(y)+\cdots
\nn \\ &=\frac{\vol S^{d-3}}{4\e(1+J_L)}(p^+)^{-1-J_L}d^{d-2}\vec{p}dp^{-}dp^{+}\delta^{d-2}(\vec{p})\delta(p^-) + \cdots
\nn \\ & =\frac{\vol S^{d-3}}{4\e(1+J_L)}d^{d}p\int d\beta \beta^{-1-J_L}\delta^d(p-\beta z)\bigg|_{z=e_+} + \cdots
\ee
Generalizing in $z$, which is possible by Lorentz invariance, we get
\be
(-2z\cdot p)^{J_L}(-p^{2})^{\frac{d-4}{2}-J_L-2}= \frac{\vol S^{d-3}}{4\e(1+J_L)}\int d\beta \beta^{-1-J_L}\delta^d(p-\beta z)+\cdots.
\ee
At $\epsilon=0$ we have $\vol S^{d-3}=2\pi$, which proves \eqref{eq:pole dist tw2}.

The result~\eqref{eq:pole2} can be used in a similar way to justify~\eqref{eq:pole2maintext}.

\section{Diagonalizing the kernel $K_\a$}
\label{sec:diagonalizingkalpha}

In this section, we diagonalize the kernel 
\be
K_\a(z_1,z_2;z_3,z_4)
&= 
\frac{\pi}{\sin\pi \a} \frac{
\p{\frac{z_{13}}{z_{23}}}^{\a}-
\p{\frac{z_{14}}{z_{24}}}^{\a}
}{z_{24}z_{13}-z_{14}z_{23}}
= \frac{1}{z_{12} z_{34}} \p{\frac{z_{14}}{z_{24}}}^\a \frac{\pi}{\sin\pi \a} \frac{u(v^{-\a}-1)}{1-v},
\ee
where $z_{ij}=-2z_i{\cdot}z_j$ and $u=\frac{z_{12}z_{34}}{z_{13}z_{24}}$, $v=\frac{z_{23}z_{14}}{z_{13}z_{24}}$ are celestial sphere cross-ratios. We perform the computation for general $\a$, though in section~\ref{sec:AppInCorrFuncs},
we only need the result for $\a=0$. The general result may be useful when studying the full space of detectors $\cH_{J_{L1},J_{L2}}$ including connected and disconnected loop corrections.

The kernel $K_\a$ is defined in any spacetime dimension $d$. However, for our applications it will suffice to determine its eigenvalues in $d=4$, which we assume henceforth. We can diagonalize it by decomposing it into projectors onto irreducible representations of the Lorentz group. Thinking of the Lorentz group as the conformal group of the celestial sphere $S^2$, this is the same as decomposing the kernel into 2-dimensional conformal partial waves:\footnote{We follow the notation and conventions of \cite{Karateev:2018oml}.}
\be
K_\a(z_1,z_2;z_3,z_4) &= \sum_{j=0}^\oo \int\limits_1^{1+i\oo} \frac{d\de}{2\pi i} A_\a(\de,j) \Psi_{\de,j}(z_i) \nn\\
\Psi_{\de,j}(z_i) &= \int D^2 z_5 \<\cP_{1-\a}(z_1) \cP_{1+\a}(z_2) \cP_{\de,j}(z_5)\>\<\cP_{\tl \de,j}(z_5) \cP_1(z_3) \cP_1(z_4)\>.
\ee
Here, $\<\cP_{\de_1}(z_1)\cdots\>$ denote standard three-point structures of fictitious operators $\cP_{\de_i}(z_i)$ in the embedding space $z_i\in \R^{3,1}$, see eq.~\eqref{standard 3pt}.
The shadow dimension $\tl \de$ is given by $\tl \de=2-\de$, and the transverse spin indices of $\cP_{\de,j}(z_5)$ and $\cP_{\tl \de,j}(z_5)$ are implicitly contracted. For fictitious embedding-space scalars, we use the shorthand $\cP_{\de,0}=\cP_{\de}$.

Remarkably, it turns out that $K_\a$ can be completely decomposed into partial waves with vanishing transverse spin ($j=0$). These are given by
\be
\Psi_{\de,0}(z_i) &= \frac{1}{z_{12}z_{34}}\p{\frac{z_{14}}{z_{24}}}^{\a}\p{S^{34}_{\tl \de,0}G_{\de,0}(z,\bar z) + S^{12}_{\de,0} G_{\tl \de,0}(z,\bar z)},\nn\\
S^{12}_{\de,0} &=\frac{\pi\G(\de-1)\G(\frac{\tl\de}{2}-\a)\G(\frac{\tl\de}{2}+\a)}{\G(2-\de)\G(\frac{\de}{2}-\a)\G(\frac{\de}{2}+\a)},\\
S^{34}_{\de,0} &= \frac{\pi \G(1-\frac{\de}{2})^2 \G(\de-1)}{\G(2-\de)\G(\frac{\de}{2})^2},
\ee
where $u=z\bar z$, $v=(1-z)(1-\bar z)$, and $G_{\de,0}(z,\bar z)$ are 2d scalar conformal blocks:
\be
G_{\de,0}(z,\bar z) &= k_{\de}(z) k_{\de}(\bar z)\,, \qquad\qquad
k_{\de}(z) = z^{\de/2} {}_2F_1\p{\tfrac{\de}{2}+\a,\tfrac{\de}{2},\de,z}.
\ee
Specifically, we find
\be
K_\a(z_1,z_2;z_3,z_4) &= \int\limits_{1}^{1+i\oo} \frac{d\de}{2\pi i} A_\a(\de) \Psi_{\de,0}(z_i),\label{eq:kerCPWPt1}
\nn\\
A_\a(\de) &\equiv \frac{\de-1}{\pi}\cos(\tfrac{\pi \de}{2})\G(1-\tfrac{\de}{2})^2\G(\tfrac{\de}{2}+\a)\G(\tfrac{\de}{2}-\a).
\ee
Note the shadow symmetry $A_J(\de)S^{34}_{\tl\de,0}=A_J(\tl\de)S^{12}_{\tl\de,0}$. To verify (\ref{eq:kerCPWPt1}), we can plug in the expression for $\Psi_{\de,0}(z_i)$ and use shadow symmetry to remove the $G_{\tl \de,0}$ term and extend the integral from $1-i\oo$ to $1+i\oo$. Closing the $\de$-contour to the right, we pick up residues of poles in $\de$, giving
\be
\label{eq:kernelblock}
\frac{\pi}{\sin \pi \a}\frac{u (v^{-\a}-1)}{1-v}&=\sum_{n=1}^{\oo}\frac{(-1)^{n+1}\G(n)^2\G(n+\a)\G(n-\a)}{\G(2n)\G(2n-1)}G_{2n,0}(z,\bar z).
\ee
We have verified (\ref{eq:kernelblock}) by expanding to high orders in $z,\bar z$.

The kernel $K_\a$ is diagonalized when acting on celestial three-point functions. Since the decomposition (\ref{eq:kernelblock}) contains only scalar blocks, the only nonzero eigenvalues come from acting on three-point functions with vanishing transverse spin:
\be
\int D^2 z_1 D^2 z_2 \<\cP_{-J_L}(z_6)\cP_{1+\a}(z_1)\cP_{1-\a}(z_2)\> K_\a(z_1,z_2;z_3,z_4) &= \pi^2\kappa_\a(J_L)\<\cP_{-J_L}(z_6)\cP_1(z_3)\cP_1(z_4)\>.
\ee
We can obtain the eigenvalue $\kappa_\a(J_L)$ from the conformal ``bubble" integral \cite{Dobrev:1977qv,Karateev:2018oml}:
\be
&\int D^2 z_1 D^2 z_2 \<\cP_{1-is}(z_6) \cP_{1+\a}(z_1)\cP_{1-\a}(z_4)\> \<\cP_{1-\a}(z_1)\cP_{1+\a}(z_2) \cP_{1+is'}(z_5)\> 
\nn\\
&= -\frac{2\pi^3}{(\de-1)^2}\,2\pi \de(s-s') \de(z_5,z_6),
\ee
where $\de=1+is=-J_L$, which together with (\ref{eq:kerCPWPt1}) gives
\be
\kappa_\a(J_L)&= -\frac{2\pi}{(J_L+1)^2} A_\a(2+J_L) = \frac{2}{J_L+1} \cos(\tfrac{\pi J_L}{2}) \G(-\tfrac{J_L}{2})^2 \G(\tfrac{J_L+2}{2}-\a) \G(\tfrac{J_L+2}{2}+\a).
\ee

\section{Coefficients $s_{ij}$}
\label{app:Scoeffs}

The coefficients $s_{ij}$ are given by
\be
	s_{11}&=4 \ptl_{\epsilon}\hat S(\de_1,\de_2,[\de_3]),\\
	s_{21}&=\p{4 \ptl_3 \ptl_{\epsilon }-4 \ptl_{\epsilon }^2-4 \ptl_2 \ptl_{\epsilon }-2 \ptl_3^2}\hat S(\de_1,\de_2,[\de_3]),\\
	s_{31}&=\p{-16 \ptl_2^2 \ptl_{\epsilon }+16 \ptl_2 \ptl_{\epsilon }^2-16 \ptl_3 \ptl_2 \ptl_{\epsilon }+8 \ptl_3 \ptl_2^2+8 \ptl_3^2 \ptl_2}\hat S(\de_1,\de_2,[\de_3]),\\
	s_{32}&=\p{8 \ptl_3 \ptl_{\epsilon }-8 \ptl_{\epsilon }^2+24 \ptl_2 \ptl_{\epsilon }-4 \ptl_3^2-8 \ptl_2 \ptl_3}\hat S(\de_1,\de_2,[\de_3]),\\
	s_{33}&=\p{4 \ptl_2-4 \ptl_{\epsilon }}\hat S(\de_1,\de_2,[\de_3]).
\ee
Here $\ptl_\e$ denotes the derivative with respect to $\e$ in $d=4-\e$, while $\ptl_i$ is the derivative with respect to $\de_i$. The above expressions are to be evaluated for
\be
	\e=0,\quad \de_1=-J_L,\quad \de_2=\de_3=1.
\ee

\section{Diagonalizing the kernel $S$}
\label{app:diagonalizes}

In this appendix, we diagonalize the kernel $S(x_1,x_2,x_3,x_4)$ defined in (\ref{eq:thekernels}), to leading order near $J=-1$.
We begin with its partial wave decomposition
\be
\label{eq:partialwavedecomp}
S(x_1,x_2,x_3,x_4) &= \int_{\frac d 2}^{\frac d 2 + i\oo} \frac{d\De}{2\pi i} \frac{C_S(\De,J)}{K_{\tl \De,J}} \Psi_{\De,J}(x_1,x_2,x_3,x_4),
\ee
where $d=4$. We follow the conventions of \cite{Simmons-Duffin:2017nub,Karateev:2018oml}.
The coefficient function $C_S(\De,J)$ was computed via the Lorentzian inversion formula in \cite{Kologlu:2019mfz}. We will only need its value at $J=-1$:
\be
C_S(\De,-1) &= \frac{N_\f^4}{2} \frac{\G(\tfrac{3-\De}{2})\G(\tfrac{\De-1}{2})^5}{\G(\De-2)^2}.
\ee
To relate $C_S(\De,J)$ to the eigenvalues of $S$, we can use the ``bubble formula" \cite{Dobrev:1977qv,Karateev:2018oml}
\be
\label{eq:bubbleformula}
\int d^{d}x_{5}d^{d}x_{6}\Psi_{\De,J}(x_1,x_2,x_5,x_6) \Psi_{\De',J'}(x_5,x_6,x_3,x_4)  &= \cB_{\De,J} 2\pi \de(s-s') \de_{JJ'} \Psi_{\De,J}(x_1,x_2,x_3,x_4),
\ee
where $\De=\frac d 2 + i s$, $\De'=\frac d 2 + is'$, and we restrict to $s,s'>0$. The left-hand side of (\ref{eq:bubbleformula}) is a composition of the kernels $\Psi_{\De,J}$ and $\Psi_{\De',J'}$ acting on pairs of points. The bubble coefficient $\cB_{\De,J}$ is given by 
\be
\cB_{\De,J} &= \frac{\pi^6}{2^{J-1} (J+1) (\De-J-3)(\De+J-1) (\De-2)^2} \nn\\
&\sim \frac{4\pi^6}{(\De-2)^4} \frac{1}{J+1},
\ee
where we have indicated its pole near $J=-1$.
Combining (\ref{eq:partialwavedecomp}) and (\ref{eq:bubbleformula}), we find that the eigenvalues of $S$ are
\be
s(\De,J) &= \frac{C_S(\De,J)}{K_{\tl \De,J}} \cB_{\De,J} \sim -\frac{\sin(\tfrac{\pi \De}{2})}{128\pi^2 (\De-2)\cos(\tfrac{\pi\De}{2})^2} \frac{1}{J+1},
\ee
where again we focus on the pole at $J=-1$.

\bibliographystyle{JHEP}
\bibliography{refs}

\providecommand{\href}[2]{#2}\begingroup\raggedright\begin{thebibliography}{10}

\bibitem{Kravchuk:2018htv}
P.~Kravchuk and D.~Simmons-Duffin, \emph{{Light-ray operators in conformal
  field theory}}, \href{http://dx.doi.org/10.1007/JHEP11(2018)102}{\emph{JHEP}
  {\bf 11} (2018) 102}, [\href{https://arxiv.org/abs/1805.00098}{{\tt
  1805.00098}}].

\bibitem{Hofman:2008ar}
D.~M. Hofman and J.~Maldacena, \emph{{Conformal collider physics: Energy and
  charge correlations}},
  \href{http://dx.doi.org/10.1088/1126-6708/2008/05/012}{\emph{JHEP} {\bf 05}
  (2008) 012}, [\href{https://arxiv.org/abs/0803.1467}{{\tt 0803.1467}}].

\bibitem{Kologlu:2019mfz}
M.~Kologlu, P.~Kravchuk, D.~Simmons-Duffin and A.~Zhiboedov, \emph{{The
  light-ray OPE and conformal colliders}},
  \href{https://arxiv.org/abs/1905.01311}{{\tt 1905.01311}}.

\bibitem{Chang:2020qpj}
C.-H. Chang, M.~Kologlu, P.~Kravchuk, D.~Simmons-Duffin and A.~Zhiboedov,
  \emph{{Transverse spin in the light-ray OPE}},
  \href{http://dx.doi.org/10.1007/JHEP05(2022)059}{\emph{JHEP} {\bf 05} (2022)
  059}, [\href{https://arxiv.org/abs/2010.04726}{{\tt 2010.04726}}].

\bibitem{Cordova:2018ygx}
C.~C\'{o}rdova and S.-H. Shao, \emph{{Light-ray Operators and the BMS
  Algebra}}, \href{http://dx.doi.org/10.1103/PhysRevD.98.125015}{\emph{Phys.
  Rev.} {\bf D98} (2018) 125015}, [\href{https://arxiv.org/abs/1810.05706}{{\tt
  1810.05706}}].

\bibitem{Christ:1972ms}
N.~H. Christ, B.~Hasslacher and A.~H. Mueller, \emph{{Light cone behavior of
  perturbation theory}},
  \href{http://dx.doi.org/10.1103/PhysRevD.6.3543}{\emph{Phys. Rev. D} {\bf 6}
  (1972) 3543}.

\bibitem{Gross:1973ju}
D.~J. Gross and F.~Wilczek, \emph{{Asymptotically Free Gauge Theories - I}},
  \href{http://dx.doi.org/10.1103/PhysRevD.8.3633}{\emph{Phys. Rev. D} {\bf 8}
  (1973) 3633--3652}.

\bibitem{Georgi:1974wnj}
H.~Georgi and H.~D. Politzer, \emph{{Electroproduction scaling in an
  asymptotically free theory of strong interactions}},
  \href{http://dx.doi.org/10.1103/PhysRevD.9.416}{\emph{Phys. Rev. D} {\bf 9}
  (1974) 416--420}.

\bibitem{Collins:1981uw}
J.~C. Collins and D.~E. Soper, \emph{{Parton Distribution and Decay
  Functions}},
  \href{http://dx.doi.org/10.1016/0550-3213(82)90021-9}{\emph{Nucl. Phys. B}
  {\bf 194} (1982) 445--492}.

\bibitem{Balitsky:1987bk}
I.~I. Balitsky and V.~M. Braun, \emph{{Evolution Equations for QCD String
  Operators}},
  \href{http://dx.doi.org/10.1016/0550-3213(89)90168-5}{\emph{Nucl. Phys. B}
  {\bf 311} (1989) 541--584}.

\bibitem{Kuraev:1977fs}
E.~A. Kuraev, L.~N. Lipatov and V.~S. Fadin, \emph{{The Pomeranchuk Singularity
  in Nonabelian Gauge Theories}}, {\emph{Sov. Phys. JETP} {\bf 45} (1977)
  199--204}.

\bibitem{Balitsky:1978ic}
I.~I. Balitsky and L.~N. Lipatov, \emph{{The Pomeranchuk Singularity in Quantum
  Chromodynamics}}, {\emph{Sov. J. Nucl. Phys.} {\bf 28} (1978) 822--829}.

\bibitem{Mueller:1994jq}
A.~H. Mueller and B.~Patel, \emph{{Single and double BFKL pomeron exchange and
  a dipole picture of high-energy hard processes}},
  \href{http://dx.doi.org/10.1016/0550-3213(94)90284-4}{\emph{Nucl. Phys. B}
  {\bf 425} (1994) 471--488}, [\href{https://arxiv.org/abs/hep-ph/9403256}{{\tt
  hep-ph/9403256}}].

\bibitem{Balitsky:1995ub}
I.~Balitsky, \emph{{Operator expansion for high-energy scattering}},
  \href{http://dx.doi.org/10.1016/0550-3213(95)00638-9}{\emph{Nucl. Phys. B}
  {\bf 463} (1996) 99--160}, [\href{https://arxiv.org/abs/hep-ph/9509348}{{\tt
  hep-ph/9509348}}].

\bibitem{Caron-Huot:2013fea}
S.~Caron-Huot, \emph{{When does the gluon reggeize?}},
  \href{http://dx.doi.org/10.1007/JHEP05(2015)093}{\emph{JHEP} {\bf 05} (2015)
  093}, [\href{https://arxiv.org/abs/1309.6521}{{\tt 1309.6521}}].

\bibitem{Hatta:2008st}
Y.~Hatta, \emph{{Relating e+ e- annihilation to high energy scattering at weak
  and strong coupling}},
  \href{http://dx.doi.org/10.1088/1126-6708/2008/11/057}{\emph{JHEP} {\bf 11}
  (2008) 057}, [\href{https://arxiv.org/abs/0810.0889}{{\tt 0810.0889}}].

\bibitem{Caron-Huot:2015bja}
S.~Caron-Huot, \emph{{Resummation of non-global logarithms and the BFKL
  equation}}, \href{http://dx.doi.org/10.1007/JHEP03(2018)036}{\emph{JHEP} {\bf
  03} (2018) 036}, [\href{https://arxiv.org/abs/1501.03754}{{\tt 1501.03754}}].

\bibitem{Vladimirov:2016dll}
A.~A. Vladimirov, \emph{{Correspondence between Soft and Rapidity Anomalous
  Dimensions}},
  \href{http://dx.doi.org/10.1103/PhysRevLett.118.062001}{\emph{Phys. Rev.
  Lett.} {\bf 118} (2017) 062001},
  [\href{https://arxiv.org/abs/1610.05791}{{\tt 1610.05791}}].

\bibitem{Mueller:2018llt}
A.~H. Mueller, \emph{{Conformal spacelike-timelike correspondence in QCD}},
  \href{http://dx.doi.org/10.1007/JHEP08(2018)139}{\emph{JHEP} {\bf 08} (2018)
  139}, [\href{https://arxiv.org/abs/1804.07249}{{\tt 1804.07249}}].

\bibitem{Dixon:2019uzg}
L.~J. Dixon, I.~Moult and H.~X. Zhu, \emph{{Collinear limit of the
  energy-energy correlator}},
  \href{http://dx.doi.org/10.1103/PhysRevD.100.014009}{\emph{Phys. Rev. D} {\bf
  100} (2019) 014009}, [\href{https://arxiv.org/abs/1905.01310}{{\tt
  1905.01310}}].

\bibitem{Brower:2006ea}
R.~C. Brower, J.~Polchinski, M.~J. Strassler and C.-I. Tan, \emph{{The Pomeron
  and gauge/string duality}},
  \href{http://dx.doi.org/10.1088/1126-6708/2007/12/005}{\emph{JHEP} {\bf 12}
  (2007) 005}, [\href{https://arxiv.org/abs/hep-th/0603115}{{\tt
  hep-th/0603115}}].

\bibitem{Costa:2012cb}
M.~S. Costa, V.~Goncalves and J.~Penedones, \emph{{Conformal Regge theory}},
  \href{http://dx.doi.org/10.1007/JHEP12(2012)091}{\emph{JHEP} {\bf 1212}
  (2012) 091}, [\href{https://arxiv.org/abs/1209.4355}{{\tt 1209.4355}}].

\bibitem{Costa:2011mg}
M.~S. Costa, J.~Penedones, D.~Poland and S.~Rychkov, \emph{{Spinning Conformal
  Correlators}}, \href{http://dx.doi.org/10.1007/JHEP11(2011)071}{\emph{JHEP}
  {\bf 11} (2011) 071}, [\href{https://arxiv.org/abs/1107.3554}{{\tt
  1107.3554}}].

\bibitem{Luscher:1974ez}
M.~Luscher and G.~Mack, \emph{{Global Conformal Invariance in Quantum Field
  Theory}}, \href{http://dx.doi.org/10.1007/BF01608988}{\emph{Commun. Math.
  Phys.} {\bf 41} (1975) 203--234}.

\bibitem{Gonzo:2020xza}
R.~Gonzo and A.~Pokraka, \emph{{Light-ray operators, detectors and
  gravitational event shapes}},
  \href{http://dx.doi.org/10.1007/JHEP05(2021)015}{\emph{JHEP} {\bf 05} (2021)
  015}, [\href{https://arxiv.org/abs/2012.01406}{{\tt 2012.01406}}].

\bibitem{Caron-Huot:2017vep}
S.~Caron-Huot, \emph{{Analyticity in Spin in Conformal Theories}},
  \href{http://dx.doi.org/10.1007/JHEP09(2017)078}{\emph{JHEP} {\bf 09} (2017)
  078}, [\href{https://arxiv.org/abs/1703.00278}{{\tt 1703.00278}}].

\bibitem{Lee:2022ige}
K.~Lee, B.~Me\c{c}aj and I.~Moult, \emph{{Conformal Colliders Meet the LHC}},
  \href{https://arxiv.org/abs/2205.03414}{{\tt 2205.03414}}.

\bibitem{Korchemsky:2021okt}
G.~Korchemsky, E.~Sokatchev and A.~Zhiboedov, \emph{{Generalizing event shapes:
  In search of lost collider time}},
  \href{https://arxiv.org/abs/2106.14899}{{\tt 2106.14899}}.

\bibitem{Mueller:1981ex}
A.~H. Mueller, \emph{{On the Multiplicity of Hadrons in QCD Jets}},
  \href{http://dx.doi.org/10.1016/0370-2693(81)90581-5}{\emph{Phys. Lett. B}
  {\bf 104} (1981) 161--164}.

\bibitem{Bolzoni:2012ii}
P.~Bolzoni, B.~A. Kniehl and A.~V. Kotikov, \emph{{Gluon and quark jet
  multiplicities at N$^3$LO+NNLL}},
  \href{http://dx.doi.org/10.1103/PhysRevLett.109.242002}{\emph{Phys. Rev.
  Lett.} {\bf 109} (2012) 242002}, [\href{https://arxiv.org/abs/1209.5914}{{\tt
  1209.5914}}].

\bibitem{BREZIN1973227}
E.~Brezin, J.~{Le Guillou}, J.~Zinn-Justin and B.~Nickel, \emph{Higher order
  contributions to critical exponents},
  \href{http://dx.doi.org/https://doi.org/10.1016/0375-9601(73)90894-3}{\emph{Physics
  Letters A} {\bf 44} (1973) 227--228}.

\bibitem{Derkachov:1997pf}
S.~E. Derkachov, J.~A. Gracey and A.~N. Manashov, \emph{{Four loop anomalous
  dimensions of gradient operators in phi**4 theory}},
  \href{http://dx.doi.org/10.1007/s100520050162}{\emph{Eur. Phys. J. C} {\bf 2}
  (1998) 569--579}, [\href{https://arxiv.org/abs/hep-ph/9705268}{{\tt
  hep-ph/9705268}}].

\bibitem{Henriksson:2022rnm}
J.~Henriksson, \emph{{The critical O(N) CFT: Methods and conformal data}},
  \href{https://arxiv.org/abs/2201.09520}{{\tt 2201.09520}}.

\bibitem{Gromov:2015wca}
N.~Gromov, F.~Levkovich-Maslyuk and G.~Sizov, \emph{{Quantum Spectral Curve and
  the Numerical Solution of the Spectral Problem in AdS5/CFT4}},
  \href{http://dx.doi.org/10.1007/JHEP06(2016)036}{\emph{JHEP} {\bf 06} (2016)
  036}, [\href{https://arxiv.org/abs/1504.06640}{{\tt 1504.06640}}].

\bibitem{Jaroszewicz:1982gr}
T.~Jaroszewicz, \emph{{Gluonic Regge Singularities and Anomalous Dimensions in
  QCD}}, \href{http://dx.doi.org/10.1016/0370-2693(82)90345-8}{\emph{Phys.
  Lett. B} {\bf 116} (1982) 291--294}.

\bibitem{Lipatov:1996ts}
L.~N. Lipatov, \emph{{Small x physics in perturbative QCD}},
  \href{http://dx.doi.org/10.1016/S0370-1573(96)00045-2}{\emph{Phys. Rept.}
  {\bf 286} (1997) 131--198}, [\href{https://arxiv.org/abs/hep-ph/9610276}{{\tt
  hep-ph/9610276}}].

\bibitem{Kotikov:2000pm}
A.~V. Kotikov and L.~N. Lipatov, \emph{{NLO corrections to the BFKL equation in
  QCD and in supersymmetric gauge theories}},
  \href{http://dx.doi.org/10.1016/S0550-3213(00)00329-1}{\emph{Nucl. Phys. B}
  {\bf 582} (2000) 19--43}, [\href{https://arxiv.org/abs/hep-ph/0004008}{{\tt
  hep-ph/0004008}}].

\bibitem{Kotikov:2002ab}
A.~V. Kotikov and L.~N. Lipatov, \emph{{DGLAP and BFKL equations in the $N=4$
  supersymmetric gauge theory}},
  \href{http://dx.doi.org/10.1016/S0550-3213(03)00264-5}{\emph{Nucl. Phys. B}
  {\bf 661} (2003) 19--61}, [\href{https://arxiv.org/abs/hep-ph/0208220}{{\tt
  hep-ph/0208220}}].

\bibitem{Kotikov:2007cy}
A.~V. Kotikov, L.~N. Lipatov, A.~Rej, M.~Staudacher and V.~N. Velizhanin,
  \emph{{Dressing and wrapping}},
  \href{http://dx.doi.org/10.1088/1742-5468/2007/10/P10003}{\emph{J. Stat.
  Mech.} {\bf 0710} (2007) P10003},
  [\href{https://arxiv.org/abs/0704.3586}{{\tt 0704.3586}}].

\bibitem{Caron-Huot:2020ouj}
S.~Caron-Huot, Y.~Gobeil and Z.~Zahraee, \emph{{The leading trajectory in the
  2+1D Ising CFT}},  \href{https://arxiv.org/abs/2007.11647}{{\tt 2007.11647}}.

\bibitem{Li:2021uki}
W.~Li, \emph{{Ising model close to $d=2$}},
  \href{https://arxiv.org/abs/2107.13679}{{\tt 2107.13679}}.

\bibitem{Caron-Huot:2020nem}
S.~Caron-Huot and J.~Sandor, \emph{{Conformal Regge Theory at Finite Boost}},
  \href{http://dx.doi.org/10.1007/JHEP05(2021)059}{\emph{JHEP} {\bf 05} (2021)
  059}, [\href{https://arxiv.org/abs/2008.11759}{{\tt 2008.11759}}].

\bibitem{Alday:2017zzv}
L.~F. Alday, J.~Henriksson and M.~van Loon, \emph{{Taming the
  $\epsilon$-expansion with large spin perturbation theory}},
  \href{http://dx.doi.org/10.1007/JHEP07(2018)131}{\emph{JHEP} {\bf 07} (2018)
  131}, [\href{https://arxiv.org/abs/1712.02314}{{\tt 1712.02314}}].

\bibitem{Kazakov:1979ik}
D.~I. Kazakov, O.~V. Tarasov and A.~A. Vladimirov, \emph{{Calculation of
  Critical Exponents by Quantum Field Theory Methods}}, {\emph{Sov. Phys. JETP}
  {\bf 50} (1979) 521}.

\bibitem{Kologlu:2019bco}
M.~Kolo\u{g}lu, P.~Kravchuk, D.~Simmons-Duffin and A.~Zhiboedov, \emph{{Shocks,
  Superconvergence, and a Stringy Equivalence Principle}},
  \href{https://arxiv.org/abs/1904.05905}{{\tt 1904.05905}}.

\bibitem{Lipatov:1985uk}
L.~N. Lipatov, \emph{{The Bare Pomeron in Quantum Chromodynamics}}, {\emph{Sov.
  Phys. JETP} {\bf 63} (1986) 904--912}.

\bibitem{Basso:2006nk}
B.~Basso and G.~P. Korchemsky, \emph{{Anomalous dimensions of high-spin
  operators beyond the leading order}},
  \href{http://dx.doi.org/10.1016/j.nuclphysb.2007.03.044}{\emph{Nucl. Phys. B}
  {\bf 775} (2007) 1--30}, [\href{https://arxiv.org/abs/hep-th/0612247}{{\tt
  hep-th/0612247}}].

\bibitem{Alday:2015eya}
L.~F. Alday, A.~Bissi and T.~Lukowski, \emph{{Large spin systematics in CFT}},
  \href{http://dx.doi.org/10.1007/JHEP11(2015)101}{\emph{JHEP} {\bf 11} (2015)
  101}, [\href{https://arxiv.org/abs/1502.07707}{{\tt 1502.07707}}].

\bibitem{Simmons-Duffin:2016wlq}
D.~Simmons-Duffin, \emph{{The Lightcone Bootstrap and the Spectrum of the 3d
  Ising CFT}}, \href{http://dx.doi.org/10.1007/JHEP03(2017)086}{\emph{JHEP}
  {\bf 03} (2017) 086}, [\href{https://arxiv.org/abs/1612.08471}{{\tt
  1612.08471}}].

\bibitem{Moch:1999eb}
S.~Moch and J.~A.~M. Vermaseren, \emph{{Deep inelastic structure functions at
  two loops}},
  \href{http://dx.doi.org/10.1016/S0550-3213(00)00045-6}{\emph{Nucl. Phys. B}
  {\bf 573} (2000) 853--907}, [\href{https://arxiv.org/abs/hep-ph/9912355}{{\tt
  hep-ph/9912355}}].

\bibitem{Cornalba:2008qf}
L.~Cornalba, M.~S. Costa and J.~Penedones, \emph{{Eikonal Methods in AdS/CFT:
  BFKL Pomeron at Weak Coupling}},
  \href{http://dx.doi.org/10.1088/1126-6708/2008/06/048}{\emph{JHEP} {\bf 06}
  (2008) 048}, [\href{https://arxiv.org/abs/0801.3002}{{\tt 0801.3002}}].

\bibitem{Derkachov:2002wz}
S.~E. Derkachov, G.~P. Korchemsky, J.~Kotanski and A.~N. Manashov,
  \emph{{Noncompact Heisenberg spin magnets from high-energy QCD. 2.
  Quantization conditions and energy spectrum}},
  \href{http://dx.doi.org/10.1016/S0550-3213(02)00842-8}{\emph{Nucl. Phys. B}
  {\bf 645} (2002) 237--297}, [\href{https://arxiv.org/abs/hep-th/0204124}{{\tt
  hep-th/0204124}}].

\bibitem{Zamolodchikov:1980mb}
A.~B. Zamolodchikov, \emph{{'FISHNET' DIAGRAMS AS A COMPLETELY INTEGRABLE
  SYSTEM}}, \href{http://dx.doi.org/10.1016/0370-2693(80)90547-X}{\emph{Phys.
  Lett. B} {\bf 97} (1980) 63--66}.

\bibitem{Gurdogan:2015csr}
O.~G\"urdo\u{g}an and V.~Kazakov, \emph{{New Integrable 4D Quantum Field
  Theories from Strongly Deformed Planar $\mathcal N = $ 4 Supersymmetric
  Yang-Mills Theory}},
  \href{http://dx.doi.org/10.1103/PhysRevLett.117.201602}{\emph{Phys. Rev.
  Lett.} {\bf 117} (2016) 201602},
  [\href{https://arxiv.org/abs/1512.06704}{{\tt 1512.06704}}].

\bibitem{Gromov:2018hut}
N.~Gromov, V.~Kazakov and G.~Korchemsky, \emph{{Exact Correlation Functions in
  Conformal Fishnet Theory}},
  \href{http://dx.doi.org/10.1007/JHEP08(2019)123}{\emph{JHEP} {\bf 08} (2019)
  123}, [\href{https://arxiv.org/abs/1808.02688}{{\tt 1808.02688}}].

\bibitem{Balitsky:2013npa}
I.~Balitsky, V.~Kazakov and E.~Sobko, \emph{{Two-point correlator of twist-2
  light-ray operators in N=4 SYM in BFKL approximation}},
  \href{https://arxiv.org/abs/1310.3752}{{\tt 1310.3752}}.

\bibitem{Dixon:2018qgp}
L.~J. Dixon, M.-X. Luo, V.~Shtabovenko, T.-Z. Yang and H.~X. Zhu,
  \emph{{Analytical Computation of Energy-Energy Correlation at Next-to-Leading
  Order in QCD}},
  \href{http://dx.doi.org/10.1103/PhysRevLett.120.102001}{\emph{Phys. Rev.
  Lett.} {\bf 120} (2018) 102001},
  [\href{https://arxiv.org/abs/1801.03219}{{\tt 1801.03219}}].

\bibitem{Braun:1999te}
V.~M. Braun, S.~E. Derkachov, G.~P. Korchemsky and A.~N. Manashov,
  \emph{{Baryon distribution amplitudes in QCD}},
  \href{http://dx.doi.org/10.1016/S0550-3213(99)00265-5}{\emph{Nucl. Phys. B}
  {\bf 553} (1999) 355--426}, [\href{https://arxiv.org/abs/hep-ph/9902375}{{\tt
  hep-ph/9902375}}].

\bibitem{Bernstein_Gelfand}
I.~M. Bernstein and S.~I. Gelfand, \emph{{Meromorphy of the function
  $P\sp{\lambda }$}}, {\emph{Funkcional. Anal. i Prilo{\v z}en} {\bf 3} (1969)
  84--85}.

\bibitem{doi:10.1002/cpa.3160230202}
M.~F. Atiyah, \emph{Resolution of singularities and division of distributions},
  \href{http://dx.doi.org/10.1002/cpa.3160230202}{\emph{Communications on Pure
  and Applied Mathematics} {\bf 23} (1970) 145--150}.

\bibitem{Gelfand:105396}
I.~M. Gelfand, G.~E. Shilov, M.~I. Graev, N.~Y. Vilenkin and I.~I.
  Pyatetskii-Shapiro, \emph{{Generalized functions}}.
\newblock AMS Chelsea Publishing. Academic Press, New York, NY, 1964.

\bibitem{Karateev:2018oml}
D.~Karateev, P.~Kravchuk and D.~Simmons-Duffin, \emph{{Harmonic Analysis and
  Mean Field Theory}},  \href{https://arxiv.org/abs/1809.05111}{{\tt
  1809.05111}}.

\bibitem{Dobrev:1977qv}
V.~K. Dobrev, G.~Mack, V.~B. Petkova, S.~G. Petrova and I.~T. Todorov,
  \emph{{Harmonic Analysis on the n-Dimensional Lorentz Group and Its
  Application to Conformal Quantum Field Theory}},
  \href{http://dx.doi.org/10.1007/BFb0009678}{\emph{Lect. Notes Phys.} {\bf 63}
  (1977) 1--280}.

\bibitem{Simmons-Duffin:2017nub}
D.~Simmons-Duffin, D.~Stanford and E.~Witten, \emph{{A spacetime derivation of
  the Lorentzian OPE inversion formula}},
  \href{http://dx.doi.org/10.1007/JHEP07(2018)085}{\emph{JHEP} {\bf 07} (2018)
  085}, [\href{https://arxiv.org/abs/1711.03816}{{\tt 1711.03816}}].

\end{thebibliography}\endgroup

\end{document}